%% file: main.tex
\documentclass[a4paper,11pt]{article}

\input{defs.tex}

\usepackage{jinstpub} 
\usepackage{lineno}
\usepackage{comment}
\usepackage{tablefootnote}
\usepackage{textcomp}

\usepackage{scrextend}
\usepackage{hyperref}

\usepackage{subcaption}
\usepackage{standalone}
\usepackage{dcolumn}
\usepackage{bm}
\usepackage{multirow}
\usepackage{booktabs}
\usepackage{caption}
\usepackage{tablefootnote}
\usepackage{pgfplots}
\pgfplotsset{compat=1.9}
\definecolor{forestgreen}{rgb}{0.0,0.5,0.0}
\usepackage{feynmp}
\usepackage{adjustbox}
\usepackage{float}
\usepackage{orcidlink}
\usepackage{calrsfs}
\usepackage[capitalise]{cleveref}
\DeclareMathAlphabet{\pazocal}{OMS}{zplm}{m}{n}

\usepackage{xspace}

\makeatletter
\newcommand\footnoteref[1]{\protected@xdef\@thefnmark{\ref{#1}}\@footnotemark}
\makeatother

\title{\boldmath Beam-Beam Backgrounds for the Cool Copper Collider}

\author[a,b,1]{Dimitrios Ntounis\orcidlink{0009-0008-1063-5620},\note{Corresponding authors.}}

\author[a]{Laith Gordon\orcidlink{0009-0004-5959-5392},}

\author[c]{Lindsey Gray\orcidlink{0000-0002-6408-4288},}

\author[d]{Elias Mettner\orcidlink{0000-0003-1267-8460},}

\author[b]{Tim Barklow\orcidlink{0000-0002-7709-037X},}

\author[a,b]{Emilio A. Nanni\orcidlink{0000-0002-1900-0778}}

\author[a,b,1]{Caterina Vernieri\orcidlink{0000-0002-0235-1053}}

\affiliation[a]{Stanford University, 450 Jane Stanford Way, Stanford, CA 94305, U.S.A.}
\affiliation[b]{SLAC National Accelerator Laboratory, 2575 Sand Hill Rd., Menlo Park, CA 94025, U.S.A.}
\affiliation[c]{Fermi National Accelerator Laboratory, Kirk Road and Pine Street, Batavia, Illinois 60510, U.S.A.}
\affiliation[d]{University of Wisconsin--Madison, 500 Lincoln Drive, Madison, WI 53706, U.S.A.}
\emailAdd{dntounis@stanford.edu}
\emailAdd{caterina@slac.stanford.edu}

\abstract{In this paper, we present a comprehensive characterization of beam-beam backgrounds for the Cool Copper Collider (\CCCnospace), a proposed linear \ee collider designed for precision Higgs studies at center-of-mass energies of 250 and 550 GeV. Using a simulation pipeline based on the \keyhep framework, we evaluate incoherent pair production and hadron photoproduction backgrounds through the SiD detector for baseline, power-efficiency, and high-luminosity \CCC operating scenarios. The occupancy induced by the beam-beam background is evaluated for each scenario, validating the compatibility of the existing SiD detector design with operations at \CCC without substantial modifications. At the same time, the modular simulation framework and analysis methodology presented in this paper offer a versatile toolkit for background studies in future collider proposals, contributing to a common platform for different machine designs.}


\begin{document}

\maketitle
\flushbottom

\section{Introduction}
\label{sec:intro}

Beam-beam backgrounds pose a fundamental constraint for proposed future $e^+e^-$ colliders. The intense electromagnetic (EM) fields of colliding bunches produce beamstrahlung photons, leading to the production of secondary particles~\cite{Schulte:1996,Yokoya:1991qz}. The dominant mechanisms for $e^+e^-$ colliders in the Higgs factory beam-parameter regime, ordered by magnitude of impact, are $e^+e^-$ incoherent pair creation (IPC) and photon-photon ($\gamma\gamma$) hadron photoproduction (HPP).

Incoherent pairs, arising from collisions of real and/or virtual photons, dominate at moderate energies, producing larger detector occupancies despite being restricted to small transverse momenta (\ptnospace). Photon-photon interactions can also produce quark-antiquark pairs that have significantly broader \pt distributions and hadronize into collimated sprays of hadrons called mini-jets. While less frequent than pair production, these higher-\pt hadronic final states contribute to central calorimeter occupancy and contaminate reconstructed jets from physics processes of interest.

Accurate modeling requires a multi-step simulation chain that includes beam-beam interactions, event generation, and the propagation of background particles through the detector volume, ensuring the proper estimation of the resulting occupancies and impact on detector performance. The future collider software ecosystem based on \keyhepnospace~\cite{Sailer:2020fah} provides a modular, scalable framework for these studies, having previously been validated through end-to-end pipelines combining beam-beam generators, hadronic event generators, and \geant simulations for the International Linear Collider (ILC), the Compact Linear Collider (CLIC) and the Future Circular Collider (FCC)~\cite{Ganis:2021vgv}.

In this paper, we present the methodology used to evaluate beam-beam backgrounds for the Cool Copper Collider (\CCCnospace), with an approach extendable to other $e^+e^-$ collider environments. A precise characterization of these backgrounds is crucial, not only for improving detector performance and ensuring reliable operations, but also for optimizing accelerator parameters to maximize luminosity and physics reach. 

\section{Cool Copper Collider and its beam parameters}
\label{sec:c3}

\CCC is a linear \ee collider concept that utilizes compact, high-gradient, normal-conducting accelerators with distributed coupling, operated at around $ 80 \ \mathrm{K}$~\cite{vernieri2023cool,dasu2022strategy}. 
The design of \CCC follows a comprehensive optimization strategy, integrating considerations of the main linacs, collider subsystems, and beam dynamics to achieve the necessary luminosity at minimal overall cost. It is specifically tailored for a physics program at center-of-mass (CoM) energies of 250 and 550~GeV, enabling precision Higgs studies and measurements of Higgs self-coupling. The \CCC beam and machine parameters are selected to match the luminosity profile of ILC~\cite{ILC_Snowmass,c3_lumi} and the proposed Linear Collider Facility (LCF) at CERN~\cite{LinearColliderVision:2025hlt,LinearCollider:2025lya}. The entire facility spans 8~km, sufficient to support both energy stages. The initial 250~GeV operation can be upgraded to 550~GeV by incorporating additional radio frequency (RF) power sources into the main linac. This upgrade strategy is feasible because the increased gradient and power demand are counterbalanced by modifications to the beam format, ensuring a constant beam-loading fraction and preserving RF efficiency. By refining the cavity geometry, optimizing the RF distribution, and leveraging the enhanced conductivity of copper at liquid nitrogen temperatures, the peak power demands on high-power RF sources are substantially reduced~\cite{nanni2023status}. This enables a high beam-loading fraction, approaching 50\%~\cite{Bane:2018fzj}, resulting in a compact and efficient collider design. Through the implementation of distributed-coupling and cryogenic copper technology, peak RF power requirements are reduced by a factor of six~\cite{Bane:2018fzj,Grudiev2010DesignOT}, comparing a cold distributed coupling structure to the effective shunt impedance of a traveling wave high-gradient linac, significantly improving overall efficiency. 

The \CCC machine parameters have recently been optimized in \cite{c3_lumi,C3_ESPPU} with the purpose of delivering high luminosity at the Interaction Point (IP) without increasing site power. The current optimized parameters for \CCC are listed in Table \ref{tab:C3_params} and consist of a baseline scenario, which delivers the same instantaneous luminosity as ILC~\cite{ILC_Snowmass}, a sustainability-update scenario that maintains luminosity while reducing site power by around $30\%$ by halving the bunch spacing and doubling the number of bunches, as well as a high-luminosity scenario.

The beam parameters at the IP follow the initial Parameter Set 1 (PS1) for the baseline and sustainability-update scenarios at 250 GeV, which is summarized in \cref{tab:C3_beam_params_PS1_PS2}. For the other scenarios, the revised parameter set, referred to as Parameter Set 2 (PS2), is used, which achieves a $\sim40\%$ increase in total luminosity, while ensuring that beamstrahlung-induced backgrounds remain at acceptable levels. The changes with respect to PS1 are a reduction in the vertical emittance $\epsilon_{y}^{*}$ from 20 nm to 12 nm and the introduction of a vertical waist shift ($w_y$) of 80 \textmu m. Additionally, a moderate increase in the horizontal emittance ($\epsilon^*_x$) from 900 nm to 1000 nm effectively controls beam-beam interaction effects, thus maintaining a manageable beam-induced background in the detector. The feasibility of an emittance reduction has been studied through a main-linac beam dynamics analysis in~\cite{Tan:2025rss}. More details on the optimization are given in~\cite{c3_lumi}.

\begin{table}[htbp]
  \centering
  \caption{Machine-level and beam-beam background related parameters for different \CCC operating scenarios: baseline (BL), sustainability update (s.u.), and high-luminosity (high-$\mathcal{L}$) at CoM energies $\sqrt{s}=250$ and 550 GeV. The bunch charge is 1~nC in all cases, and the crossing angle is 14~mrad, compensated by crab crossing~\cite{adolphsen2013internationallinearcollidertechnical}. For the beam-beam related quantities, the photon-photon luminosity $\mathcal{L}_{\gamma \gamma}$ per bunch crossing (BX) is given, as well as the average number $N_{\mathrm{IPC}}$ of IPC particles produced and the average number $N_{\mathrm{HPP}}$ of HPP events per BX and for an entire bunch train. Each HPP event includes, on average, 7 particles at 250 GeV, rising to 16 particles at 550 GeV.}
  \label{tab:C3_params}
  \begin{tabular}{l ccc ccc}
    \toprule
    & \multicolumn{3}{c}{\CCCtwo} & \multicolumn{3}{c}{\CCCfive}\\
    \cmidrule(r){2-4}\cmidrule(l){5-7}
    Scenario & BL & s.\,u.\ & high‑$\mathcal{L}$ & BL & s.\,u.\ & high‑$\mathcal{L}$\\
    \midrule
    Gradient [MeV/m]               & \multicolumn{3}{c}{70} & \multicolumn{3}{c}{120}\\
    Bunches / train                & 133 & 266 & 532 & 75 & 150 & 300\\
    Rep.\ rate [Hz]                & 120 & 60 & 120 & 120 & 60 & 60\\
    Bunch spacing [ns]             & 5.26 & 2.63 & 2.63 & 3.50 & 1.75 & 1.75\\
    Luminosity [$10^{34}\,\mathrm{cm^{-2}s^{-1}}$]
                                   & 1.3 & 1.3 & 7.6 & 2.4 & 2.4 & 4.8\\
    Site power [MW]                & ${\sim}150$ & ${\sim}110$ & ${\sim}180$
                                   & ${\sim}175$ & ${\sim}125$ & ${\sim}180$\\
    Beam parameter set & \multicolumn{2}{c}{PS1} & PS2 &  \multicolumn{3}{c}{PS2}  \\ 
    \midrule

    $\mathcal{L_{\gamma \gamma}}/\mathrm{BX}$ [\textmu b$^{-1}$]      & \multicolumn{2}{c}{0.20} & 0.23 & \multicolumn{3}{c}{0.95}\\    
    $N_{\mathrm{IPC}}/\mathrm{BX}$ [$10^{4}$]               & \multicolumn{2}{c}{4.7} & 5.9 & \multicolumn{3}{c}{15.5}\\
    $N_{\mathrm{HPP}}/\mathrm{BX}$                & \multicolumn{2}{c}{0.059} & 0.065 & \multicolumn{3}{c}{0.29}\\
    $N_{\mathrm{IPC}}/\mathrm{train}$ [$10^{6}$]              & 6.3 & 12.5 & 25.0 & 11.6 & 23.3 & 46.5\\
    $N_{\mathrm{HPP}}/\mathrm{train}$                & 7.8 & 15.7 & 34.6 & 21.8 & 43.5 & 87.0\\

    \bottomrule
  \end{tabular}
\end{table}

\begin{table}[htbp]
  \centering
  \caption{Main parameters at the IP for the two C$^{3}$ beam-dynamics working points. Adapted from~\cite{c3_lumi}.}
    \label{tab:C3_beam_params_PS1_PS2}
  \begin{tabular}{lccrr}
    \toprule
    Beam Parameter & Symbol & Unit & PS1 & PS2 \\[2pt]
    \midrule
    RMS bunch length           & $\sigma_{z}^{*}$ & \textmu m  & \multicolumn{2}{c}{100} \\
    Horizontal beta function           & $\beta_{x}^{*}$ &  mm  & \multicolumn{2}{c}{12}\\
    Vertical beta function           & $\beta_{y}^{*}$ & mm  & \multicolumn{2}{c}{0.12} \\

    Vertical waist shift       & $w_{y}$      & \textmu m & 0   & 80  \\
    Norm.\ horiz.\ emittance   & $\varepsilon_{x}^{*}$ & nm & 900  & 1000 \\
    Norm.\ vert.\ emittance    & $\varepsilon_{y}^{*}$ & nm & 20   & 12  \\
    \bottomrule
  \end{tabular}
\end{table}

\section{Simulation framework}
\label{sec:sim_pipeline}

\begin{figure}[htbp]
  \centering
  \hspace*{-13mm}
  \includegraphics[width=1.1\linewidth]{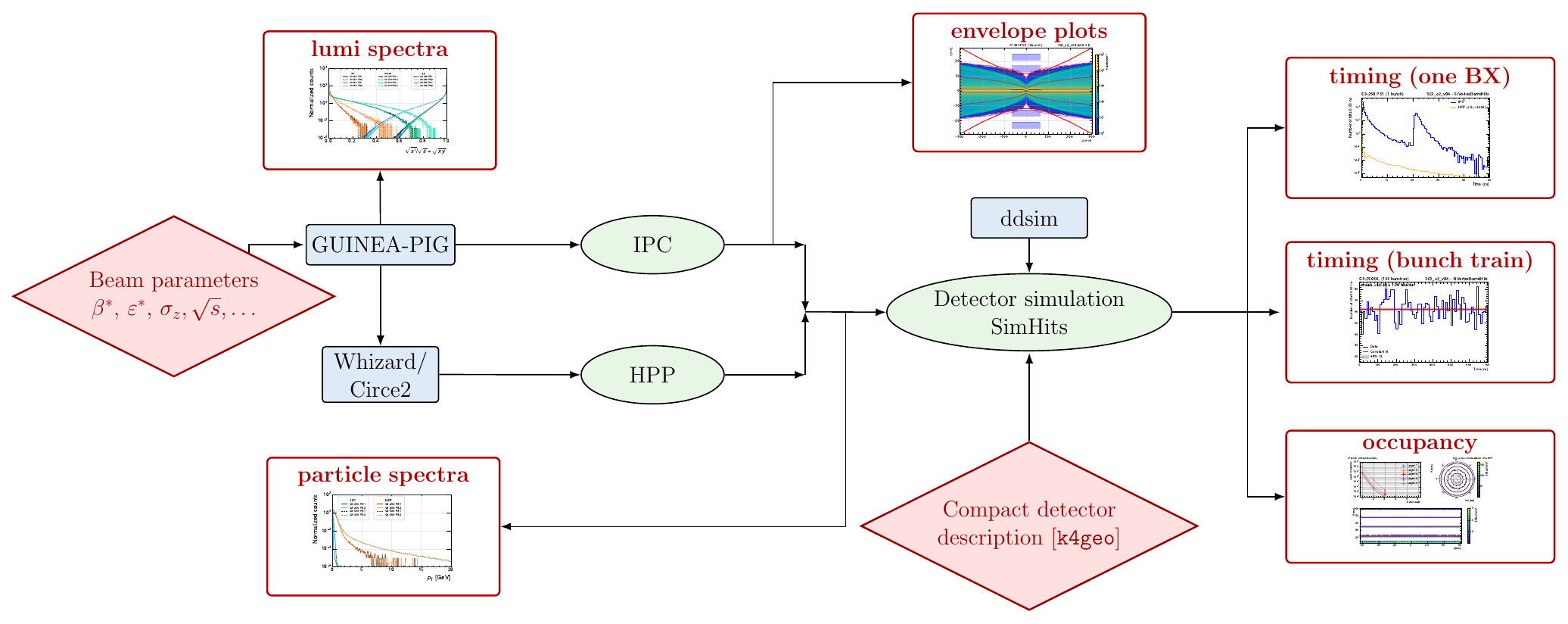}
  \caption{Simulation pipeline used in this paper from the background generation to the detector hits simulation. Nodes in red, blue and green represent simulation inputs, tools and outputs, respectively. Typical diagnostic plots obtained at various stages of the pipeline are also provided for reference.}
  \label{fig:pipeline_flowchart}
\end{figure}

Our simulation pipeline integrates established tools into a reproducible end-to-end chain. Beam-beam interactions are simulated using \GPnospace~\cite{Schulte:1996,Schulte:1999} or its modern C++ variant \GPPPnospace~\cite{Rimbault:2007wfy}, capturing non-linear pinch effects, beamstrahlung, and coherent/incoherent processes. Photon-induced hadronic backgrounds are generated by interfacing with event generators such as \whiznospace~\cite{Kilian:2007gr} and \pythianospace~\cite{Sj_strand_2006}, using photon spectra obtained from \circenospace~\cite{Ohl:1996fi}. The detector description is performed with \ddhepnospace~\cite{Frank:2014zya}, which is interfaced with \geantnospace~\cite{Agostinelli:2002hh} for the full propagation of particles through the detector volume. Using the \texttt{ddsim}~\cite{Petric:2017psf} toolkit, we obtain the simulated hits (SimHits) that follow the \edmhep~\cite{Gaede:2021izq} Event Data Model. Figure~\ref{fig:pipeline_flowchart} summarizes the simulation workflow and the diagnostics we retain at each stage.

This workflow is parameter–driven and modular. Beam parameters, such as energy, bunch size, and charge, enter through a single configuration, and the beam–beam stage may accept multiple inputs: while we use \GP as default, alternative codes such as \textsc{CAIN}~\cite{CAIN} and \textsc{WARPX}~\cite{warpx} can be accommodated with minimal adjustments. On the detector side, the pipeline works with any \ddhep geometry -- \texttt{k4geo} models for SiD, ILD, CLD, IDEA, ALLEGRO, and other detector concepts are available~\cite{k4geo}; hits are persisted in \edmhep so downstream timing and occupancy studies remain identical across detector choices and beam parameter scans.

All steering cards, interface scripts, and post-processing used in this work have been made available online to facilitate similar future studies by the community. See ``Code availability'' for further information.

\subsection{Detector simulation}
\label{subsec:detector_sim}

For the studies presented in this work, the SiD detector concept~\cite{Aihara:2009ad}, originally developed for the ILC, is used. SiD features all-silicon vertex and tracker systems, a silicon-tungsten electromagnetic calorimeter (ECAL), and a scintillator-steel hadronic calorimeter (HCAL), all enclosed within a $5\,$T solenoid and muon spectrometers interleaved with the flux return steel. It also includes two forward calorimetry systems, the luminosity calorimeter (LumiCal) and the beam calorimeter (BeamCal), intended for luminosity and beam condition monitoring. The main SiD parameters are given in~\cref{app:SiD}.

For the purposes of this study, we have assumed the \texttt{SiD\_o2\_v04} geometry description, as implemented in \texttt{k4geo}~\cite{k4geo_sid_o2_v04}. This description keeps geometry, materials, and sensitive surfaces in a single hierarchy and guarantees bit‑identical simulation between fast prototyping studies and the final digitization/reconstruction chain. The machine-detector interface (MDI) region, including the beampipe and forward shielding, is an integral part of the \texttt{k4geo} geometry description and is simulated together with all other detector components. The beampipe geometry follows the SiD baseline design~\cite{behnke2013internationallinearcollidertechnical} and consists of a thin-walled central beryllium cylinder that transitions to conical sections extending toward the forward calorimetry. The central section, with an inner radius of 12~mm, is positioned to clear the IPC background envelopes shown in~\cref{fig:envelope_plots}.

To study the effect on the detector occupancy, two critical parameters must additionally be defined: the cell sizes in each subdetector and the energy thresholds to count hits. These are given in~\cref{tab:SiD_thresholds_cell_size}. 

For the cell sizes, notably, silicon pixels with 10 \textmu m pitch based on the Monolithic Active Pixel Sensor (MAPS) technology are assumed for the vertex, whereas for the tracker and ECAL, large-area MAPS are also considered. Simulation studies for the SiD ECAL~\cite{Brau:2022sxr} demonstrate the superior EM energy resolution and shower separation of MAPS compared to the target in the ILC Technical Design Report~\cite{behnke2013internationallinearcollidertechnical}. Significant progress has been made in the development of MAPS sensors for applications in vertex and tracking detectors, as exemplified by the ALICE detector upgrades~\cite{ALICE:2023udb, Groettvik:2024onw}.

\begin{table}[h]
    \centering

    \caption{Assumed cell size, sensor thickness, and energy threshold for the various SiD subdetectors. }
    \label{tab:SiD_thresholds_cell_size}

\resizebox{\textwidth}{!}{
\begin{tabular}{cccccc}
    
\toprule  \textbf{Barrel Detector} & cell size [mm$^2$] & sensor thickness [mm] & $E_{\text{MIP}}$ (mode) [MeV] & $E_{\text{thr}}$ [MeV] \\
\midrule Vertex  & $0.010 \times 0.010$  & 0.02   & 0.0058  & 0.0006 \\
Tracker & $0.025 \times 0.100$  & 0.30\tablefootnote{\label{2ndtablefoot}The tracker sensor thickness is currently fixed in the \texttt{k4geo} simulation geometry. For MAPS-based implementations, thinner sensors ($<100$ \textmu m) would be used in practice.}  & 0.092 & 0.03  \\
ECAL & $0.025 \times 0.100$   & 0.32  & 0.11 & 0.05 \\
HCAL & $30 \times 30$ & 3.00  &  0.49& 0.24  \\
Muon system &  $L \times 41$\tablefootnote{\label{1sttablefoot}The muon system employs two orthogonal strip planes per layer with pitch 41mm and length $L\!\approx\!5.5$ m ($z$ view) and $L\!\approx\!2.9$–$4.7$ m ($r-\phi$ view) in the barrel, and $L\!\approx\!1.8$ m (vertical) and $L\!\approx\!5.5$ m (horizontal) in the endcap.} & 3.00 & 0.52 & 0.005  \\
\toprule  \textbf{Endcap Detector} &  &  &  & \\
\midrule Vertex Endcap & $0.010 \times 0.010$  & 0.02  & 0.0047 & 0.0005\\
Vertex Forward & $0.010 \times 0.010$ & 0.02  & 0.0044 & 0.0004 \\
Tracker & $0.025 \times 0.100$  & 0.30   & 0.087 & 0.03 \\
ECAL & $0.025 \times 0.100$   & 0.32  & 0.097& 0.05 \\
HCAL & $30 \times 30$  & 3.00 & 0.46 & 0.23 \\
Muon system & $L \times 41$\textsuperscript{\ref{1sttablefoot}}  & 3.00 &  0.47 & 0.005 \\
LumiCal & $2.5 \times 2.5$  & 0.32  &  0.088  & 0.04 \\
BeamCal & $5.0 \times 5.0$  & 0.32  &  0.092& 0.05 \\
\bottomrule
\end{tabular}%
}

\end{table}

 For each subdetector, we determine the characteristic energy deposit from a minimum-ionizing particle (MIP) by computing the mode of the energy-loss distribution for single MIPs traversing the sensitive silicon or scintillator layer of thickness given in \cref{tab:SiD_thresholds_cell_size}. The adopted threshold $E_{\text{thr}}$ is then set to $1-50\%$ of this modal value $E_{\text{MIP}}$, providing a comfortable signal-to-noise margin while retaining ${\gtrsim}99\%$ hit efficiency for MIPs. The MIP energy losses and corresponding thresholds are given in \cref{tab:SiD_thresholds_cell_size}. Silicon systems, namely the vertex, tracker, ECAL, and forward calorimeters, have thresholds clustered around 0.5 keV---or  $0.1$ MIP---for the vertex pixels and 30-50 keV---or $0.3$–$0.5$ MIP---for the thicker $300$ \textmu m pixels. For the HCAL tiles, thresholds up to 240 keV---corresponding to  $0.5$ MIP---are used. Finally, for the outermost muon system, a threshold of $0.01$ MIP is assumed, since it experiences much lower particle fluxes, enabling operations at lower thresholds without becoming overwhelmed by noise hits.

\begin{figure}[h]
    \centering
    \includegraphics[width=\linewidth]{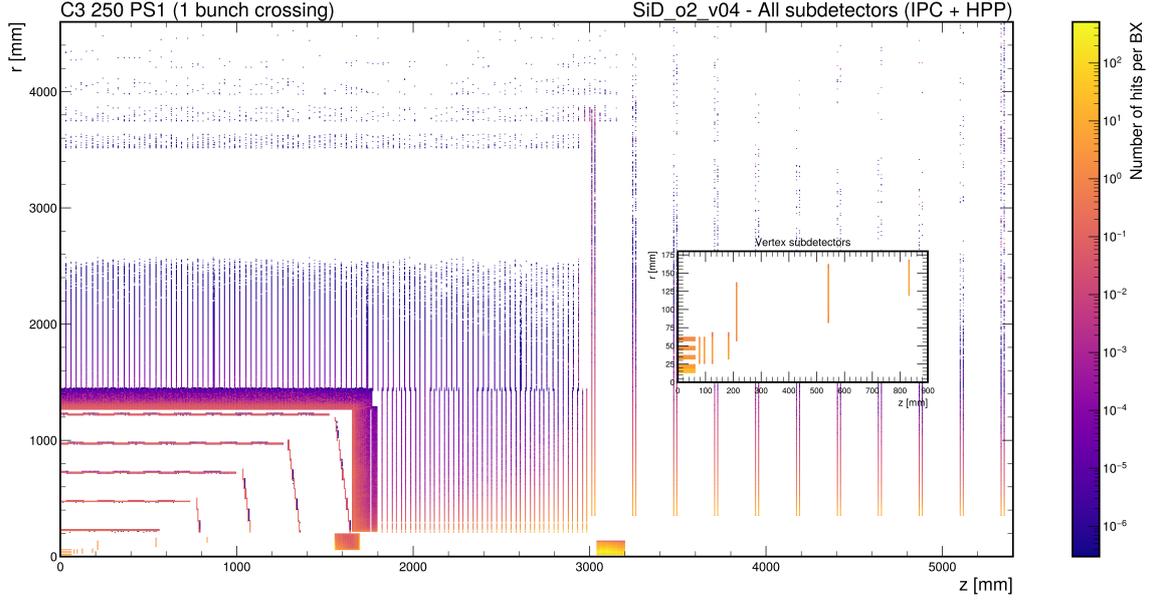}
    \caption{Quarter view of the SiD detector in the $r-z$ plane showing the expected number of hits from the IPC and HPP backgrounds in each subdetector system per bunch crossing, assuming the \CCCtwo PS1 parameter scenario. The inset provides an enlarged view of the vertex region.}
    \label{fig:full_det_view}
\end{figure}

A depiction of the entire SiD detector is given in~\cref{fig:full_det_view}, which shows the $r-z$ distribution of the expected number of hits from background particles coming from the IPC and HPP processes for one bunch crossing of the \CCCtwo PS1 beam scenario. As can be seen, larger hit fluxes are observed in the endcap, compared to the barrel, subdetectors, indicating the forward production nature of the majority of these background particles.

\section{Beam-beam backgrounds}
\label{sec:BS_BBB}

\subsection{Beamstrahlung}
\label{subsec:BS}

The EM interactions between the high-charge-density colliding bunches at \CCC generate intense synchrotron-like radiation, called beamstrahlung. This process not only degrades the nominal CoM energy but also serves as the primary source of photons that drive subsequent background production mechanisms.

The key quantity for the characterization of beamstrahlung is the parameter

\begin{equation}
    \langle \Upsilon \rangle = \frac{5}{6} \frac{N_{e} r_{e}^{2} \gamma}{\alpha (\sigma_{x}^{*}+\sigma_{y}^{*})\sigma_{z}^{*}} 
    \label{eq:upsilon}
\end{equation}

\noindent where $N_{e}$ is the bunch population, $\gamma$ is the Lorentz factor of the beam particles, and $\sigma_{x,y}^{*}$ and $\sigma_{z}^{*}$ are the transverse RMS bunch sizes and the bunch length, respectively. It expresses the average field strength experienced by beam particles in units of the Schwinger critical field and is a crucial parameter that must be kept at as low levels as possible in order to suppress photon production at the IP. For \CCCnospace, $\langle \Upsilon \rangle$ takes values of about 0.06 at 250 GeV and 0.20 at 550 GeV, indicating that beamstrahlung is significant but still below the regime where quantum corrections dominate. The average fractional energy loss is roughly 3\% at 250 GeV and increases to 9\% at 550 GeV~\cite{c3_lumi}, while the luminosity spectrum remains sufficiently narrow as required for precision physics measurements.

\begin{figure}[htbp]
    \centering
    \includegraphics[width=\linewidth]{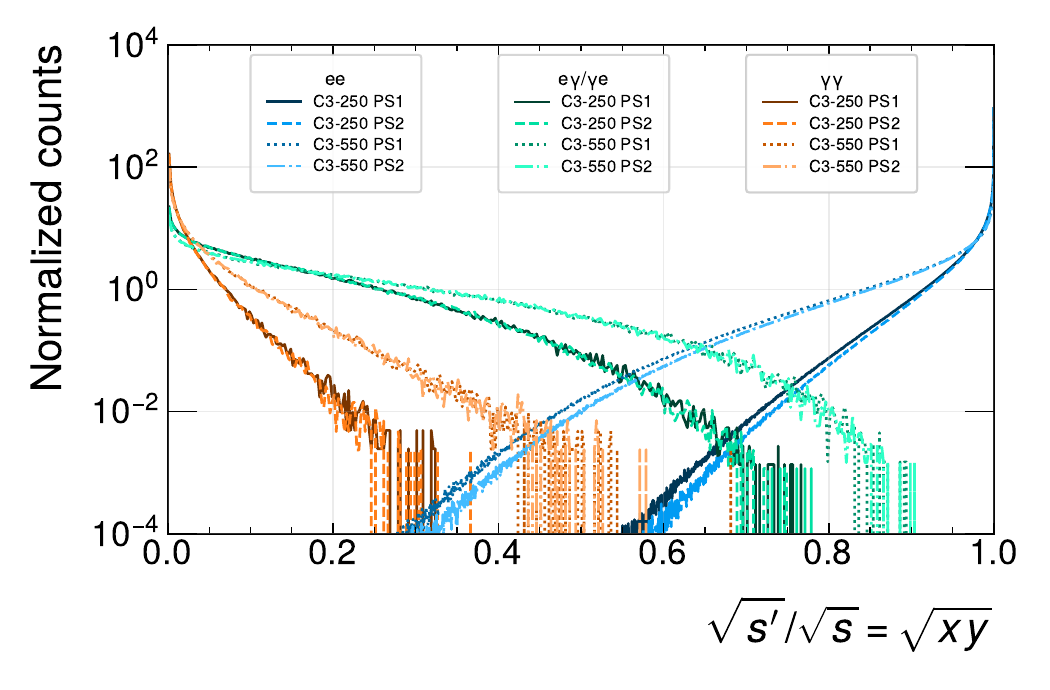}
    \caption{Normalized luminosity spectra of the relative center-of-mass energy $\sqrt{s'}/\sqrt{s}=\sqrt{xy}$ where $x,y$ the relevant energy fractions compared to the nominal beam energy of the colliding particles in the electron and positron beam, respectively,  for $e^{+}e^{-}$, $e^{-}\gamma/\gamma e^{+}$, and $\gamma \gamma$ collisions for the four different \CCC parameter sets of~\cref{tab:C3_beam_params_PS1_PS2}.}
    \label{fig:lumi_spectra}
\end{figure}

The effect of beamstrahlung on the luminosity spectra is computed using \GP simulations that fully account for beam-beam effects, and is shown in \cref{fig:lumi_spectra} for the various \CCC operating scenarios. In this figure, the differential luminosity spectra are shown as a function of the CoM energy $\sqrt{s'}$ of the colliding particles (electrons, positrons, or photons) normalized to the nominal CoM energy $\sqrt{s}$ of the beams (250 or 550 GeV). The spectra for $e^{+}e^{-}$ collisions exhibit the characteristic peak near the nominal CoM energy with an asymmetric tail extending to lower energies due to beamstrahlung losses. At 250 GeV, about $70\%$ of the luminosity remains within $1 \%$ of the nominal energy, decreasing to $50 \%$ at 550 GeV due to the increased beam-beam effects at higher energies. In contrast, the spectra for $\gamma \gamma$ and $e^{\pm}\gamma$ collisions peak at zero, with sharper distributions at 250 GeV than at 550 GeV.

\subsection{Incoherent pair production}
\label{subsec:incoh_pairs}

IPC is the primary beam–beam background at \CCCnospace, arising from quantum electrodynamic (QED) scatterings between beamstrahlung photons and virtual photons from the opposing bunch, rather than from the collective bunch field. Note that, additional beamstrahlung-induced processes, such as trident cascades and coherent pair production, have a negligible cross-section for \CCC parameters and are not further analyzed in this study~\cite{c3_lumi,Esberg:2014}. IPC consists of three main subprocesses, ordered by relative contribution~\cite{Rimbault:2006,Esberg:2014}: (i) Bethe–Heitler (BH), $\gamma\gamma^\ast\to e^+e^-$; (ii) linear Breit–Wheeler (BW), $\gamma\gamma\to e^+e^-$; and (iii) Landau–Lifshitz (LL), $\gamma^\ast\gamma^\ast\to e^+e^-$.

The IPC background is simulated using the Particle-In-Cell code \GPPPnospace. Beams are represented by macro-particles that are longitudinally sliced and distributed onto a 2D mesh of cells; the code advances the bunches in time-steps, deposits the macro-particle charges onto the mesh, solves the fields on the cell nodes, and pushes the particles forward accordingly. During the collision, \GPPPnospace: (i) emits beamstrahlung photons, (ii) constructs equivalent-photon fluxes, and (iii) generates $e^+e^-$ pairs from each of the three incoherent subprocesses~\cite{Rimbault:2006,Esberg:2014}.

The effective cross section for the sum of the three processes, taking into account virtuality and beam size effects, is $\mathcal{O}(10-100) \ \mathrm{mb}$~\cite{Rimbault:2006}, leading to $\mathcal{O}(10^{4}-10^{5})$ $e^\pm$ per bunch crossing. The exact values for each \CCC scenario are given in~\cref{tab:C3_params}.

\begin{figure}[h]
            \centering
            \begin{subfigure}[b]{0.497\textwidth}
                \centering
                \includegraphics[width=1.01\textwidth]{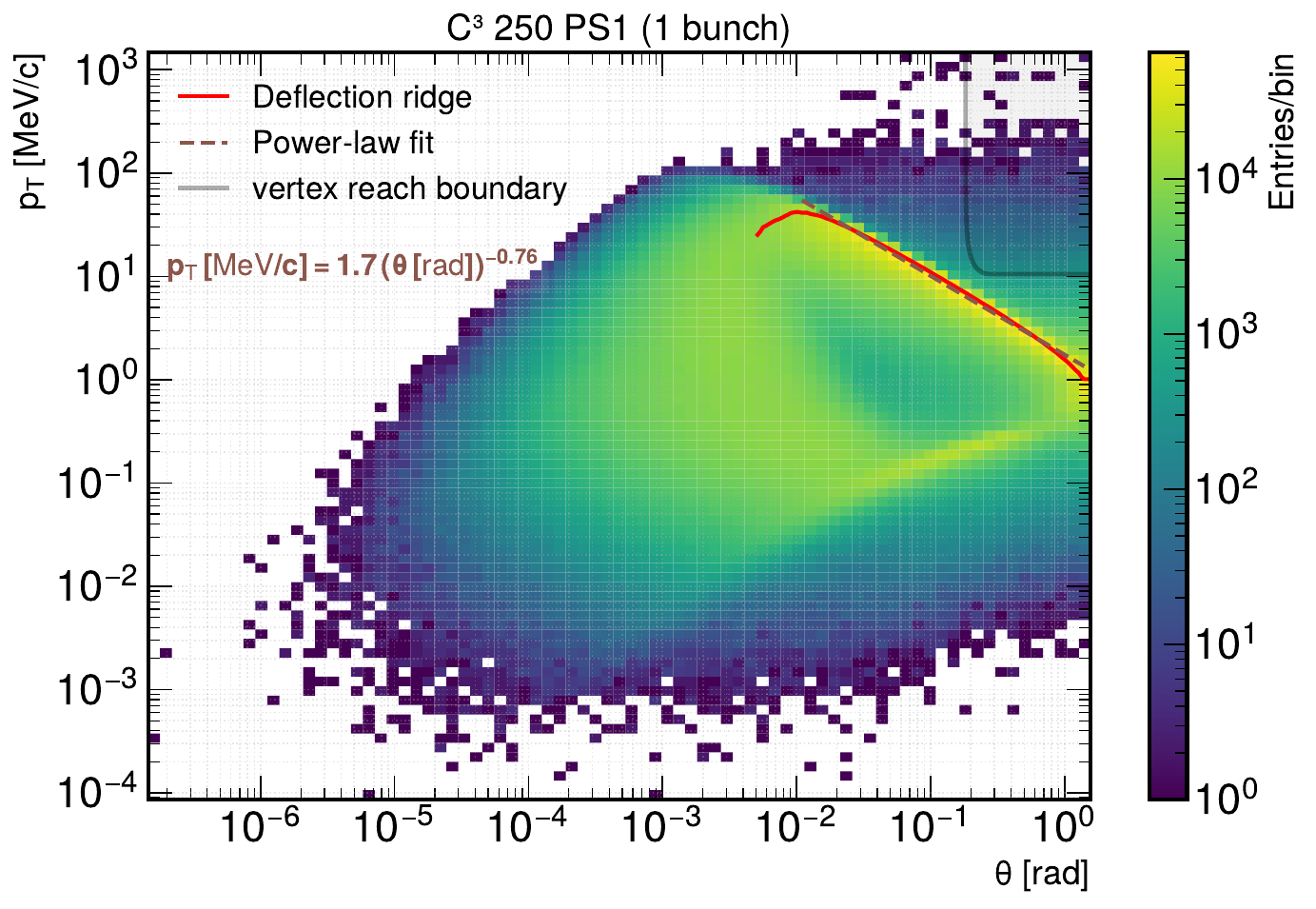}
                \caption{}
                \label{fig:C3_250_PS1_deflection}
            \end{subfigure}
            \hfill
            \begin{subfigure}[b]{0.497\textwidth}
                \centering
                \includegraphics[width=1.01\textwidth]{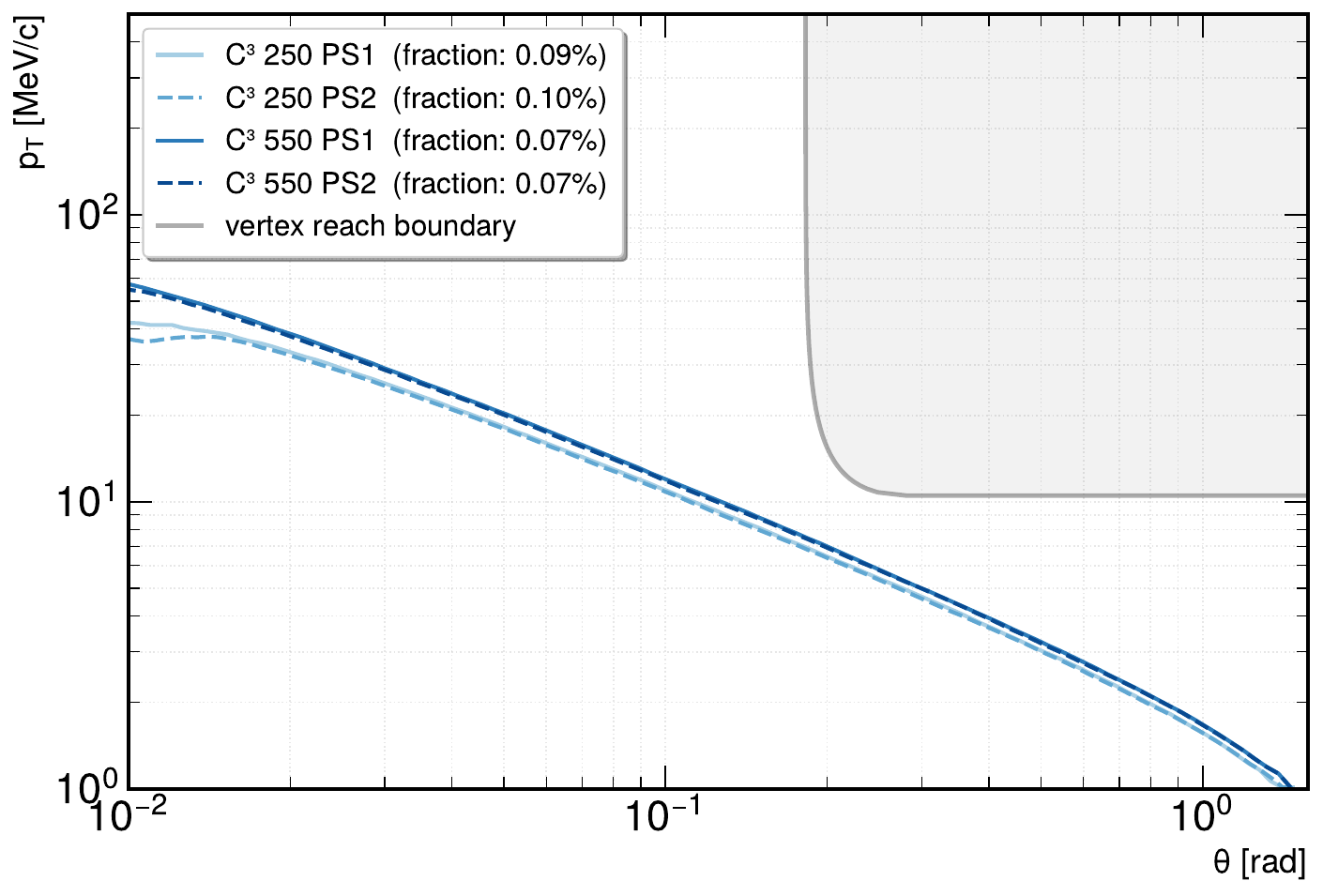}
                \caption{}
                \label{fig:deflection_all}
            \end{subfigure}
                \caption{ (a) Distribution of the IPC particles for \CCCtwo PS1 in the $p_{\mathrm{T}}-\theta$ plane. The accumulation zone due to the beam-beam deflection effect is highlighted in red and is fitted with a power law in brown. (b) Deflection zones for all \CCC beam parameter sets, restricted to larger values of $p_{\mathrm{T}}$ and $\theta$. The grey area represents the $p_{\mathrm{T}}$ and $\theta$ values necessary for the particles to reach the first vertex barrel layer. Information on how this area is derived is given in~\cref{app:reachability}.}
    \label{fig:IPC_deflection}
\end{figure}

The IPC background is mainly concentrated at small angles $\theta$ relative to the beam axis due to the relativistic boost. A small fraction of particles, however, is produced at sufficiently large $\theta$ and with sufficient transverse momenta \pt to escape the beam pipe and reach sensitive detector components, potentially impacting detector and electronics design. \cref{fig:C3_250_PS1_deflection} shows the distribution of IPC particles in the $p_{\mathrm{T}}-\theta$ plane, with two distinct regions: a low $p_{\mathrm{T}}$ region, caused by focusing from oppositely charged beam particles, and a high $p_{\mathrm{T}}$ region, where same-charge beam particles deflect the IPC pairs. This second region, known as the \emph{deflection zone}, should be kept beyond the reach of the innermost SiD detector layer. Indeed, as is verified in~\cref{fig:deflection_all}, assuming a 5 T solenoid field and a radial distance of 14 mm of the first vertex barrel layer from the IP, the deflection zone does not reach the vertex detector for any \CCC beam scenario, with fewer than one in a thousand IPC particles reaching the detector.

\begin{figure}[htbp]
    \makebox[\textwidth][l]{%
        \hspace*{-0.8cm}%
        \begin{minipage}{1.05\textwidth}
            \centering
            \begin{subfigure}[b]{0.48\textwidth}
                \centering
                \includegraphics[width=\textwidth]{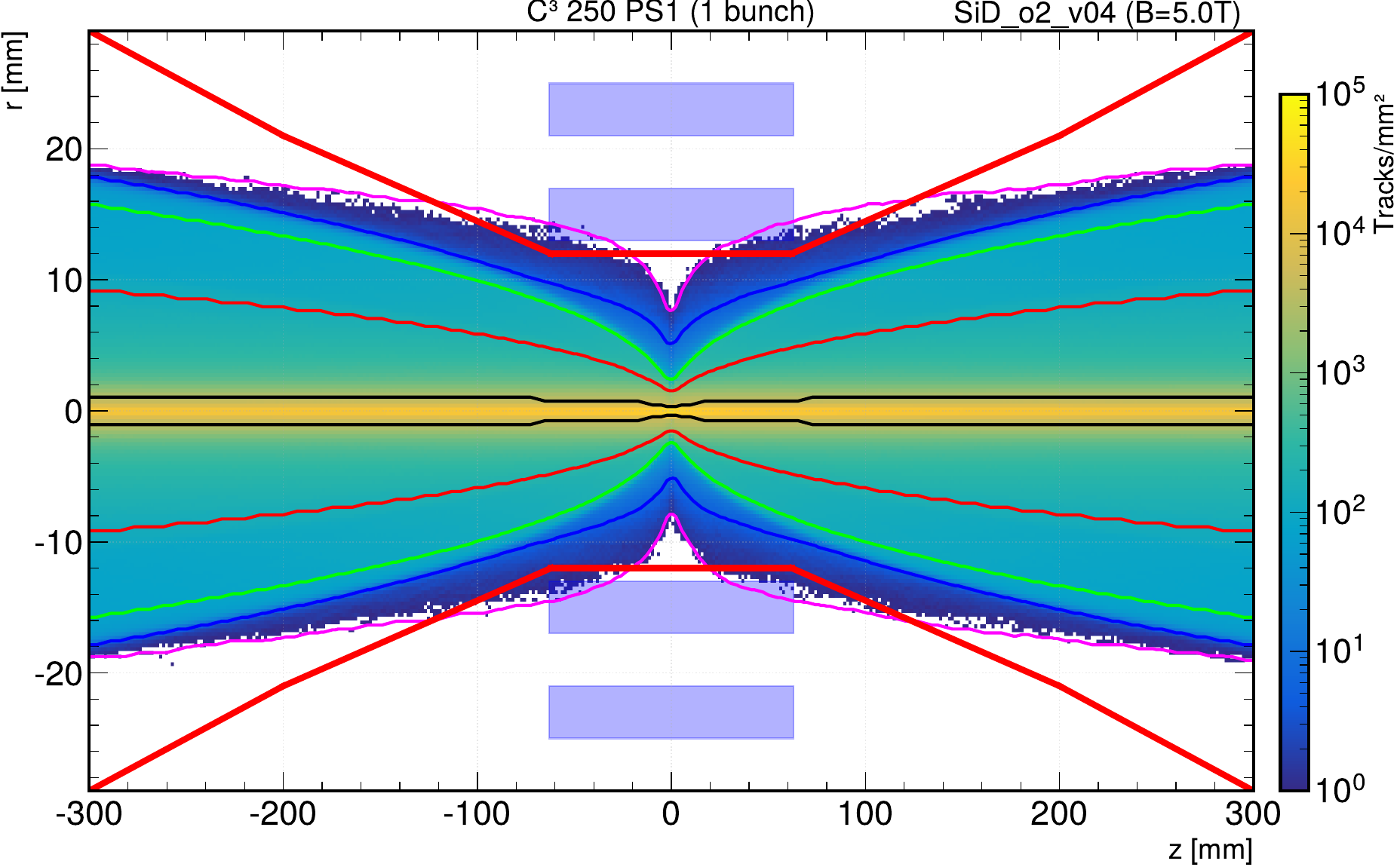}
                \caption{}
                \label{fig:C3_250_PS1}
            \end{subfigure}
            \hfill
            \begin{subfigure}[b]{0.48\textwidth}
                \centering
                \includegraphics[width=1.25\textwidth]{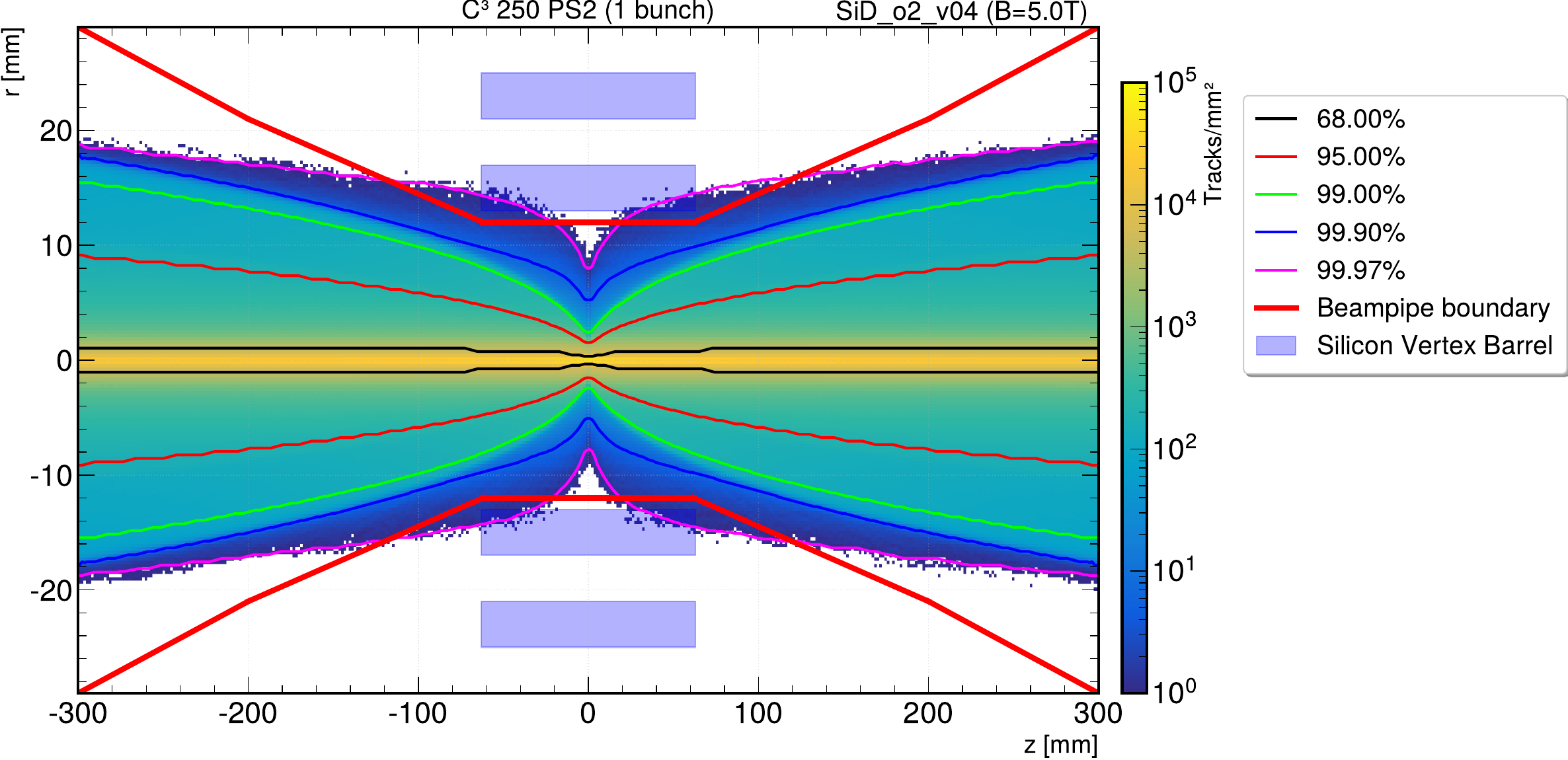}
                \caption{}
                \label{fig:C3_250_PS2}
            \end{subfigure}
            \hfill
            \begin{subfigure}[b]{0.48\textwidth}
                \centering
                \includegraphics[width=\textwidth]{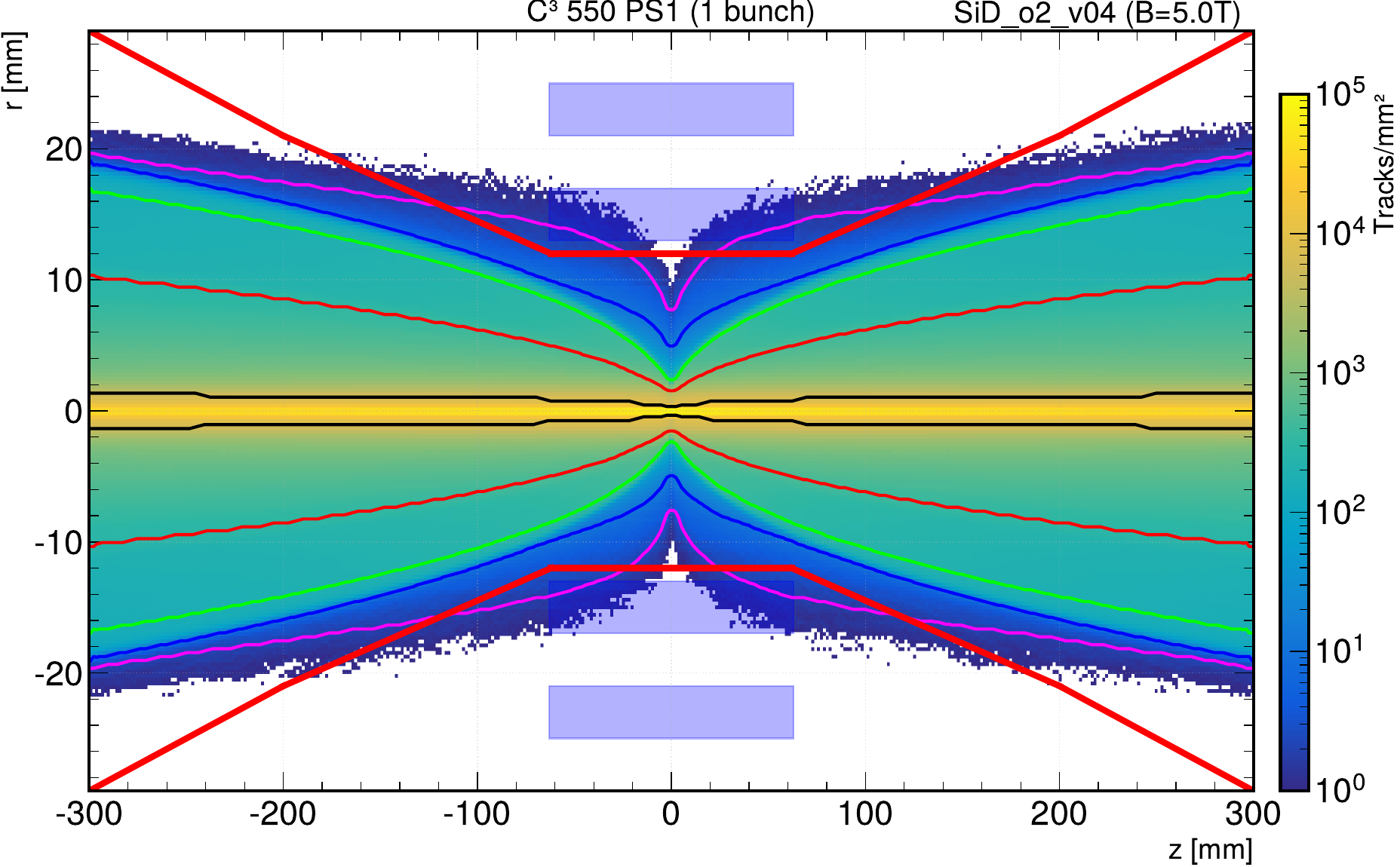}
                \caption{}
                \label{fig:C3_550_PS1}
            \end{subfigure}
            \hfill
            \begin{subfigure}[b]{0.48\textwidth}
                \centering
                \includegraphics[width=\textwidth]{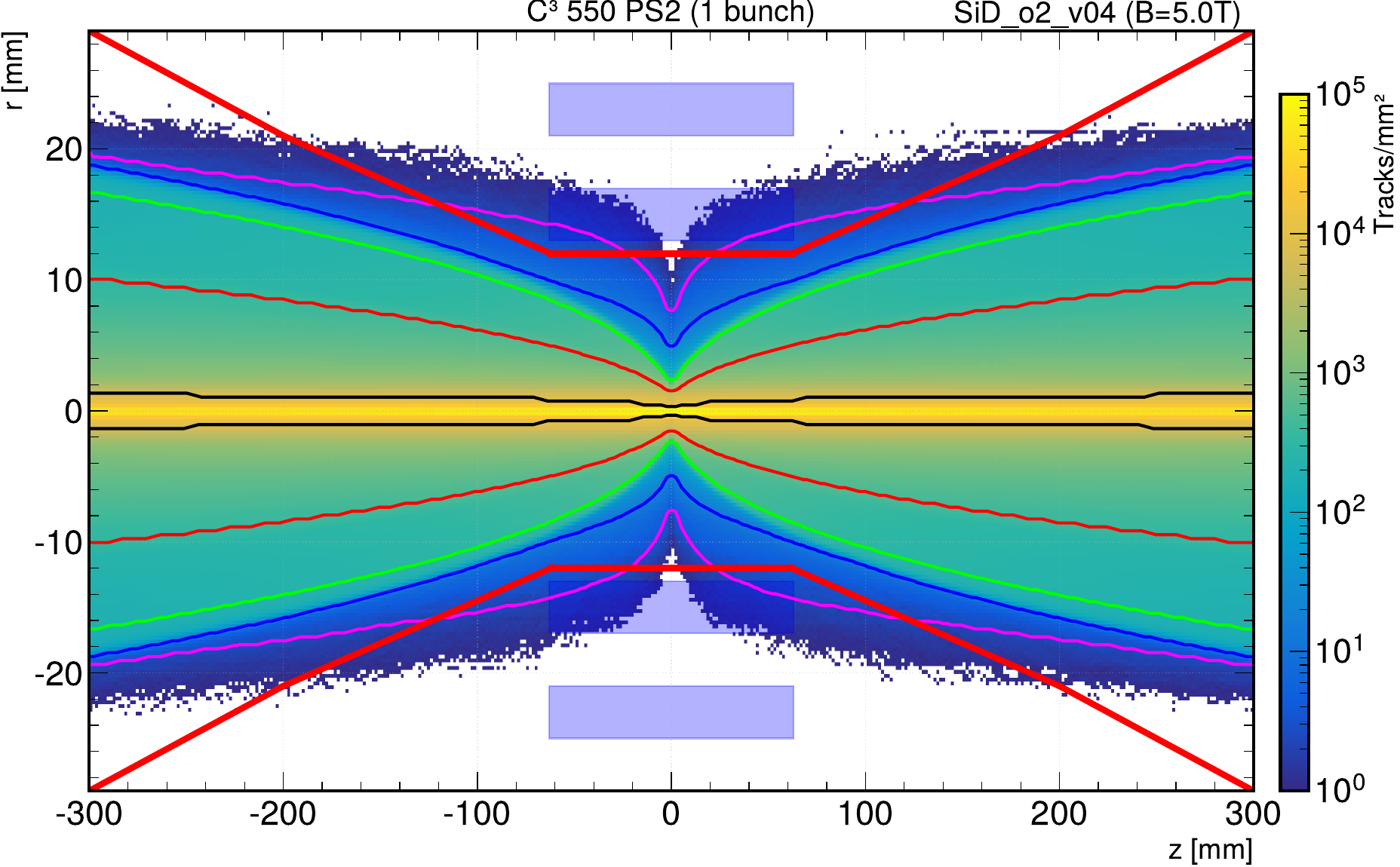}
                \caption{}
                \label{fig:C3_550_PS2}
            \end{subfigure}
        \end{minipage}%
    }
    \caption{Lab-frame $r-z$ distributions, and corresponding envelopes at various percentiles, of the trajectories of the IPC particles under the influence of the detector solenoid magnetic field for \CCC operating at 250 GeV (top) and 550 GeV (bottom) and for PS1 (left) and PS2 (right) beam parameter configurations.}
    \label{fig:envelope_plots}
\end{figure}

An alternative way to visualize the impact of this background on the vertex system is through the ``envelope plots'' in~\cref{fig:envelope_plots}, which show the helical trajectories of background electrons and positrons as they move outward from the IP under the strong solenoid magnetic field. We notice that the envelopes extend towards larger radii as the CoM energy increases from 250 to 550 GeV. However, for all beam scenarios, the envelope containing $99.9\%$ of these particles remains within the beam pipe, meaning that the vast majority of particles do not reach the vertex detector.

\subsection{Hadron photoproduction}
\label{subsec:hadron_photo}

Beyond the dominant IPC background, photon-photon interactions at \CCC generate hadronic final states through quark pair production and subsequent hadronization. Although subdominant in  cross section --- about 0.3 \textmu b --- HPP produces a more centrally distributed hadronic activity that can significantly impact calorimeter performance.
Hadron photoproduction events are generated by using  the parameterized \circetwo photon spectra to provide initial momenta to a dedicated $\gamma\gamma\!\to$ hadrons model in \whizthreenospace~\cite{Stienemeier:2021}. The implementation follows the Barklow–Chen–Peskin prescription of separating soft and jet-like components at low hadronic invariant mass, tuned with PETRA and LEP data, and then using perturbative quantum chromodynamics and a hadronization model at higher masses~\cite{Chen:1993}. The diphoton CoM energy at which this transition occurs is configurable and set to $\sqrt{s_{\gamma\gamma}}=3 \ \mathrm{GeV}$. Concretely, for  $\sqrt{s_{\gamma\gamma}}$ values below this threshold, we use a model constrained to historical $e^+e^-$/two-photon measurements, while for larger values of $\sqrt{s_{\gamma\gamma}}$ events are showered and hadronized with \pythia 6.4~\cite{Sj_strand_2006}, preserving energy-momentum and flavor correlations across the transition~\cite{Bierlich:2022}. 

The $\gamma \gamma$ luminosity spectrum at \CCCnospace, as extracted from \GPPPnospace, is parameterized in \circetwo~\cite{Ohl:1996fi} as a function of the energy fractions $x,y$ of the colliding photons originating from the electron and positron beams, respectively. This approach captures the full spectrum of virtual and real photon interactions, including beamstrahlung-enhanced contributions at low photon energies. The differential photon luminosity as a function of the photon-photon CoM energy, normalized to the nominal beam center-of-mass energy $\sqrt{s_{\gamma \gamma }}/\sqrt{s}$, is shown in~\cref{fig:lumi_spectra} for different \CCC beam parameter sets at both $\sqrt{s}=$ 250 and 550 GeV.

These luminosity spectra are integrated with the fixed-energy  $\gamma \gamma \rightarrow \mathrm{hadrons}$ cross section from \whizthree to calculate the effective cross-section for the HPP process:

\begin{equation}
   \sigma_{\mathrm{eff}}\left(e^{+} e^{-} \rightarrow \gamma \gamma \rightarrow \mathrm{hadrons}\right)= \int{\mathrm{d}x \mathrm{d}y \frac{\mathrm{d} \mathcal{L}_{\gamma \gamma}}{\mathrm{d}x \mathrm{d}y}(x,y) \hat{\sigma}_{\gamma \gamma \rightarrow \mathrm{hadrons}}\left(\sqrt{xy} \sqrt{s}\right) }
\end{equation}

\noindent which is found to be 0.295 (0.284) \textmu b for \CCCtwo PS1 (PS2) and  0.309 \textmu b   for \CCCfive PS1 and PS2, with a relative uncertainty of $0.5\%$. Multiplied by the integrated per-bunch-crossing luminosity, we estimate 0.059 (0.065) events per bunch crossing for \CCCtwo PS1 (PS2), rising to 0.29 at 550 GeV. The total number of events per train is obtained by multiplying these values with the number of bunches per train and is listed in the last row of~\cref{tab:C3_params} for the various scenarios. Overall, up to $\sim 35$ (87) HPP events are expected on average for \CCCtwo (550).

\begin{figure}[htbp]
     \centering
     \begin{subfigure}[b]{0.497\textwidth}
         \centering
         \includegraphics[width=1.03\textwidth]{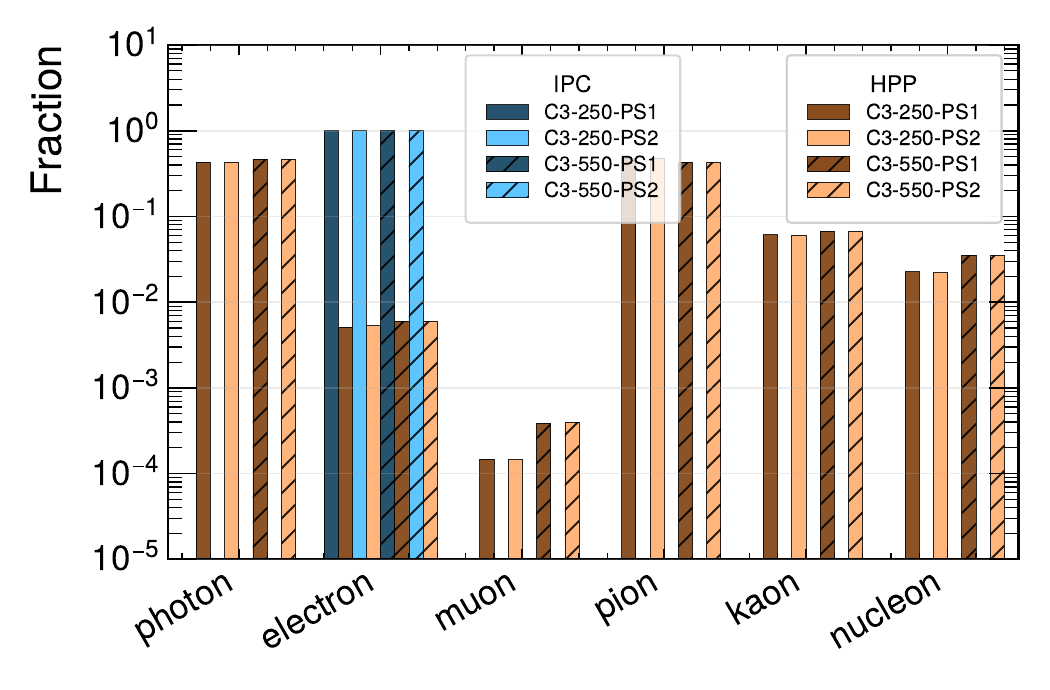} 
         \caption{}
         \label{fig:bkg_type}
     \end{subfigure}
     \hfill
     \begin{subfigure}[b]{0.497\textwidth}
         \centering
         \includegraphics[width=1.03\textwidth]{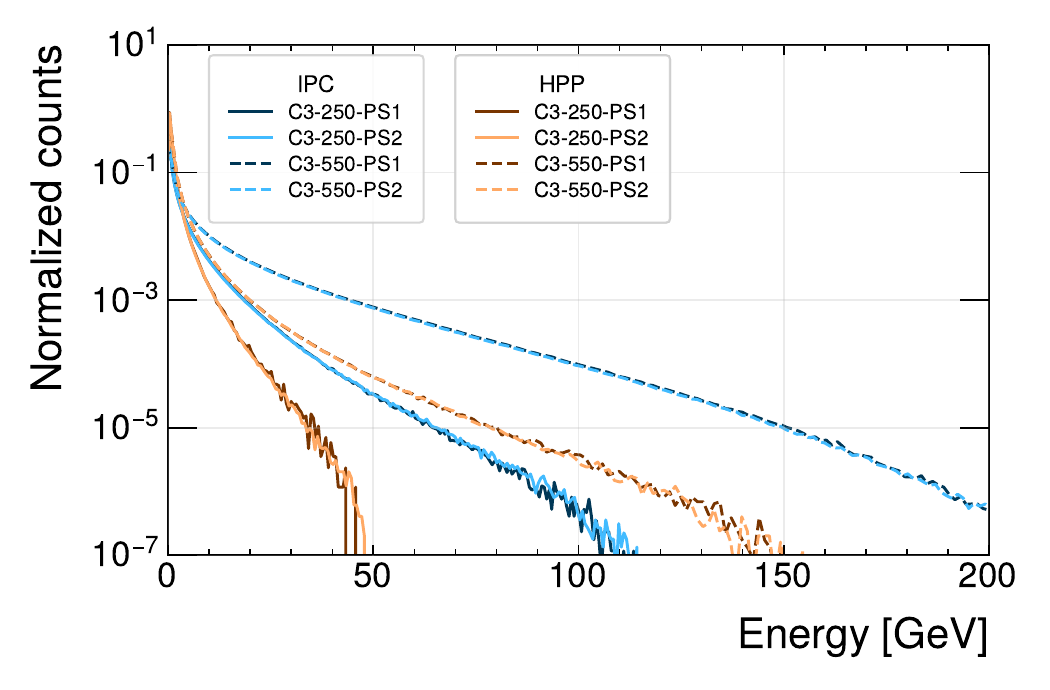}
         \caption{}
         \label{fig:bkg_energy}
     \end{subfigure}
     \hfill
     \begin{subfigure}[b]{0.497\textwidth}
         \centering
         \includegraphics[width=1.03\textwidth]{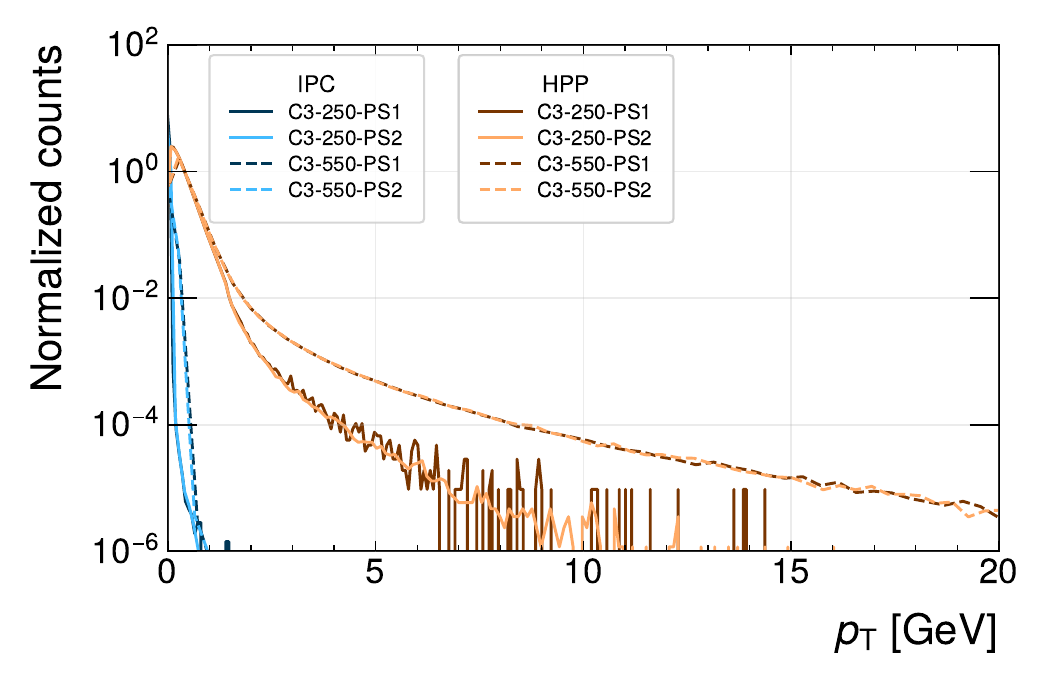} 
         \caption{}
         \label{fig:bkg_pt}
     \end{subfigure}
     \hfill
     \begin{subfigure}[b]{0.497\textwidth}
         \centering
         \includegraphics[width=1.03\textwidth]{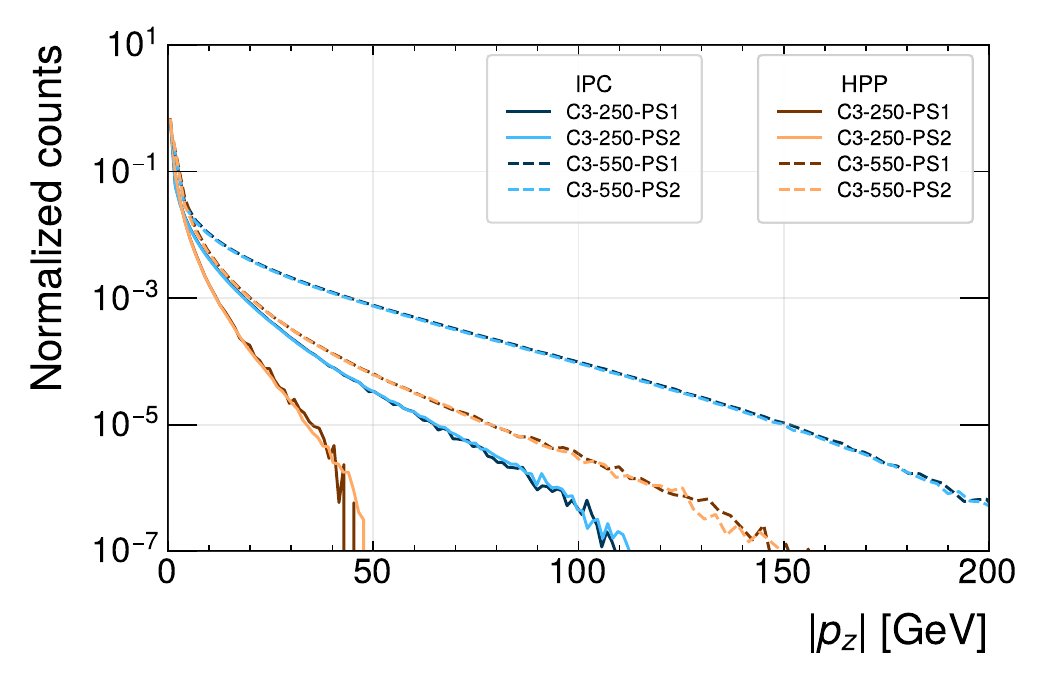} 
         \caption{}
         \label{fig:bkg_pz}
     \end{subfigure}
        \caption{Distributions of the (a) particle type, (b) total energy, (c) transverse momentum \ptnospace, and (d) longitudinal momentum $p_{z}$ of all final-state particles for the IPC and HPP background processes and for all four \CCC beam parameter scenarios.}
        \label{fig:IPC_vs_HPP_E_pt_pz}
\end{figure}

The particle species, energy and momentum distributions of the IPC and HPP background particles are overlaid in~\cref{fig:IPC_vs_HPP_E_pt_pz}. As is shown in~\cref{fig:bkg_type}, the significant fraction of hadrons in HPP events implies higher penetration rates in the calorimeters compared to IPC particles. Additionally, HPP particles are produced with significantly higher transverse momenta than $e^{+}e^{-}$ pairs from IPC, as shown in~\cref{fig:bkg_pt}, meaning that they more easily reach the barrel detectors and contribute to their occupancy. Overall, wider energy and momenta spectra are observed when increasing the CoM energy from 250 to 550 GeV, with no significant differences between the PS1 and PS2 parameter sets.

When additionally taking into account the dependency of the production rates on the CoM energy, both IPC and HPP backgrounds increase significantly from 250 to 550~GeV, as shown in~\cref{tab:C3_params}. The IPC rate increases by a factor of three per bunch crossing, reflecting the enhanced beamstrahlung and stronger beam-beam interactions at higher energy. The HPP component shows even stronger energy scaling, with a five-fold increase, due to the compounding effects of the inclusive $\gamma\gamma\!\to$ hadrons cross-section being folded with the harder photon spectra at 550~GeV, as given in~\cref{fig:lumi_spectra}, as well as the higher per-bunch-crossing luminosity at 550~GeV.

\section{Analysis of the results}
\label{sec:results}

Having established the simulation framework and characterized the beam-beam background sources, we now analyze their impact on detector performance. The IPC and HPP backgrounds generated using the methods described in Sections \ref{subsec:incoh_pairs} and \ref{subsec:hadron_photo} are propagated through the full SiD detector simulation outlined in Section \ref{subsec:detector_sim}, producing SimHits in each subdetector system. These SimHits encode both the temporal and spatial distribution of background particles, enabling comprehensive assessment of their effects on detector operation. In the following subsections, we present detailed analyses of three critical aspects: the temporal structure of backgrounds and their evolution over bunch trains, the spatial distribution and resulting channel occupancies across detector subsystems, and finally the implications for detector design and potential mitigation strategies. Together, these analyses demonstrate that \CCCnospace's background rates remain well within manageable bounds for precision physics measurements using existing detector technologies.

\subsection{Time profiles}
\label{subsec:time}

\begin{figure}[h]
     \centering
     \begin{subfigure}[b]{0.45\textwidth}
         \centering
         \includegraphics[width=\textwidth]{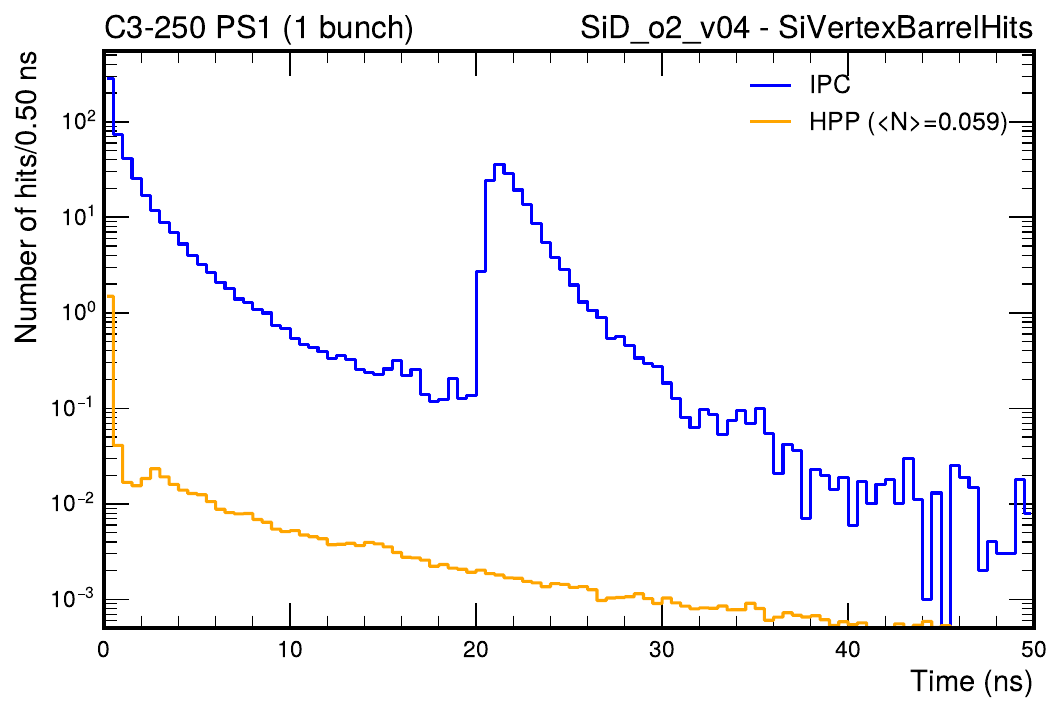}
         \caption{}
        \label{fig:C3_250_PS1_one_BX_timing_SiVertexBarrelHits}
     \end{subfigure}
     \hfill
     \begin{subfigure}[b]{0.45\textwidth}
         \centering
         \includegraphics[width=\textwidth]{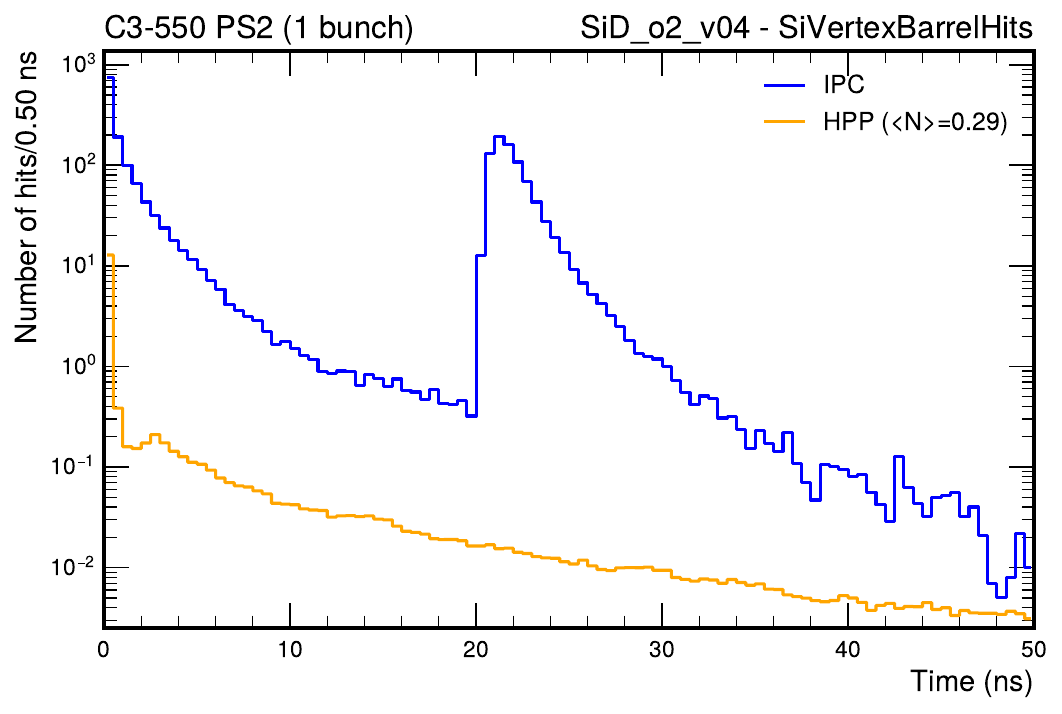}
         \caption{}
         \label{fig:C3_550_PS2_one_BX_timing_SiVertexBarrelHits}
     \end{subfigure}
     
     \begin{subfigure}[b]{0.45\textwidth}
         \centering
         \includegraphics[width=\textwidth]{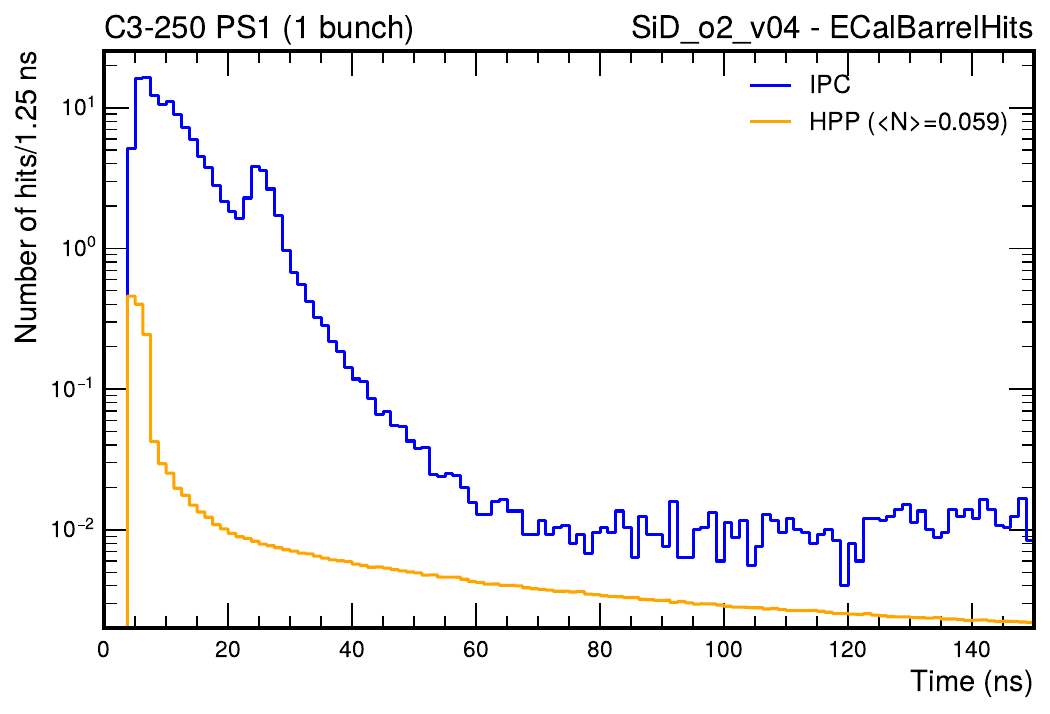}
         \caption{}
         \label{fig:C3_250_PS1_one_BX_timing_ECalBarrelHits}
     \end{subfigure}
     \hfill
     \begin{subfigure}[b]{0.45\textwidth}
         \centering
         \includegraphics[width=\textwidth]{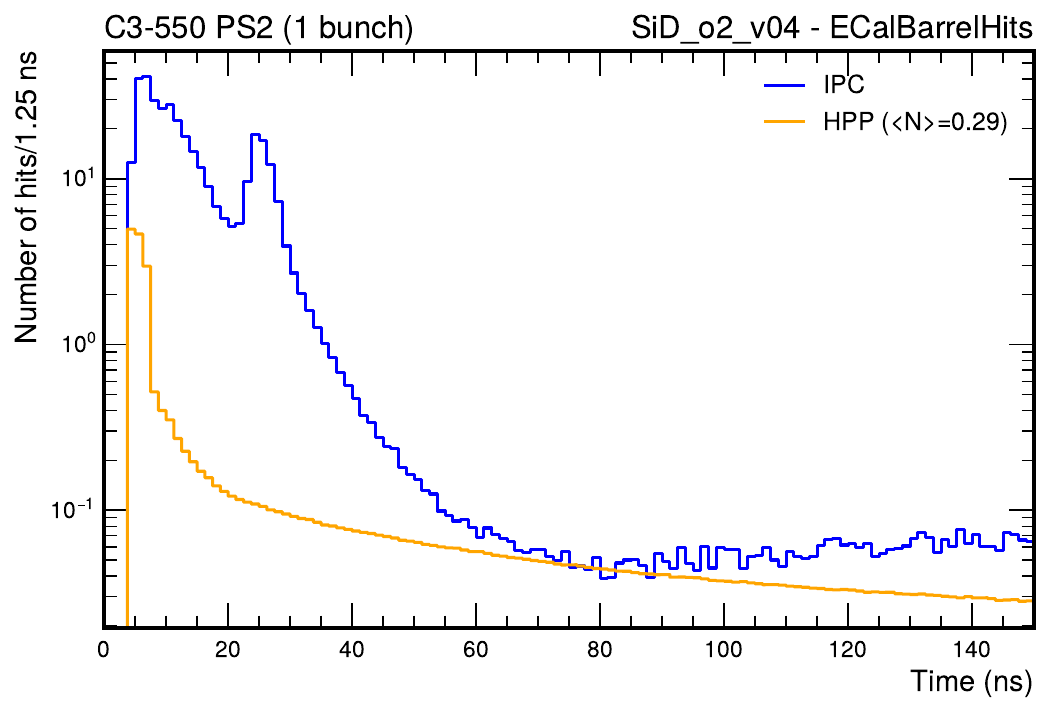}
         \caption{}
         \label{fig:C3_550_PS2_one_BX_timing_ECalBarrelHits}
     \end{subfigure}

     \begin{subfigure}[b]{0.45\textwidth}
         \centering
         \includegraphics[width=\textwidth]{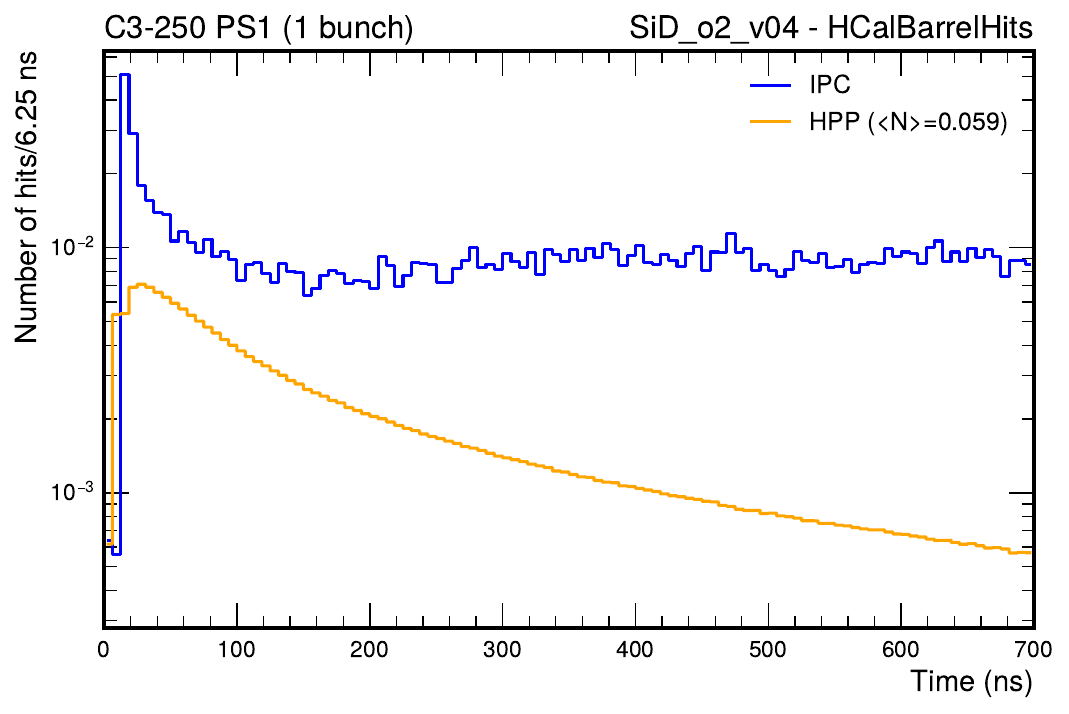}
         \caption{}
         \label{fig:C3_250_PS1_one_BX_timing_HCalBarrelHits}
     \end{subfigure}
     \hfill
     \begin{subfigure}[b]{0.45\textwidth}
         \centering
         \includegraphics[width=\textwidth]{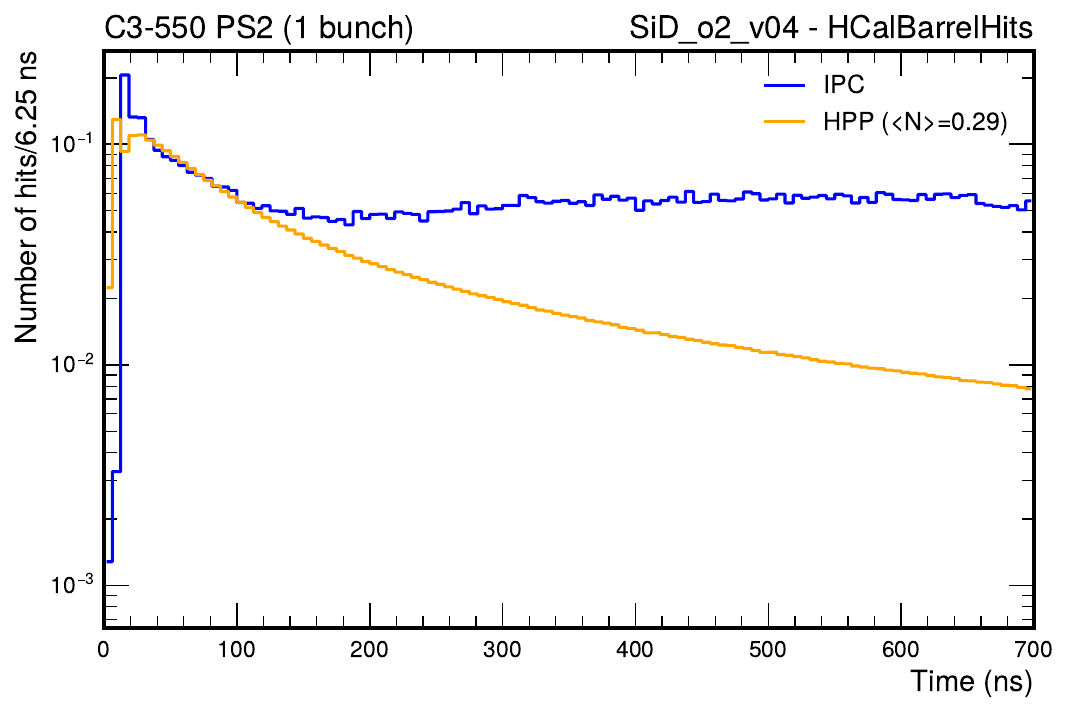}
         \caption{}
         \label{fig:C3_550_PS2_one_BX_timing_HCalBarrelHits}
     \end{subfigure}

        \caption{Time distributions of hits from the IPC and HPP backgrounds corresponding to one bunch crossing for the \CCCnospace-250 PS1 (\emph{left}) and \CCCnospace-550 PS2 (\emph{right}) beam parameters and for various barrel subdetectors: (a),(b) vertex, (c),(d) ECAL and (e),(f) HCAL.}
        \label{fig:one_BX_timing_barrel}
\end{figure}

\begin{figure}[h]
     \centering
     \begin{subfigure}[b]{0.45\textwidth}
         \centering
         \includegraphics[width=\textwidth]{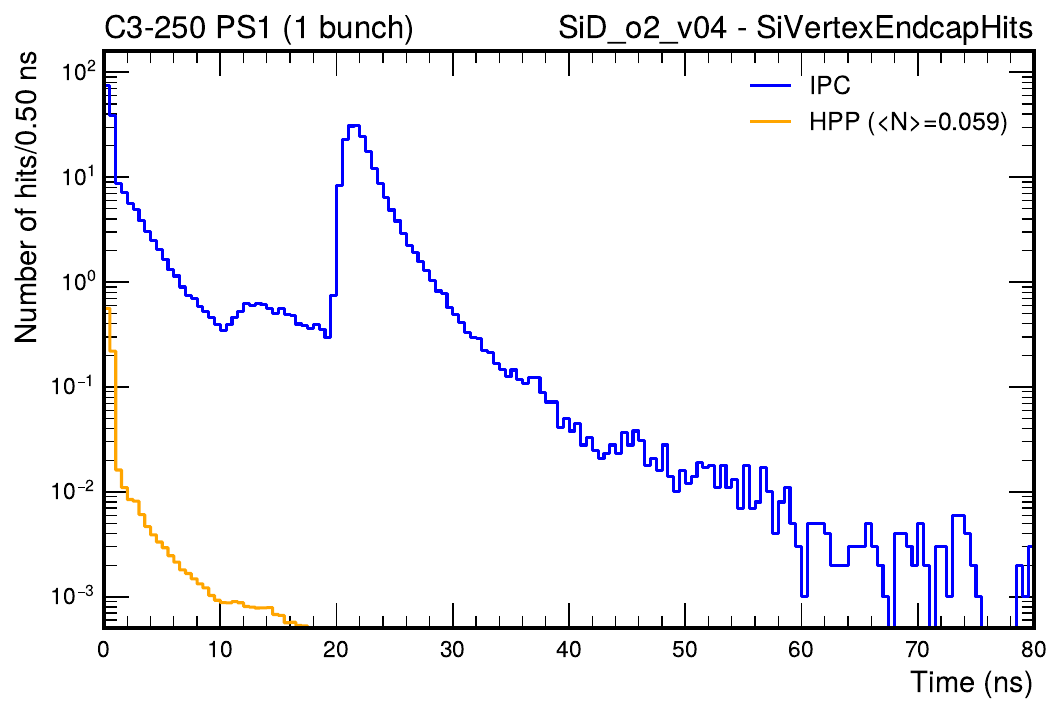}
         \caption{}
        \label{fig:C3_250_PS1_one_BX_timing_SiVertexEndcapHits}
     \end{subfigure}
     \hfill
     \begin{subfigure}[b]{0.45\textwidth}
         \centering
         \includegraphics[width=\textwidth]{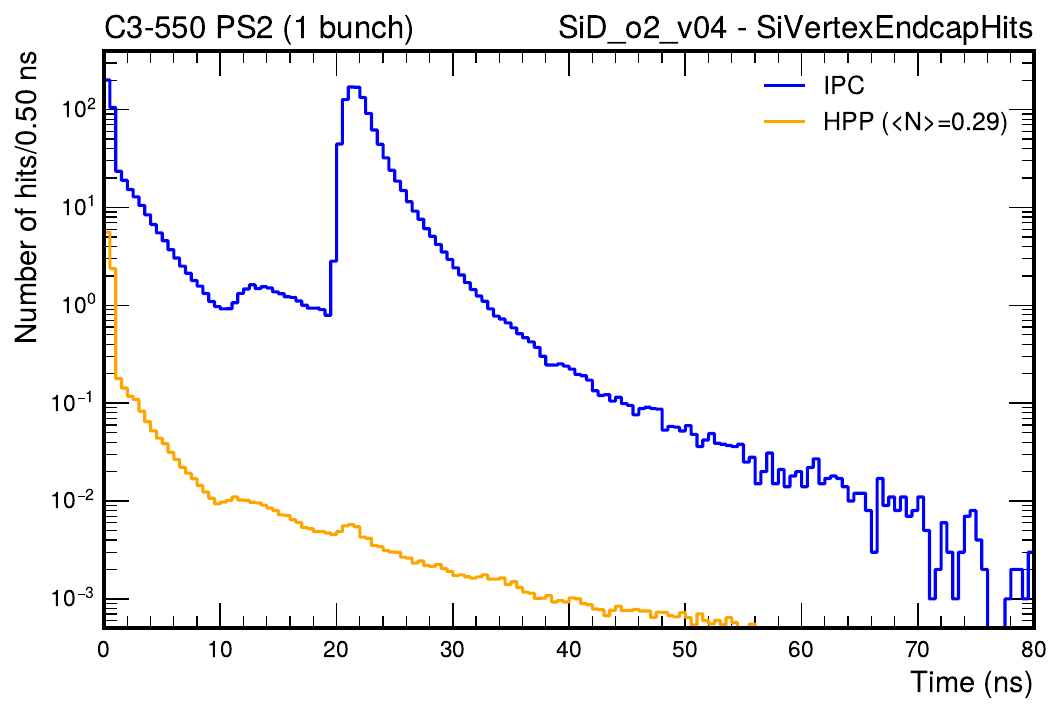}
         \caption{}
         \label{fig:C3_550_PS2_one_BX_timing_SiVertexEndcapHits}
     \end{subfigure}
     
     \begin{subfigure}[b]{0.45\textwidth}
         \centering
         \includegraphics[width=\textwidth]{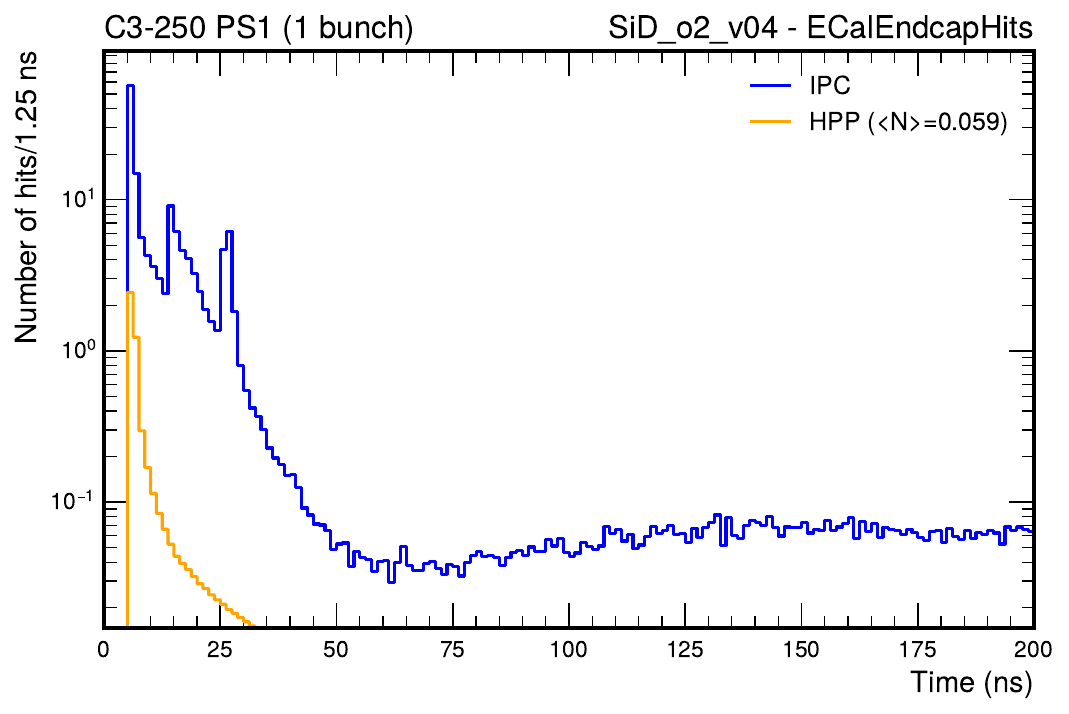}
         \caption{}
         \label{fig:C3_250_PS1_one_BX_timing_ECalEndcapHits}
     \end{subfigure}
     \hfill
     \begin{subfigure}[b]{0.45\textwidth}
         \centering
         \includegraphics[width=\textwidth]{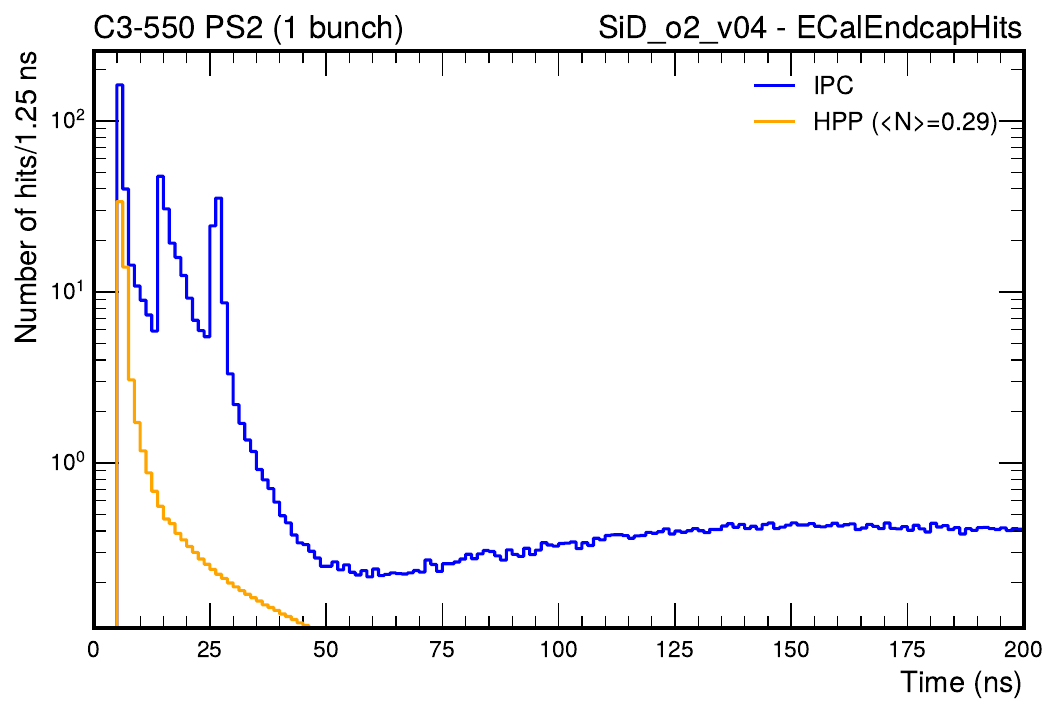}
         \caption{}
         \label{fig:C3_550_PS2_one_BX_timing_ECalEndcapHits}
     \end{subfigure}

     \begin{subfigure}[b]{0.45\textwidth}
         \centering
         \includegraphics[width=\textwidth]{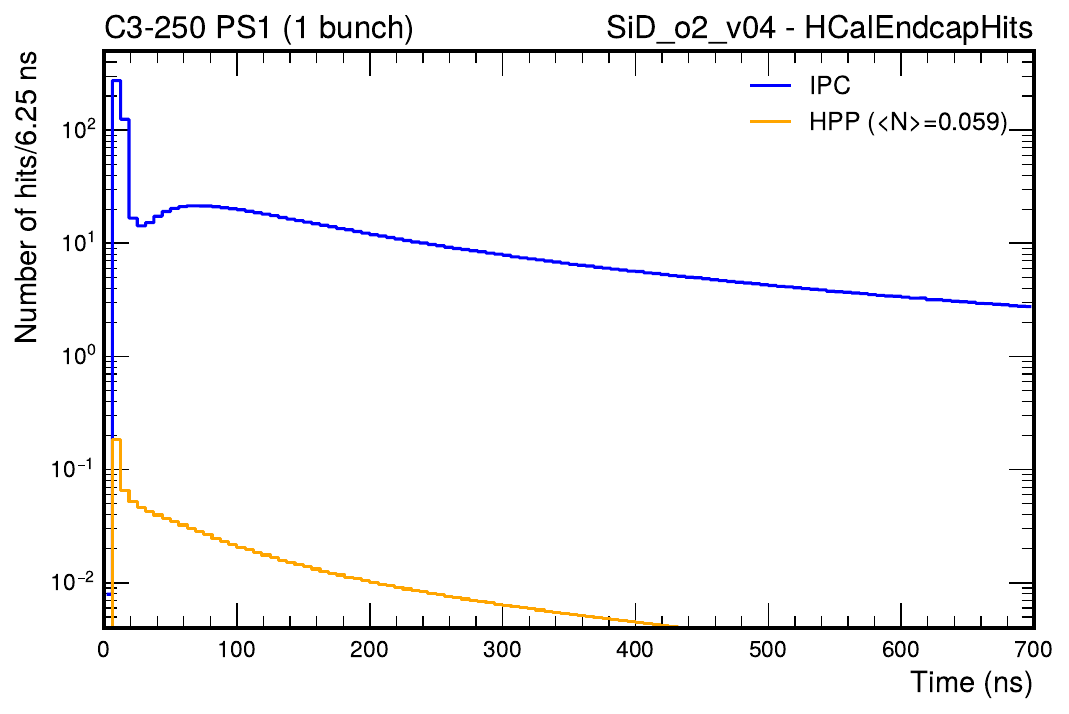}
         \caption{}
         \label{fig:C3_250_PS1_one_BX_timing_HCalEndcapHits}
     \end{subfigure}
     \hfill
     \begin{subfigure}[b]{0.45\textwidth}
         \centering
         \includegraphics[width=\textwidth]{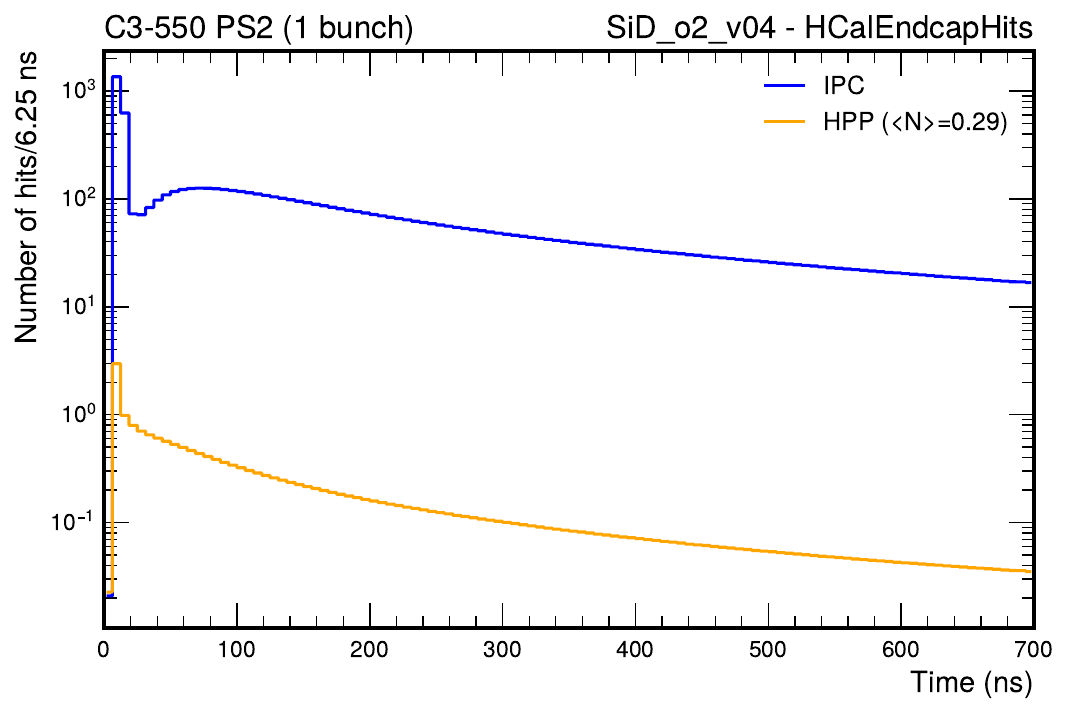}
         \caption{}
         \label{fig:C3_550_PS2_one_BX_timing_HCalEndcapHits}
     \end{subfigure}

        \caption{Time distributions of hits from the IPC and HPP backgrounds corresponding to one bunch crossing for the \CCCnospace-250 PS1 (\emph{left}) and \CCCnospace-550 PS2 (\emph{right}) beam parameters and for various endcap subdetectors: (a),(b) vertex, (c),(d) ECAL and (e),(f) HCAL.}
        \label{fig:one_BX_timing_endcap}
\end{figure}

The time distribution of hits per bunch crossing from IPC and HPP backgrounds for the \CCCtwo PS1 and \CCCfive PS2 beam parameters in the vertex, ECAL, and HCAL systems is given in~\cref{fig:one_BX_timing_barrel,fig:one_BX_timing_endcap} for the barrel and endcap detectors, respectively. Overall, the IPC background dominates over HPP, due to its larger effective cross-section.

The time profiles of the IPC and HPP backgrounds exhibit quite different characteristics. Incoherent pairs exhibit sharp peaks synchronized with bunch crossings, as well as delayed structures from backsplash off forward instrumentation. For instance, in the vertex barrel, a secondary peak appears at around 20~ns: pairs stream out along the beamline and interact with the very forward BeamCal, roughly 3~m downstream. The scattered particles and produced secondaries then travel towards the IP. The round-trip time is consistent with $t \simeq 2L/c \simeq 20$~ns, with additional spread due to material interactions and the angular spread of the outgoing particles. This backsplash signature becomes more pronounced at 550~GeV due to the higher IPC flux impinging on the BeamCal. Similar secondary peaks are present in the tracker and ECAL.

The HPP background, on the contrary, does not exhibit this sharp backscatter effect. Instead, its production is more transversely distributed, without a single forward target, and the hadronic shower development smears late activity into a long exponential tail after the primary interaction. This arises from multiple mechanisms: shower development in calorimeters, thermal neutron and proton propagation, and nuclear de-excitation processes. This tail means that while the majority of hadronic energy deposits occur within the first few ns, the remaining fraction spreads across subsequent bunch crossings with decreasing intensity that accumulates with the length of the bunch train. For a single hadronic event, measurable energy deposits persist for hundreds of ns, affecting the entire bunch train.

Additionally, since IPC pairs are highly forward (larger $|p_{z}|/p$ and small $p_{\mathrm{T}}$ compared to HPP), IPC dominates the timing distributions in both barrel and endcap detectors, with the effect being particularly pronounced in the endcaps where the forward-peaked pairs are concentrated. HPP, with comparatively higher $p_{\mathrm{T}}$, tends to populate the central region; at 550~GeV the HPP contribution becomes visible in the ECAL and HCAL barrels, though IPC still dominates overall.

The superposition of these temporal patterns creates a complex pile-up structure when integrating over an entire bunch train. Within a single bunch crossing window, the instantaneous background comprises: (i) prompt pairs from the current collision, (ii) backscattered pairs from specific previous crossings determined by the detector geometry and bunch spacing, (iii) the exponentially decaying tail from all previous hadronic interactions within the train. For the sustainability and high-luminosity scenarios of~\cref{tab:C3_params} with shorter bunch spacings, the temporal overlap of these components becomes more significant, making their distinction more challenging.

Overall, the integrated effect over a full bunch train shows that IPC pairs contribute approximately 99\% of the total hit count for the endcap and forward detectors, whereas hadronic events, due to their more central nature and longer temporal evolution, affect primarily the barrel detectors, contributing up to $\mathcal{O}(10\%)$ of the hits in the ECAL, and approaching comparable levels as IPC in the HCAL at the highest energies. This makes the hadronic background particularly relevant for calorimeter performance and jet reconstruction accuracy.

\subsection{Average hit rates}
\label{subsec:av_hit_rates}

\begin{figure}[h]
     \centering
     \begin{subfigure}[b]{0.495\textwidth}
         \centering
         \includegraphics[width=\textwidth]{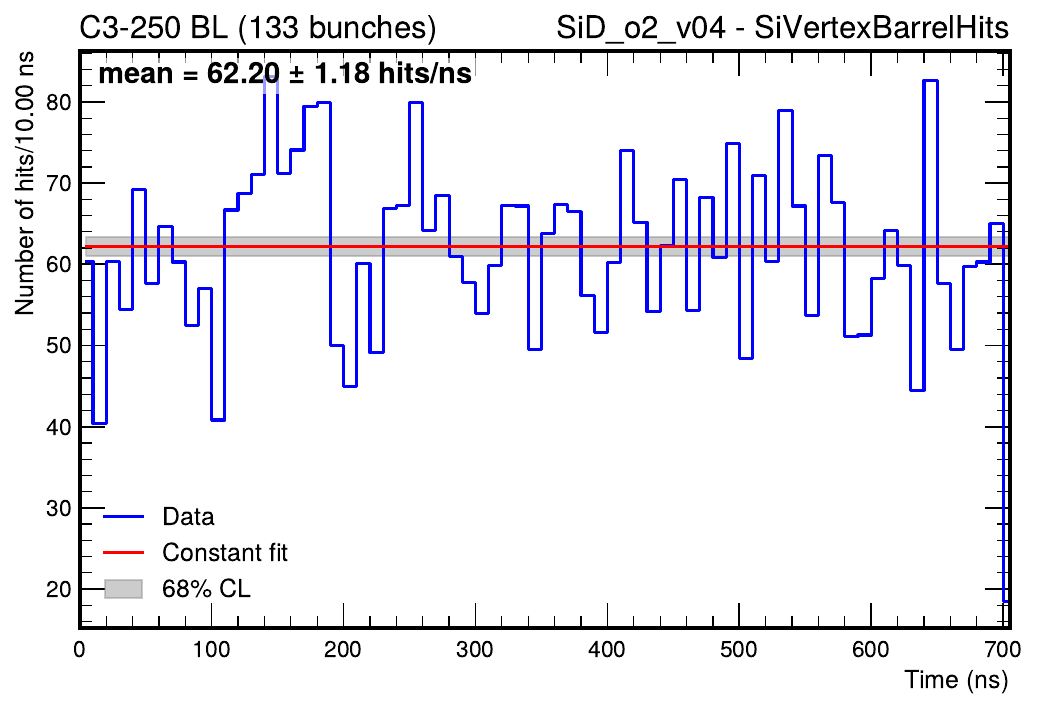}
         \caption{}
         \label{fig:C3_250_BL_timing_all_BX_SiVertexBarrelHits}
     \end{subfigure}
     \hfill
     \begin{subfigure}[b]{0.495\textwidth}
         \centering

         \includegraphics[width=\textwidth]{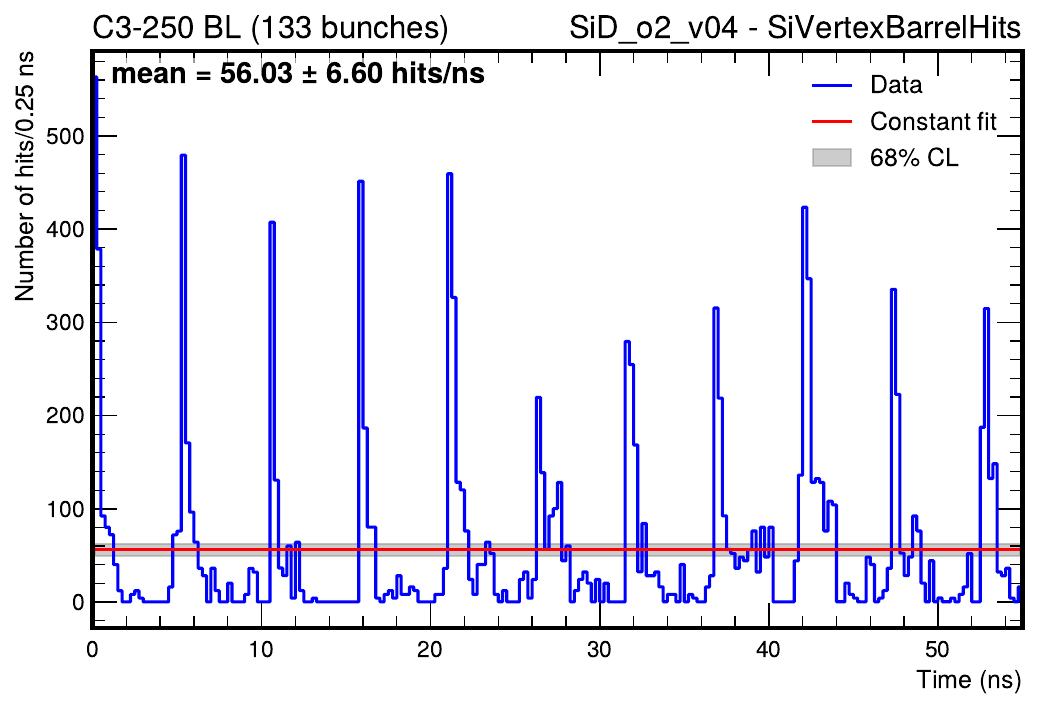}
         \caption{}
         \label{fig:C3_250_BL_timing_all_BX_SiVertexBarrelHits_zoomed_in}

     \end{subfigure}

        \caption{Time distributions of hits from the added contributions of the IPC and HPP backgrounds in the vertex barrel detector corresponding to one full bunch train for the \CCCtwo BL scenario and for different integrated times: (a) for an entire bunch train of 133 bunches and (b) for a duration of $54 \ \mathrm{ns}$ which corresponds to roughly 11 bunch crossings.}
        \label{fig:C3_timing_all_BX_SiVertexBarrel}
\end{figure}

When the single-bunch-crossing timing profiles are overlaid across an entire train,  narrow features are smeared out and a quasi-stationary baseline appears in the vertex detector, as shown in~\cref{fig:C3_250_BL_timing_all_BX_SiVertexBarrelHits}. The resulting per-time-bin hit densities are well described by flat plateaus; their level encodes the effective overlay load per nanosecond. Finer time discretization, cf.~\cref{fig:C3_250_BL_timing_all_BX_SiVertexBarrelHits_zoomed_in}, reveals the characteristic peaks due to the arrival of subsequent bunch crossings, separated by the corresponding bunch spacing.

\begin{figure}[htbp]
     \centering
     \begin{subfigure}[b]{\textwidth}
         \centering
         \includegraphics[width=\textwidth]{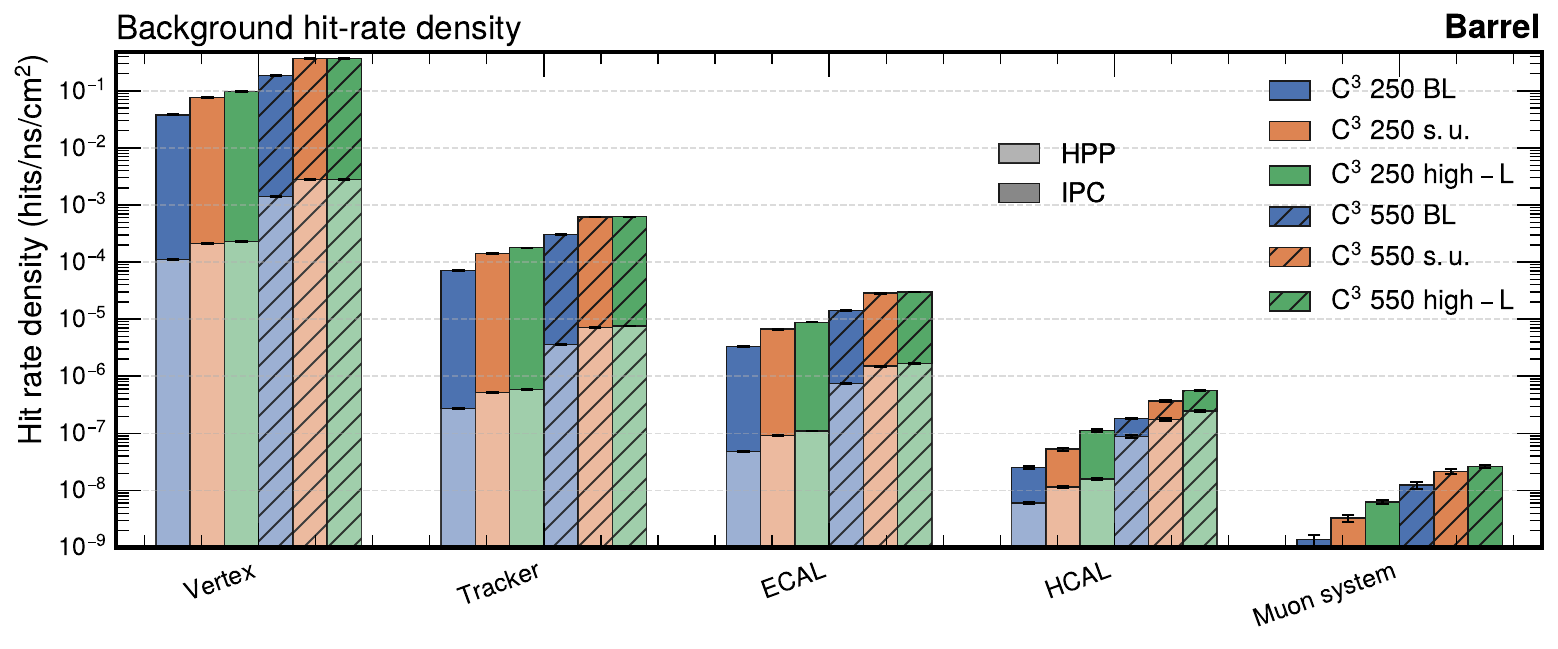}
         \caption{}
         \label{fig:hit_rates_barrel_density_IPC_HPP}
     \end{subfigure}
     \hfill
     \begin{subfigure}[b]{\textwidth}
         \centering

         \includegraphics[width=\textwidth]{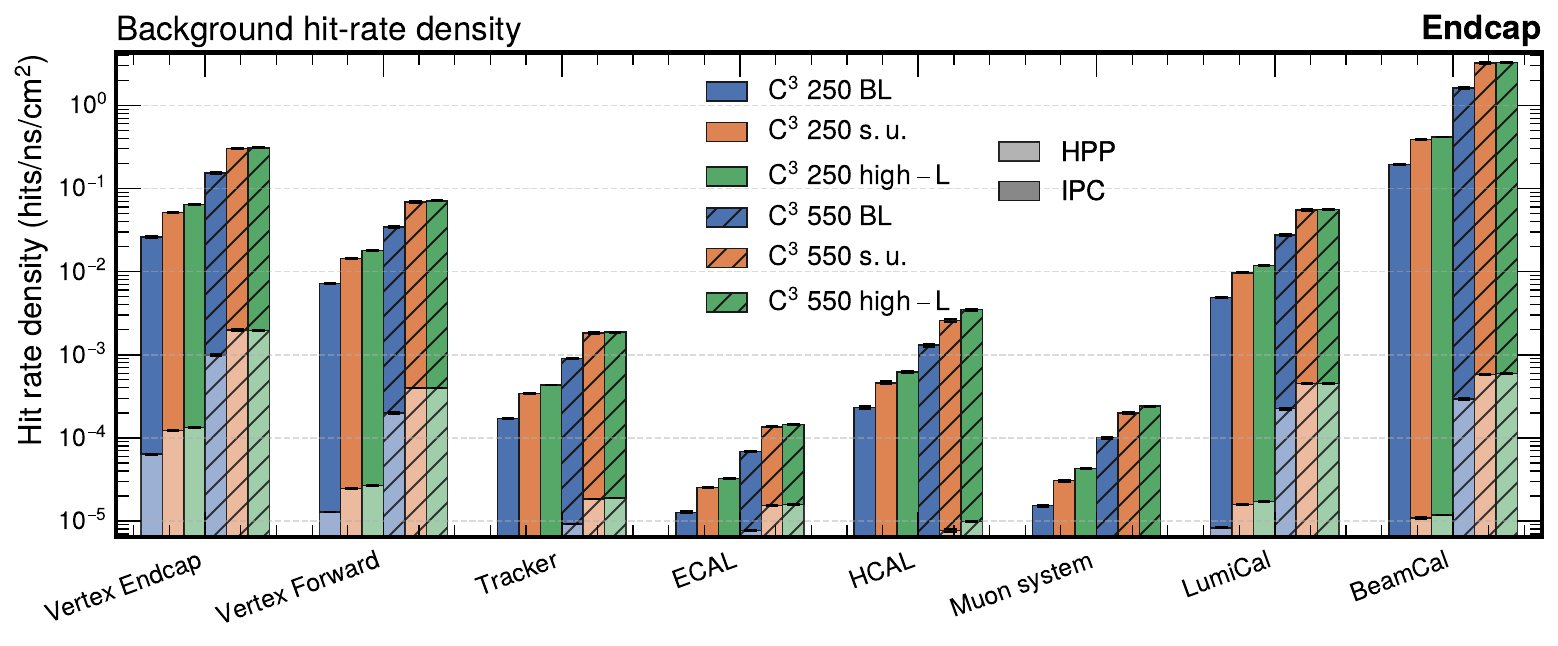}
         \caption{}
         \label{fig:hit_rates_endcap_density_IPC_HPP}

     \end{subfigure}

        \caption{Background hit densities in the (a) barrel and (b) endcap subdetectors, averaged over an entire bunch train, for various \CCC running scenarios. The HPP and IPC contributions are shown as stacked segments, indicated by lighter and darker hues, respectively, whereas collider energy is encoded by hatching. The exact numerical values for the hit rates are provided in~\cref{app_subsec:bkg_hit_rates}.}
        \label{fig:hit_rate_densities}
\end{figure}

The average hit-rate densities, obtained by dividing the average hit rates by the total surface area of the corresponding subdetector, are summarized in~\cref{fig:hit_rate_densities} for all six \CCC operating scenarios and reveal systematic dependencies on bunch spacing, train length, and CoM energy that provide useful insights into the background composition and evolution.

The most obvious feature is the robust scaling with bunch spacing. Comparing baseline (BL) to sustainability (s.\,u.) configurations, where the bunch spacing is halved from 5.26\,ns to 2.63\,ns at 250\,GeV and from 3.50\,ns to 1.75\,ns at 550\,GeV, we observe a near-universal doubling of hit-rate densities across all subsystems, as the time-averaged rates scale as $\rho_{\rm time} \approx N_{\rm hits/BX}/\Delta t_{\rm BX}$. This confirms that train-overlay plateaus are governed primarily by per-bunch-crossing yields and bunch spacing. This clean scaling relationship holds because \CCCnospace's bunch trains are sufficiently short that the long-lived hadronic shower evolution effects—which develop over hundreds of ns—do not yet dominate the accumulated background. For significantly longer bunch trains, these slow hadronic processes would build up over time, causing deviations from the simple linear scaling with train length and bunch spacing.

The transition from s.\,u.\ to high-$\mathcal{L}$ configurations, which extends the train length while maintaining bunch spacing, reveals a critical distinction between prompt and slow background components. The HCAL exhibits 35-100\% increases in average hit-rate densities at both energies when doubling the train length. This is due to its occupancy being affected by hadronic processes. In contrast, pair-dominated tracking systems show more modest increases. This difference in response reaffirms that the long-lived hadronic afterglow accumulates more significantly over extended trains compared to prompt IPC deposits.

The energy dependence between 250 and 550\, GeV reveals significant increases across all detector systems, reflecting the enhanced beam-beam interactions at higher CoM energies. The background hit densities in the tracking systems scale substantially with energy, with a five (four) -fold increase for the vertex (tracker) barrel in the baseline configuration. This strong energy dependence arises from the combined effects of increased IPC production rates and the harder photon spectra at 550\, GeV, which produce pairs with higher transverse momenta capable of reaching the detector volume.

Similarly, the ECAL barrel hit density increases by a factor of four at higher energies, driven by both the enhanced IPC flux and the five-fold increase in HPP events at higher energy. The background density in the barrel HCAL experiences an eight-fold enhancement, reflecting the growth in hadronic activity as the $\gamma\gamma$ CoM energy increases. This stems both from the increased HPP cross-section and the larger transverse momenta of produced hadrons, which tend to populate the central detector region at 550\,GeV.

The forward systems demonstrate the most extreme energy scaling, with the BeamCal hit density increasing by nearly an order of magnitude between 250 and 550\,GeV. This reflects the forward-peaked nature of pair production combined with the higher total IPC yield at increased beam energy. The endcap HCAL and muon systems show similar increases, confirming that forward instrumentation is affected the most by the larger background rates at higher collision energies.

\subsection{Spatial distribution, detector impact, and mitigation strategies}
\label{subsec:occupancy}

The time distributions and average hit densities presented in the previous section provide valuable metrics for understanding the relative contributions and scaling behavior of the IPC and HPP backgrounds. However, a comprehensive assessment of their impact on detector performance requires quantification of the spatial distribution and resulting channel occupancy. The main challenge is the finite granularity of detector readout: each sensing element or pixel can record only a limited number of hits during a bunch train before saturating. When background particles occupy these channels, they become unavailable for recording signals from hard-scatter events of physics interest, and in turn degrade reconstruction efficiency and measurement precision.

\paragraph{Occupancy metrics and analysis.}
The spatial distribution of background hits directly determines detector channel occupancy, 
which we quantify through two complementary metrics. First, we calculate the \emph{average occupancy} in each subdetector for a full bunch train, defined as: 

\begin{equation}
    \langle\mathrm{Occupancy}\rangle_{D}=S_{D} \cdot C_{D} \cdot \frac{\# \text{ of hits in subdetector }D}{\# \text{ of cells in subdetector }D}
    \label{eq:av_occ}
\end{equation}

\noindent where the number of hits in each subdetector $D$ can be calculated by multiplying the values of~\cref{tab:hit_rates} in~\cref{app_subsec:bkg_hit_rates} with the bunch train duration for each scenario, and the number of cells in $D$ is obtained from the detector geometry outlined in~\cref{subsec:detector_sim} and assuming the cell sizes of~\cref{tab:SiD_thresholds_cell_size}. Finally, $S_{D}$ is a safety factor used to account for potential occupancy increase due to mismodeling effects, both at the background generation and the detector simulation level, and $C_{D}$ is a cluster size factor to account for charge sharing effects. For the purposes of this study, and following similar assumptions previously made in the literature~\cite{Arominski:2704642,Boscolo:2023grv}, we adopt $S_{D}=2$ for all subdetectors, and $C_{D}=3$ for the vertex detector, with $C_{D}=1$ everywhere else\footnote{ For MAPS with $\simeq 10$ \textmu m pitch operated at nominal thresholds, we assume a mean cluster size of 3, consistent with ALPIDE beam-test characterizations~\cite{Rinella:2022htp,Andronic:2025hcm}. For all other subdetectors, no charge sharing is assumed.}.

The average occupancy for the vertex and tracker detectors are gathered in~\cref{tab:av_occupancy} for the various \CCC parameter scenarios. Following previous ILC studies~\cite{Schutz:2018ynd,behnke2013internationallinearcollidertechnical}, an occupancy threshold of $10^{-4}$ is defined as the design target for the tracking detectors. This threshold has been validated through extensive simulation studies for the ILC detector concepts and was found to preserve hit efficiency at levels not significantly lower than the intrinsic sensor efficiency, thereby maintaining robust tracking performance essential for precision vertexing and heavy-flavor tagging. It thus serves as a practical benchmark to enable direct comparison with other linear collider background studies.

Using this threshold, we notice that the average occupancy for the vertex exceeds $10^{-4}$ for all \CCC operating scenarios, rising to $\mathcal{O}(10^{-3})$ for the highest-luminosity scenario at 550 GeV. On the contrary, the average occupancy for the tracker barrel and endcap detectors remains well below the $10^{-4}$ limit. With respect to the baseline \CCC scenario at 250 GeV, the occupancy increases approximately by two (five)-fold for the s.u. (high-$\mathcal{L}$) configuration at 250 GeV, whereas for the higher energy runs at 550 GeV, we observe on average a two-, four-, and eight-fold increase for the baseline, s.u. and high-$\mathcal{L}$ scenarios, respectively.

\begin{table}[h]
    \centering

    \caption{Average occupancy in units of $10^{-4}$ for the vertex and tracker detectors and for all \CCC parameter scenarios under consideration. Underlined values in parentheses express the occupancy ratio for each subdetector with respect to the occupancy value for the corresponding subdetector in the baseline \CCCtwo scenario.}
    \label{tab:av_occupancy}
    
    \begin{tabular}{ccccccc}

    \multicolumn{7}{c}{\textbf{Average Occupancy [$10^{-4}$]}} \\
    \toprule
    & \multicolumn{3}{c}{\CCCtwo} & \multicolumn{3}{c}{\CCCfive}\\
    \cmidrule(r){2-4}\cmidrule(l){5-7}
    Scenario & BL & s.\,u.\ & high‑$\mathcal{L}$ & BL & s.\,u.\ & high‑$\mathcal{L}$\\
    \midrule

\multicolumn{7}{c}{Barrel Detector}    \\
\midrule

Vertex & 1.6 & 3.3  & 8.3 & 3.0 & 6.0 & 12.1 \\
& ($\underline{1.0\times}$) & ($\underline{2.1\times}$)  & ($\underline{5.2\times}$) & ($\underline{1.9\times}$) & ($\underline{3.8\times}$) & ($\underline{7.6\times}$) \\
Tracker &  0.030 &  0.060 &  0.150 & 0.050 & 0.100  &  0.200 \\
& ($\underline{1.0\times}$) & ($\underline{2.0\times}$)  & ($\underline{5.0\times}$) & ($\underline{1.7\times}$) & ($\underline{3.3\times}$) & ($\underline{6.7\times}$) \\
\midrule  \multicolumn{7}{c}{Endcap Detector} \\
\midrule  
Vertex Endcap & 1.1 & 2.2  & 5.4 & 2.5  & 5.0 & 9.9 \\
& ($\underline{1.0\times}$) & ($\underline{2.0\times}$)  & ($\underline{4.9\times}$) & ($\underline{2.3\times}$) & ($\underline{4.5\times}$) & ($\underline{9.0\times}$) \\
Vertex Forward &  0.31 & 0.62  & 1.55 & 0.57 & 1.15 &  2.29 \\
& ($\underline{1.0\times}$) & ($\underline{2.0\times}$)  & ($\underline{5.0\times}$) & ($\underline{1.8\times}$) & ($\underline{3.7\times}$) & ($\underline{7.4\times}$) \\
Tracker & 0.061 & 0.123  &  0.307 & 0.127 & 0.254 & 0.508 \\
& ($\underline{1.0\times}$) & ($\underline{2.0\times}$)  & ($\underline{5.0\times}$) & ($\underline{2.1\times}$) & ($\underline{4.2\times}$) & ($\underline{8.3\times}$) \\

\bottomrule
\end{tabular}

\end{table}

Although the average occupancy is a useful metric, it doesn't take into account the geometric distribution of hits within each subdetector. For instance, while the average occupancy for the vertex detector is above $10^{-4}$, it is easy to imagine that most of hits are in the first (closest to the IP) layer of the vertex barrel and at the edges of the detector along the longitudinal direction, due to the forward nature of the IPC background. For this reason, we also calculate the \emph{layer occupancy} by accumulating hits over each subdetector layer instead of the entire subdetector. This is achieved by using the detector geometry description from \texttt{k4geo} and overlaying a pixelated grid on top of each module, assuming the cell sizes of~\cref{tab:SiD_thresholds_cell_size}. Within each layer, we then compute the layer occupancy as:

\begin{equation}
    \text{layer occupancy}(\ell_{D},\mathrm{BD})=S_{D}\cdot C_{D}\cdot \frac{\# \text{ of cells in } \ell_{D} \text{ with} \geq \mathrm{BD} \text{ hits}}{\# \text{ of cells in } \ell_{D}}
    \label{eq:layer_occ}
\end{equation}

\noindent where $\ell_{D}$ denotes the layer of subdetector $D$ and $\mathrm{BD}$ is the so-called buffer depth, which represents the on-sensor memory, i.e., the maximum number of hits that can be stored per channel before readout at the end of each train. For $S_{D},C_{D}$ we follow the same assumptions as in~\cref{eq:av_occ}.

Therefore, the layer occupancy expresses the fraction of readout channels that receive a number of background hits equal to or exceeding their buffer depth after an entire bunch train. Cells that have received a number of hits from background processes equal to or exceeding their buffer depth are considered \emph{dead cells}, since they don't have the capacity to register additional hits from a potential hard-scatter process occurring in the same bunch train.

\begin{figure}[htbp]
     \centering
     \begin{subfigure}[b]{\textwidth}
         \centering
         \includegraphics[width=\textwidth]{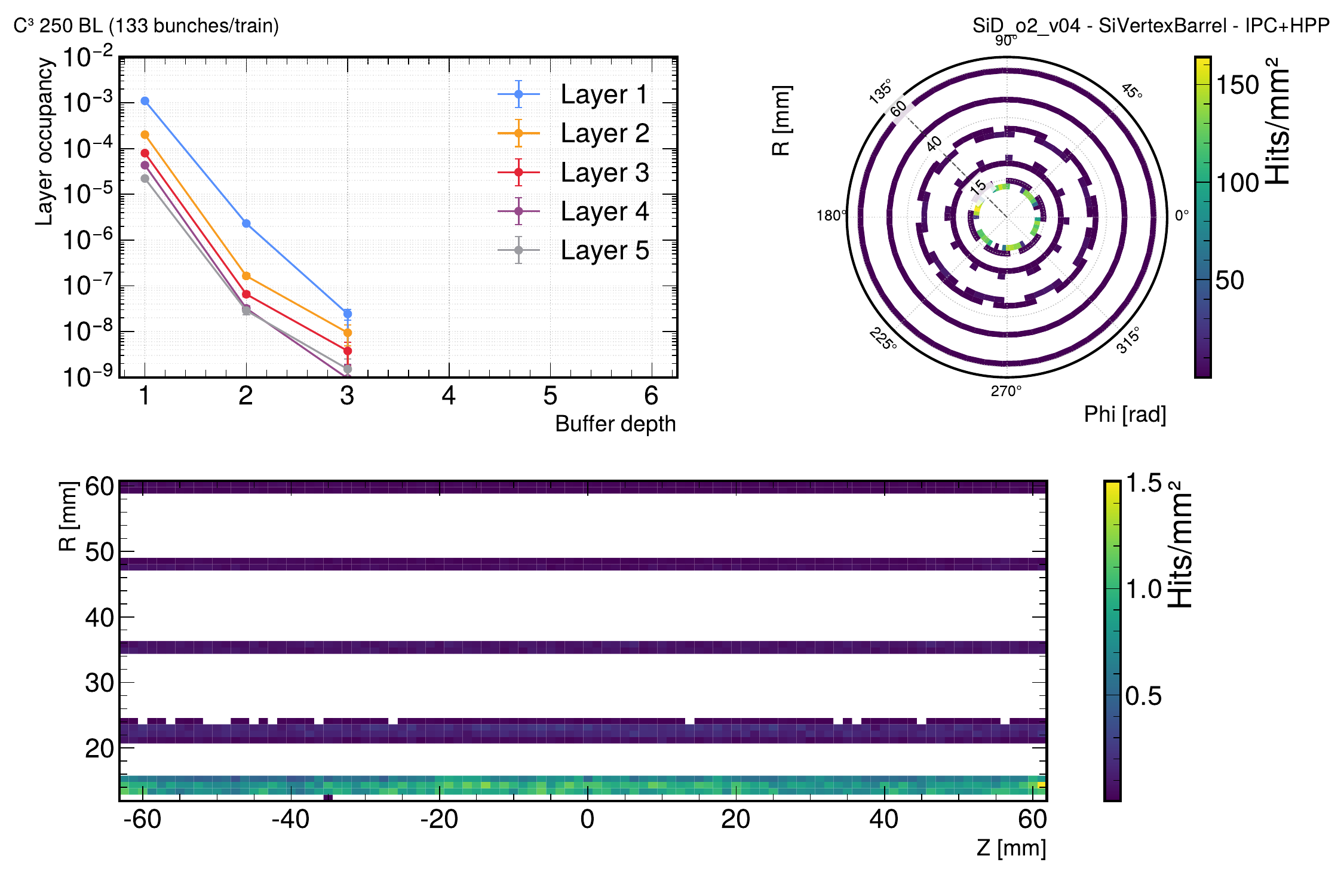}
         \caption{}
         \label{fig:SiVertexBarrel_train133_SUM_C3_250_occupancy}
     \end{subfigure}
     \hfill
     \begin{subfigure}[b]{\textwidth}
         \centering

         \includegraphics[width=\textwidth]{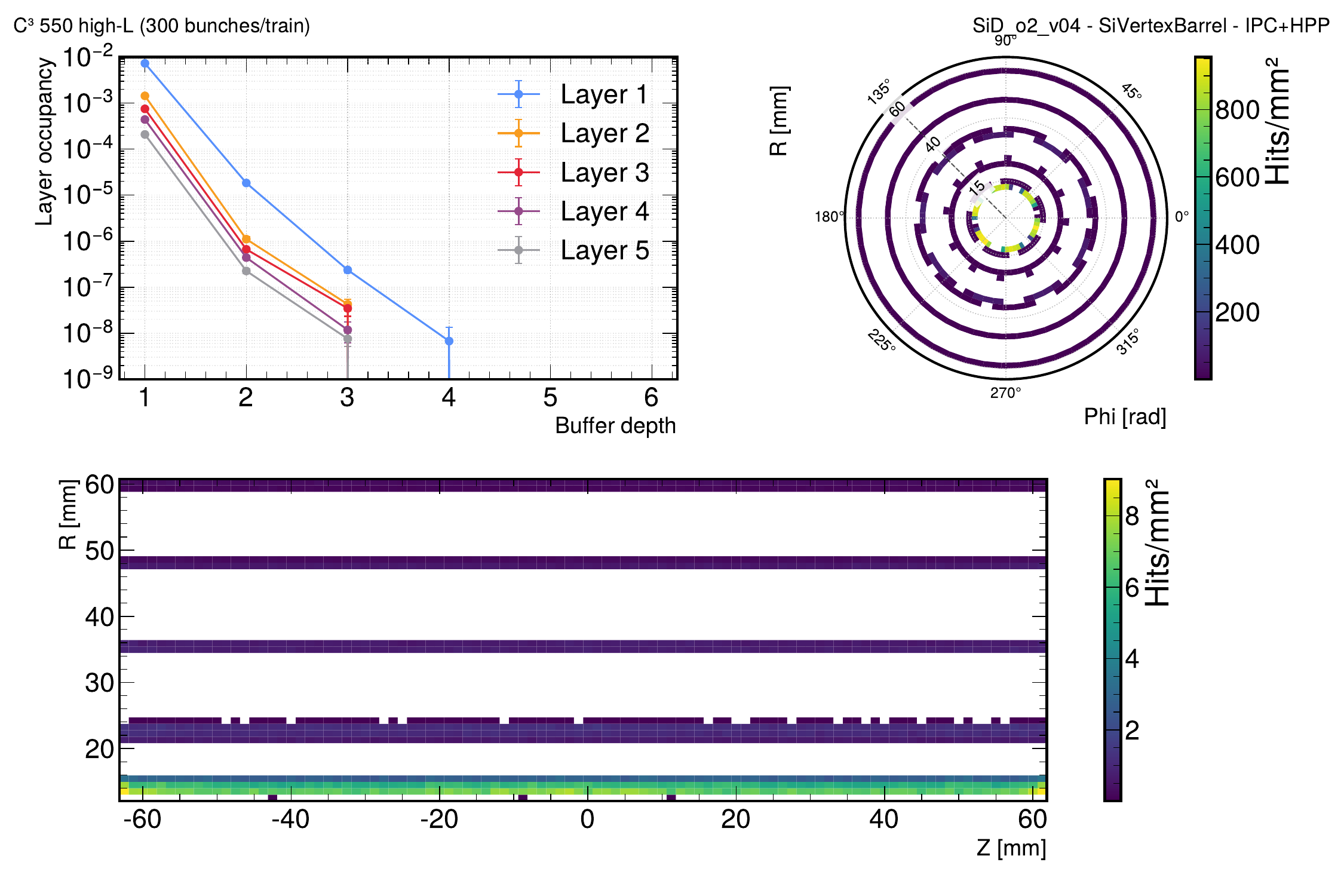}
         \caption{}
         \label{fig:SiVertexBarrel_train300_SUM_C3_550_occupancy}

     \end{subfigure}

        \caption{Occupancy (\emph{upper left}) and hit density distributions in the $R-\phi$ (\emph{upper right}) and $Z-R$ (\emph{lower}) planes for the summed contribution of the IPC and HPP backgrounds integrated over an entire bunch train in the SiD vertex barrel detector for the (a) baseline \CCC scenario at 250 GeV and (b) high-$\mathcal{L}$ scenario at 550 GeV.}
        \label{fig:occupancy}
\end{figure}

Figure~\ref{fig:occupancy} shows the spatial distribution and the resulting layer occupancy of the background hits in the vertex barrel for the most conservative, \CCCtwo BL, and aggressive, \CCCfive high-$\mathcal{L}$, beam configurations. As expected, hotspots with high background rates are concentrated on the first vertex barrel, which exceeds the $10^{-4}$ occupancy threshold in the case of a buffer depth equal to one. The remaining layers maintain occupancies mostly below that limit for the \CCCtwo BL scenario but exceed it at the most aggressive \CCCfive high-$\mathcal{L}$ one.

Following the same analysis for all other combinations of tracking subdetector and \CCC scenario, we can calculate the buffer depth required to maintain the maximum layer occupancy below $10^{-4}$ for each detector layer. The buffer depth requirements are summarized in Table~\ref{tab:buffer_depth}, with the corresponding occupancy plots for the vertex barrel and endcap subdetectors given in~\cref{app_subsec:Occupancy_plots}.

\begin{table}[h]
    \centering

    \caption{Required buffer depth to maintain a maximum layer occupancy below $10^{-4}$ for all subdetector layers of the vertex and tracker detectors and for all \CCC running scenarios under consideration.}
    \label{tab:buffer_depth}
    
    \begin{tabular}{ccccccc}

    \multicolumn{7}{c}{\textbf{Buffer Depth}} \\
    \toprule
    & \multicolumn{3}{c}{\CCCtwo} & \multicolumn{3}{c}{\CCCfive}\\
    \cmidrule(r){2-4}\cmidrule(l){5-7}
    Scenario & BL & s.\,u.\ & high‑$\mathcal{L}$ & BL & s.\,u.\ & high‑$\mathcal{L}$\\
    \midrule

    \multicolumn{7}{c}{Barrel Detector}    \\
    \midrule  
    Vertex & 2 & 2  & 2 & 2 & 2 & 2 \\
    Tracker &  1 &  1 &  1 & 1 & 1  &  1 \\
    \midrule  \multicolumn{7}{c}{Endcap Detector} \\
    \midrule  
    Vertex Endcap & 2 & 2 & 3 & 2 & 2 & 3 \\
    Vertex Forward &  1 & 1  & 2 & 1 & 2 &  2 \\
    Tracker & 1 & 1  &  1 & 1 & 1 & 1 \\
    \bottomrule
    \end{tabular}
    
\end{table}

\paragraph{Implications for detector design and readout architecture.}

These occupancy results reveal that the tracker barrel and endcap remain within occupancy limits without multi-buffer operation. The vertex detector, while most exposed to beam–beam backgrounds, requires only modest local buffer depths, two to three. This could be addressed by employing a dedicated readout with on-pixel memory capable of storing two to three hits locally during the bunch train. This approach would provide sufficient buffering to maintain occupancies below $10^{-4}$ throughout the entire tracking system and across all \CCC scenarios, thus ensuring that all detector hits can be successfully recorded. This is within limits of the buffer depth of four previously found to be necessary to ensure that the maximum layer occupancy for the SiD vertex barrel does not exceed the occupancy threshold under the ILC beam conditions~\cite{Schutz:2018ynd}.

The power-pulsed operation, unique to linear colliders, enables further optimization of the readout architecture. During the 8.3 (16.7) \,ms inter-train period at 120 (60) \,Hz operation, the entire detector content can be read-out without dead time, eliminating the need for complex trigger systems potentially required at circular colliders. Advanced zero-suppression algorithms operating at the front-end level, exploiting the sparse nature of the data in the tracking detectors ( $<$1\% occupancy even in high-background regions), reduce the size of the data to be read-out by several orders of magnitude, bringing them to manageable levels.

Once all hits collected during a bunch train are safely recorded and read-out, discriminating beam-beam background hits from those stemming from hard-scatter processes becomes the next challenge. For \CCCnospace, even with the smaller bunch spacing of the sustainability and high-luminosity scenarios --- 2.63 ns at 250 GeV and 1.75 ns at 550 GeV ---, the low integrated occupancies mean that precise bunch-crossing identification is not critical for background rejection. Rather than investing in sub-ns timing capabilities to separate individual bunch crossings, the detector could employ conventional time-stamping with timing resolution at the several ns level, accumulating hits over the entire bunch train. During offline reconstruction, a time-window selection around the hard-scatter event can then be employed to reject background hits. The resulting architecture—with simplified front-end electronics and reduced power consumption—represents a technical and cost advantage compared to designs requiring precise bunch-by-bunch separation.

It is important to note that the occupancy studies presented in this work are performed at the SimHit level, with energy thresholds and cell sizes applied during post-processing rather than through full detector digitization. A complete digitization chain would include additional effects such as electronic noise, charge collection inefficiencies, electron drift modeling, signal cross-talk, and realistic front-end readout behavior. However, the impact of several of these effects---particularly charge spreading and lateral diffusion---is expected to be small due to the thin MAPS sensors employed (20 \textmu m active thickness in the vertex), which inherently limit carrier diffusion distances. While these effects would modify the absolute occupancy values, they are expected to scale similarly across all beam parameter scenarios. Therefore, the relative comparisons between different \CCC configurations, the identification of critical detector regions, and the buffer depth requirements derived in this study remain valid. The conservative safety factors ($S_D=2$) and cluster size assumptions ($C_D=3$ for pixel detectors) included in our occupancy calculations provide additional margin to account for any remaining unmodeled digitization effects. Ultimately, a quantitative determination of tracking-performance degradation under realistic operational conditions will require full track reconstruction studies with realistic pattern-recognition algorithms, timing resolution models, and noise characteristics across the assumed occupancy scenarios.

\paragraph{Background-specific mitigation strategies.}

The distinct kinematics and temporal profiles of IPC and HPP backgrounds enable targeted mitigation at the reconstruction level. For pair backgrounds, the strategy exploits their characteristic features: low transverse momenta ($<$100\,MeV), forward peaking, and prompt timing synchronized with bunch crossings. Dedicated pair-finding algorithms operating as a first reconstruction pass can identify and flag pair candidates based on impact parameter, momentum, and timing criteria. These flagged hits can be excluded from subsequent pattern recognition, reducing combinatorial background in track finding. Additionally, due to the regular, deterministic nature of pair trajectories, machine learning models trained on the full hit pattern in an event could identify the characteristic spirals of pairs in the magnetic field with small false positive rate.

Hadronic background mitigation requires different techniques suited to their extended temporal profile and broader spatial distribution. In particular, particle flow reconstruction must account for the additional hadronic activity: typically, a few charged tracks plus comparable numbers of photons and neutral hadrons per HPP event. Time-evolution fitting of calorimeter signals could also be used to separate the prompt component from slowly varying hadronic tails, improving jet energy reconstruction and mitigating pile-up-like contamination in precision measurements.

\section{Conclusions}
\label{sec:conclusions}

This study presents the first comprehensive evaluation of beam-beam backgrounds for the Cool Copper Collider, establishing both the methodological framework and quantitative baselines essential for detector design and accelerator optimization. Through the systematic simulation of incoherent pair production and hadron photoproduction processes across multiple operational scenarios, we have demonstrated that \CCCnospace's background environment, while distinct from other proposed Higgs factories, remains well within manageable bounds for precision physics measurements.

This conclusion was specifically validated for the SiD detector concept, which was shown to be compatible with the \CCC operating parameters, and is generally applicable to detectors optimized for the ILC/LCF beam conditions. While the per-bunch-crossing background yields at \CCC are comparable to ILC/LCF, the significantly shorter bunch trains---133 bunches for baseline parameters versus ILC/LCF's 1312 bunches---reduce integrated backgrounds per train by an order of magnitude. This  reduced background, notably, implies that the first vertex barrel layer can remain at the ILC/LCF design radius, preserving the vertexing and impact parameter resolution targets.

The compatibility of \CCC with existing ILC/LCF detector concepts, combined with comparable integrated luminosity targets, implies that \CCC can achieve equivalent precision physics reach as other linear Higgs factory designs. Furthermore, the simulation pipeline developed for this work, built on the \keyhep ecosystem and integrating \GPnospace, \whiznospace/\circenospace, and \geantnospace/\ddhep tools, provides a flexible framework that can also be adapted to circular \ee collider proposals, such as the FCC-ee. By making this infrastructure publicly available, we hope to contribute to a common platform for future collider detector development, accelerating the path toward construction readiness, regardless of the collider design ultimately chosen.

\newpage 

\acknowledgments

The authors express their gratitude to Martin Breidenbach, Chris Damerell, Sergo Jindariani, Kevin Pedro, and Lorenzo Rota for their insightful discussions and comments on earlier versions of this draft. The work of the authors is supported by the U.S. Department of Energy under Contract No. DE-AC02-
76SF00515.

\paragraph{Code availability}
\label{par:code_availability}

The complete analysis framework and all relevant code supporting the results of this study are publicly accessible at \url{https://github.com/dntounis/Beam_Beam_Backgrounds}. The datasets generated for this paper are available from the corresponding author upon reasonable request.

\newpage 

\appendix

\section{Additional information for the SiD detector concept}
\label{app:SiD}

\cref{tab:SiD_parameters} reproduces the key radii, longitudinal extents, and technologies of the various SiD subsystems. Relative to the Detailed Baseline Design~\cite{behnke2013internationallinearcollidertechnical}, the main functional difference is the transition from Resistive Plate Chambers (RPCs) to scintillator tiles read out by Silicon photomultiplier (SiPMs) in the HCAL; all absorber thicknesses and support material fractions follow the updated study of ~\cite{Breidenbach:2021sdo}.

\begin{table}[h]
    \centering

    \caption{Main parameters of the SiD concept. The changes with respect to the baseline SiD design~\cite{behnke2013internationallinearcollidertechnical} are explained in the text.}
    \label{tab:SiD_parameters}

    \resizebox{0.83\textwidth}{!}{
    \begin{tabular}{cccccc}
\toprule  \textbf{Barrel Detector} & Technology & $r_{\text{inner}}$ [cm]  & $r_{\text{outer}}$ [cm] & $z$-range [cm]  \\
\midrule Vertex & Silicon pixels & 1.4 & 6.0 & $\pm 6.25$  \\
Tracker & Silicon {strips} & 21.7 & $122.1 $ & $\pm 152.2$  \\
ECAL & Silicon pixels-Tungsten & 126.5 & $140.9 $ & $\pm 176.5$  \\
HCAL & {Scintillator}-Steel & 141.7 & $249.3$ & $\pm 301.8$ \\
Solenoid & 5 Tesla superconducting & 259.1 & $339.2$  & $\pm 298.3$ \\
Muon system & Scintillator-Steel & 340.2 & $604.2 $ & $\pm 303.3$   \\
\toprule \textbf{Endcap Detector} & Technology & $z_{\text{inner}}$ [cm] & $z_{\text{outer}}$ [cm] & $r_{\text{outer}}$ [cm] \\
\midrule Vertex Endcap & Silicon pixels & 7.6 & 18.0 & 7.1  \\
Vertex Forward & Silicon {pixels} & 21.1 & 83.4 & 16.6  \\
Tracker & Silicon {strips} & 77.0 & 164.3 & 125.5  \\
ECAL & Silicon pixels-Tungsten & 165.7 & 180.0 & 125.0  \\
HCAL & Scintillator-Steel & 180.5 & 302.8 & 140.2 \\
Muon system & Scintillator-Steel & 303.3 & 567.3 & 604.2  \\
LumiCal & Silicon-Tungsten & 155.7 & 170.0 & 20.0 \\
BeamCal & Semiconductor-Tungsten & 277.5 & 300.7 & 13.5 \\
\bottomrule
\end{tabular}
}
\end{table}

\begin{figure}[h!]
    \centering
    \includegraphics[width=0.4\linewidth]
    {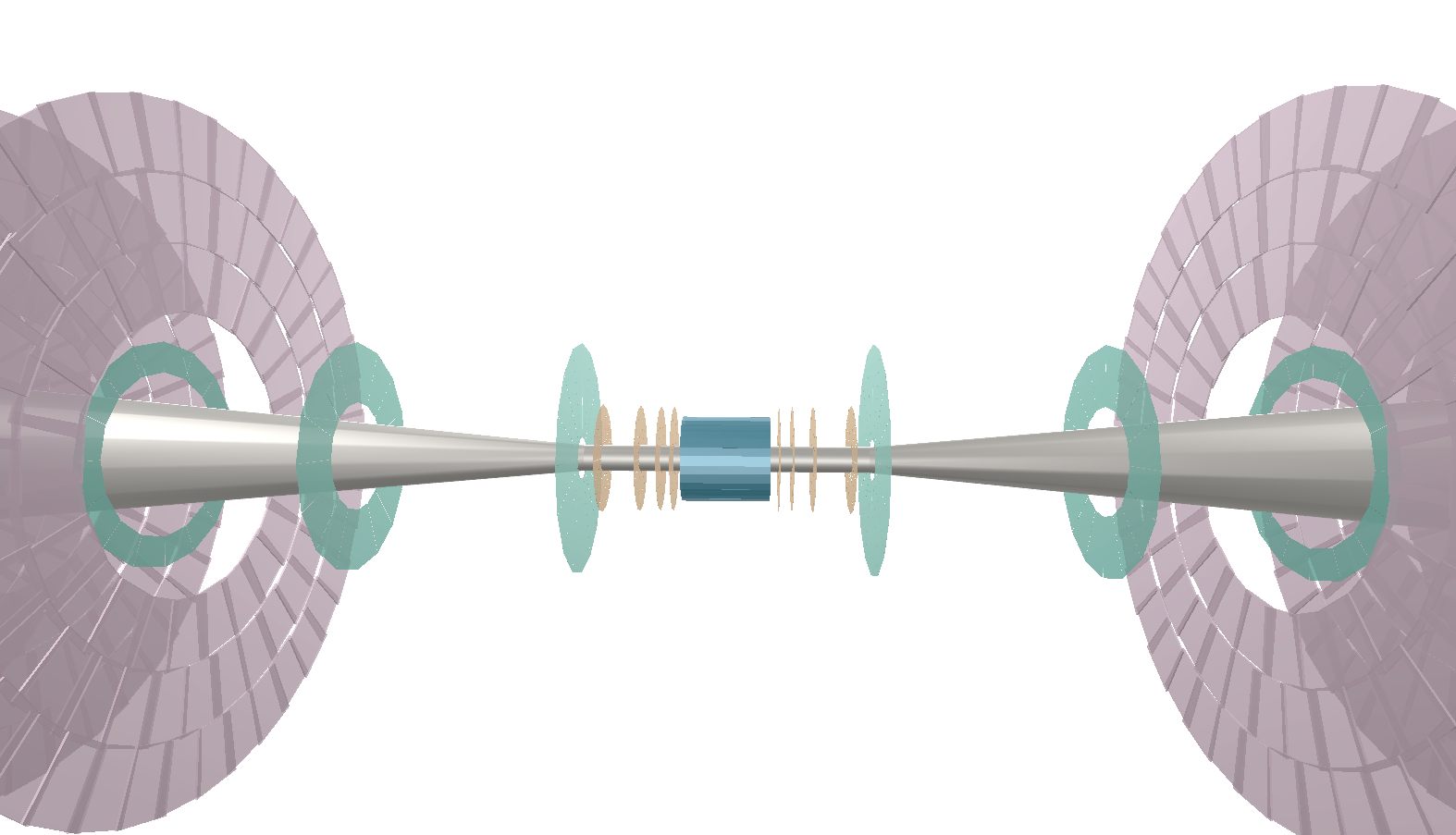}
    \caption{Schematic of the SiD inner detector near the interaction point. The vertex detector is colored in blue, orange, and green for the barrel, endcap, and forward subsystems correspondingly. The first tracker endcap disks are also shown in magenta for reference. 
}
    \label{fig:sid_vertex_layout}
\end{figure}

For the vertex system specifically, SiD employs a barrel-disk geometry with three functionally distinct regions, illustrated in~\cref{fig:sid_vertex_layout}. The five-layer \emph{vertex barrel} provides precise tracking in the central region, while four \emph{vertex endcap} disks at intermediate $z$ ensure hermetic coverage and smooth pattern recognition in the transition to forward angles. Three additional \emph{forward vertex} disks extend coverage to $\cos\theta \approx 0.984$~\cite{behnke2013internationallinearcollidertechnical}, bridging the gap to the tracker. 

\newpage

\section{Reachability condition for IPC particles}
\label{app:reachability}

The reachability criterion of IPC particles to the vertex detector can be derived from the trajectory equation $r(z)$ of a charged relativistic particle in a uniform axial magnetic field $\vec{B} = B_0 \hat{z}$. We start with the relativistic Lorentz force equation:

\vspace*{-0.17cm}
$$\frac{d\vec{p}}{dt} = q(\vec{v} \times \vec{B}) \ ,\vec{p} = \gamma m \vec{v} \Rightarrow \frac{d\vec{v}}{dt}= \vec{v} \times \vec{\omega}_{c} $$

\noindent where $\vec{\omega}_{c} = \frac{q\vec{B}}{\gamma m}$ is the cyclotron frequency and we used the fact that the energy of the particle is conserved under $\vec{B}$, hence the Lorentz factor $\gamma$ is constant.

The above equation can be expressed in cylindrical coordinates $(r, \phi, z)$ and transported to the complex plane by setting $u(t)=\dot{r}+ir\dot{\phi}$, which can be shown to satisfy $u_{\mathrm{T}}(t) \coloneq e^{i\phi}u(t)=u_{\mathrm{T},0}e^{-i\omega_{c}t}=\dot{x}+i\dot{y}$. Expressed in Cartesian coordinates, the equations of motion then turn out to be: 

\vspace*{-0.80cm}
$$x(t) = R [\sin(\omega_c t - \chi) + \sin\chi] + x_0 \ , \ y(t) = R[\cos(\omega_c t - \chi)-\cos{\chi}] + y_0 \ , \ z(t) = v_z t + z_0$$

\vspace*{-0.10cm}

\noindent where $R=p_{\mathrm{T}}/qB$ is the Larmor radius, $p_{T}$ the (conserved) transverse momentum of the particle and $(x_0, y_0, z_0)$ the coordinates of its production vertex, which can be assumed to be well approximated by the location (0,0,0) of the IP. Hence, eliminating time by setting $\omega_c t =\frac{z\tan{\theta}}{R}$, where $\theta$ the angle of the momentum of the particle with respect to the $z$ axis, we obtain the desired trajectory

\vspace*{-0.17cm}
 $$r(z) = 2R\left|\sin\left(\frac{z \tan\theta}{2R}\right)\right|$$

Modeling the first SiD vertex barrel layer as a cylindrical shell of radius $r_{\mathrm{det}}$ and longitudinal extent $|z| \leq z_{\mathrm{max}}$, we deduce that, for a given $p_{\mathrm{T}}$, the minimum value of $\theta$ for which the particle intersects the layer is obtained when  it reaches $r=  r_{\mathrm{det}}$ when $z=\pm |z_{\mathrm{zmax}}|$, and so the reachability boundary is given by

\vspace*{-0.25cm}
\[2R\left|\sin\left(\frac{z_{\mathrm{max}} \tan\theta}{2R}\right)\right|=r_{\mathrm{det}} \Rightarrow \boxed{p_{\mathrm{T}} \left [\sin{\left ( \frac{\tan{\theta}}{p_{\mathrm{T}}}\cdot \frac{qBz_{\mathrm{max}}}{2}\right)} \right ]=\frac{qBr_{\mathrm{det}}}{2}}\]

\begin{figure}[h]
    \centering
    \includegraphics[width=0.43\linewidth]{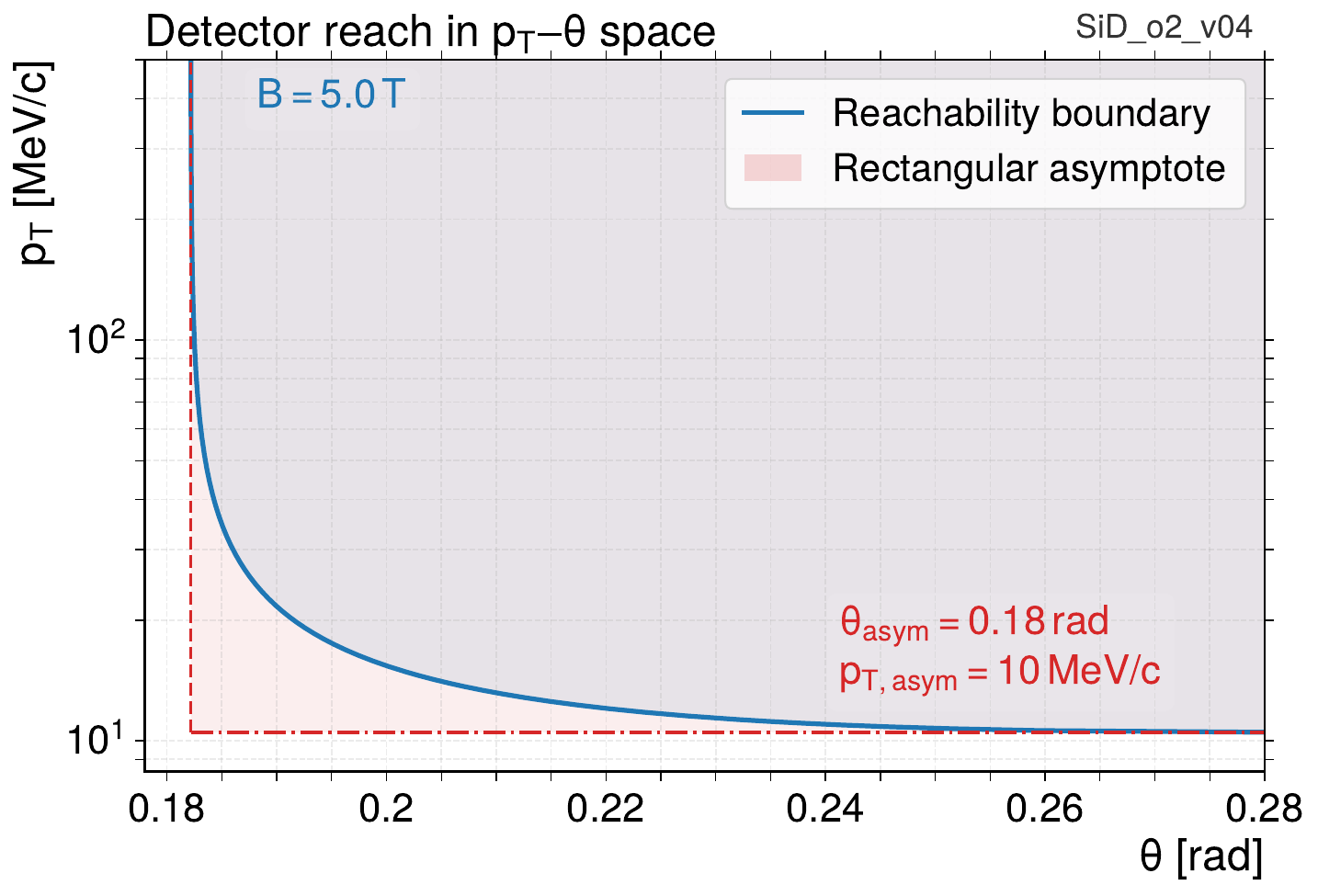}
    \caption{The reachability boundary in $p_{\mathrm{T}}-\theta$ for the SiD innermost vertex barrel layer is given in blue. The corresponding asymptotic values of $p_{\mathrm{T}}$ and $\theta$ are given in red.}
    \label{fig:reachability_boundary}
\end{figure}

The numerical solution of this transcedental equation substituting the SiD values $B=5$ T, $r_{\mathrm{det}} = 14$ mm, and $z_{\mathrm{max}} = 76$ mm is given in~\cref{fig:reachability_boundary} and corresponds  to $p_{\mathrm{T}}\gtrsim 10 \ \mathrm{MeV}$ and $\theta \gtrsim 10.4^{\circ}$.

\vspace*{-0.4cm}

\newpage

\section{Background hit rates}
\label{app_subsec:bkg_hit_rates}

The background hit rates used to derive the hit densities of~\cref{fig:hit_rate_densities} are given in~\cref{tab:hit_rates}.

\begin{table}[h]
    \centering

    \caption{Number of hits per ns from the combined contribution of the HPP and IPC backgrounds, averaged over an entire bunch train in each case. The errors given correspond to the $68\%$ CL uncertainties of the constant value fits.}
    \label{tab:hit_rates}
    
    \resizebox{\textwidth}{!}{
    \begin{tabular}{ccccccc}

    \multicolumn{7}{c}{\textbf{Background Hits/ns}} \\

    \toprule
    & \multicolumn{3}{c}{\CCCtwo} & \multicolumn{3}{c}{\CCCfive}\\
    \cmidrule(r){2-4}\cmidrule(l){5-7}
    Scenario & BL & s.\,u.\ & high‑$\mathcal{L}$ & BL & s.\,u.\ & high‑$\mathcal{L}$\\
    \midrule

\multicolumn{7}{c}{Barrel Detector}    \\
\midrule  
Vertex & 62.2 ± 1.2 & 124.2 ± 1.9  & 158.2 ± 1.4 & 301.4 ± 7.9 & 597.2 ± 10.6 & 600.8 ± 7.1 \\
Tracker &  56.7 ± 0.6 &  112.6 ± 1.2 &  142.9 ± 0.8 & 244.1 ± 2.8 & 490.7 ± 5.8  &  498.8 ± 3.4\\
ECAL  & 31.2 ± 0.5 & 62.5 ± 0.9 &  82.5 ± 0.7 &  133.0 ± 1.7 & 266.3 ± 3.6 & 280.5 ± 2.3  \\ 
HCAL  & 0.7 ± 0.1 & 1.6 ± 0.1 & 3.3 ± 0.1 & 5.4 ± 0.3 & 11.0 ± 0.5 & 16.9 ± 0.5 \\ 
Muon system  & 0.03 ± 0.01 & 0.06 ± 0.01  &  0.12 ± 0.01 & 0.23 ± 0.03 & 0.41 ± 0.04 & 0.50 ± 0.03 \\  
\midrule  \multicolumn{7}{c}{Endcap Detector} \\
\midrule  
Vertex Endcap & 34.4 ± 0.6 & 68.0 ± 1.0  & 84.5 ± 0.8 & 202.1 ± 4.8 & 399.8 ± 8.2 & 407.9 ± 4.5 \\
Vertex Forward &  27.2 ± 0.5 & 54.4 ± 0.9  & 67.7 ± 0.6 & 130.6 ± 2.5 & 260.8 ± 5.2 &  268.1 ± 2.7 \\
Tracker & 42.7 ± 0.7 & 85.2 ± 1.4  &  108.4 ± 0.9 & 227.2 ± 4.3 & 455.4 ± 9.0 & 467.8 ± 4.6 \\
ECAL  & 37.5 ± 1.0  & 74.8 ± 1.3  & 95.7 ± 1.0 &  202.0 ± 4.5  & 401.9 ± 6.7 & 430.5 ± 4.4 \\ 
HCAL  & 1220 ± 42 & 2436 ± 85 & 3290 ± 69 & 6896 ± 240  & 13718 ± 484  & 18488 ± 433 \\ 
Muon system  & 776 ± 20 & 1545 ± 41 & 2149 ± 39 & 5082 ± 111 & 10117 ± 226 & 12168 ± 195 \\  
LumiCal & 315.1 ± 4.2 & 628.8 ± 8.1  & 772.9 ± 5.2  & 1790.3 ± 30.5 & 3574.6 ± 55.0   & 3649.5 ± 28.9 \\  
BeamCal & 10139 ± 200 & 20177 ± 340 & 21670 ± 206 & 84773 ± 2610 & 168471 ± 3750 & 170871 ± 2389 \\  
\bottomrule
\end{tabular}
}

\end{table}

\newpage

\section{Occupancy plots}
\label{app_subsec:Occupancy_plots}

\begin{figure}[htbp]
     \centering
     \begin{subfigure}[b]{0.33\textwidth}
         \centering
         \includegraphics[width=\textwidth]{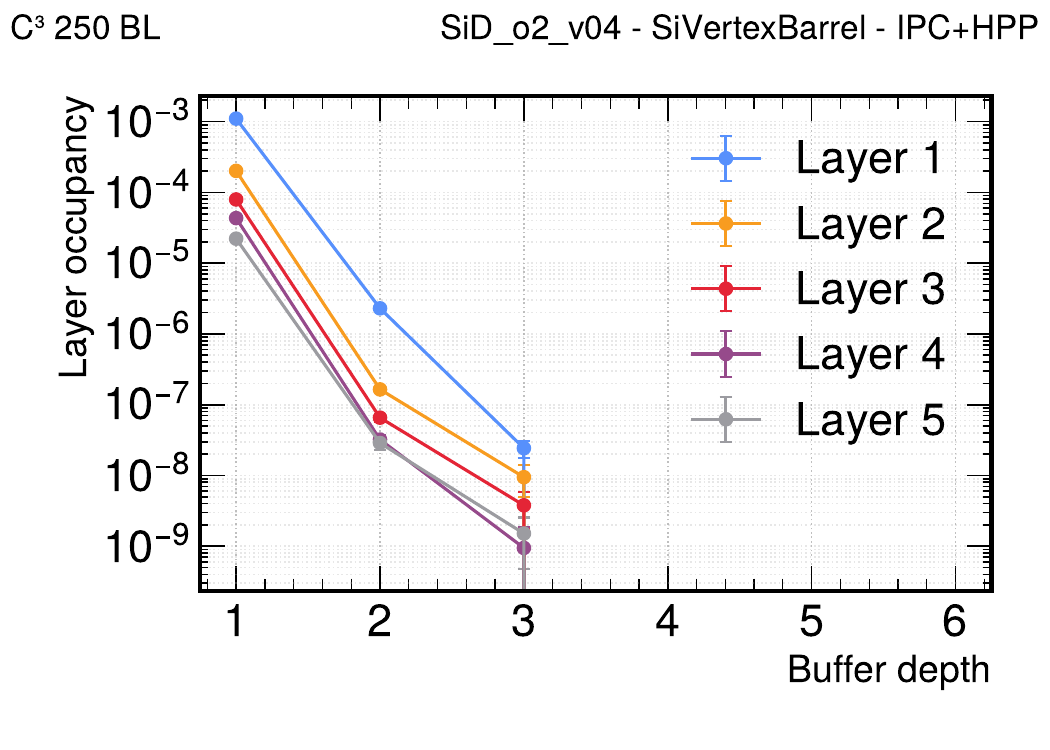}
         \caption{}
     \end{subfigure}%
     ~
     \begin{subfigure}[b]{0.33\textwidth}
         \centering
         \includegraphics[width=\textwidth]{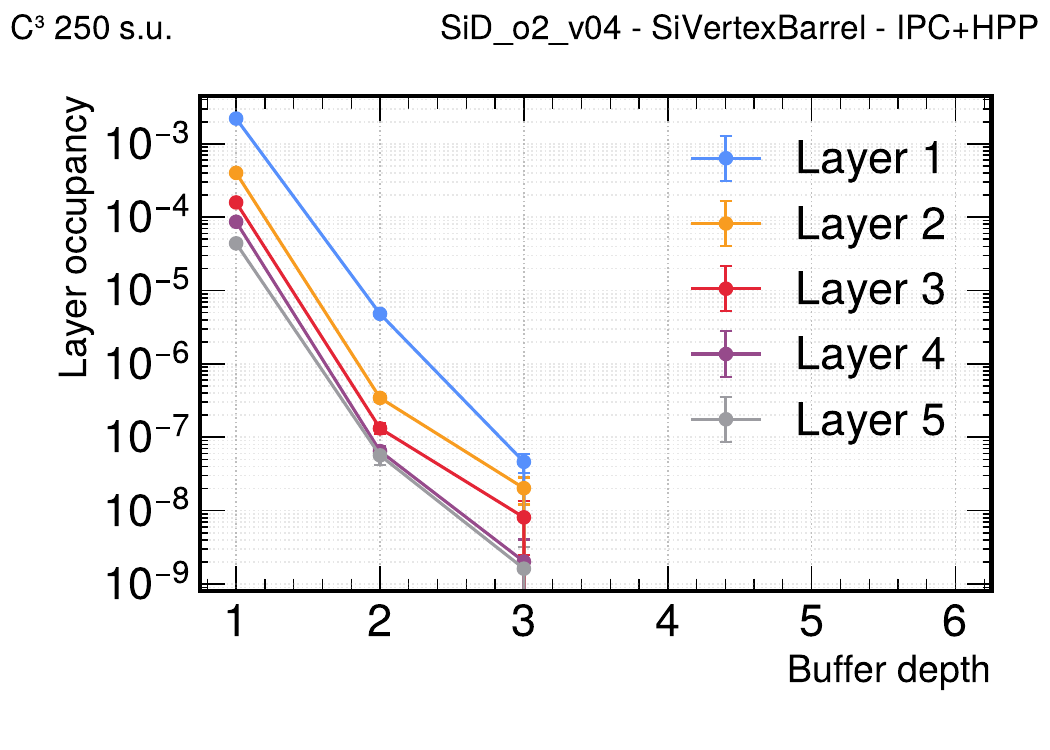}
         \caption{}
     \end{subfigure}%
    ~
     \begin{subfigure}[b]{0.33\textwidth}
         \centering

         \includegraphics[width=\textwidth]{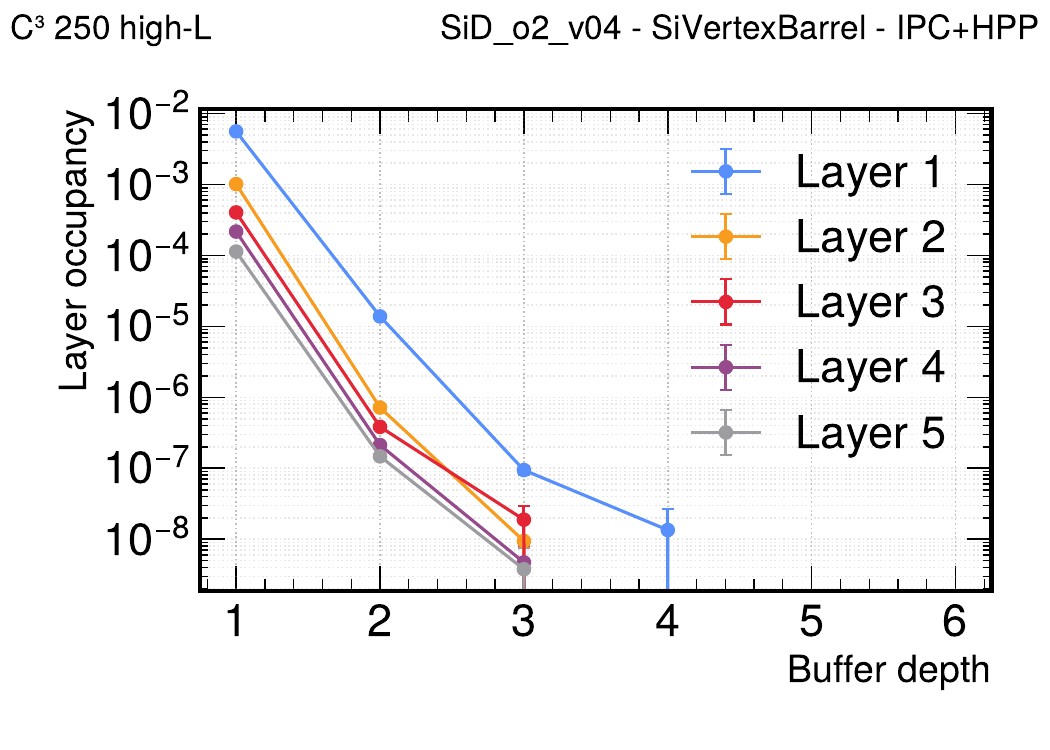}
         \caption{}
     \end{subfigure}
     \hfill
    \begin{subfigure}[b]{0.33\textwidth}
         \centering
         \includegraphics[width=\textwidth]{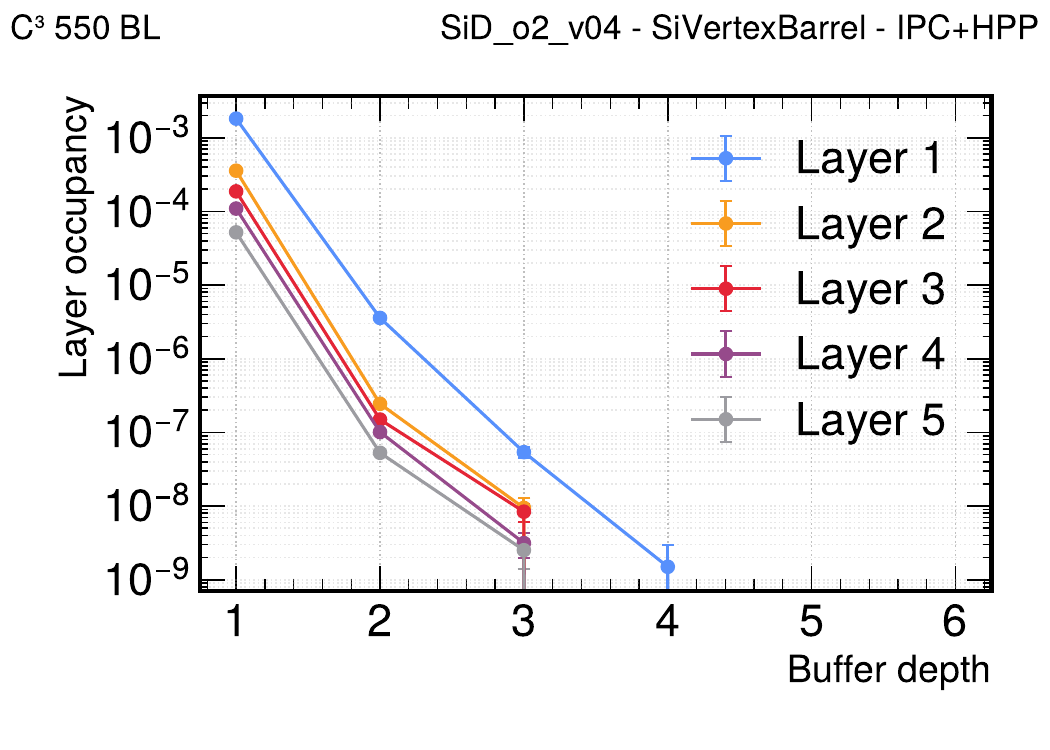}
         \caption{}
     \end{subfigure}%
     ~
     \begin{subfigure}[b]{0.33\textwidth}
         \centering

         \includegraphics[width=\textwidth]{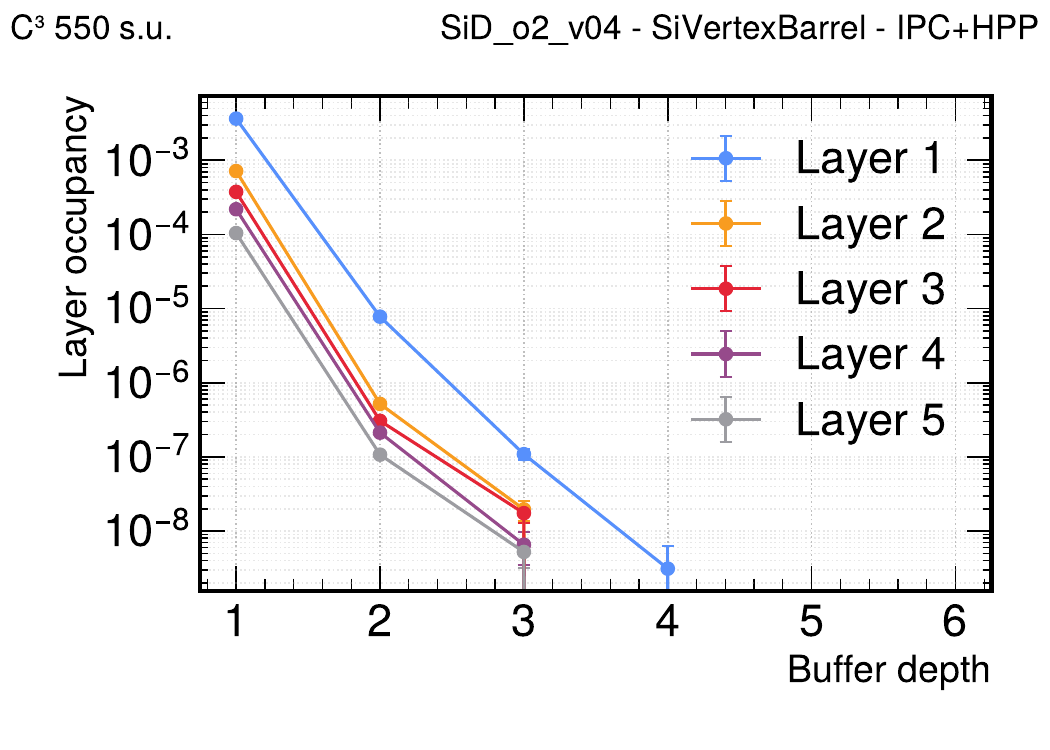}
         \caption{}
     \end{subfigure}%
     ~
    \begin{subfigure}[b]{0.33\textwidth}
         \centering

         \includegraphics[width=\textwidth]{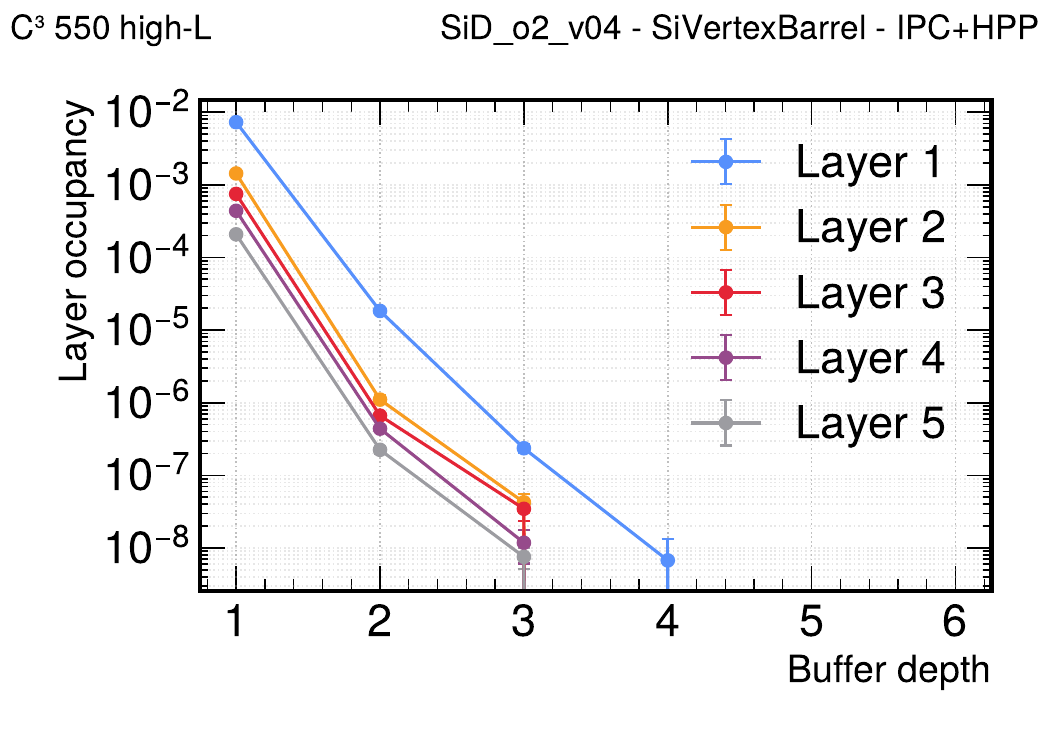}
         \caption{}
     \end{subfigure}
     \hfill
    \begin{subfigure}[b]{0.33\textwidth}
             \centering
             \includegraphics[width=\textwidth]{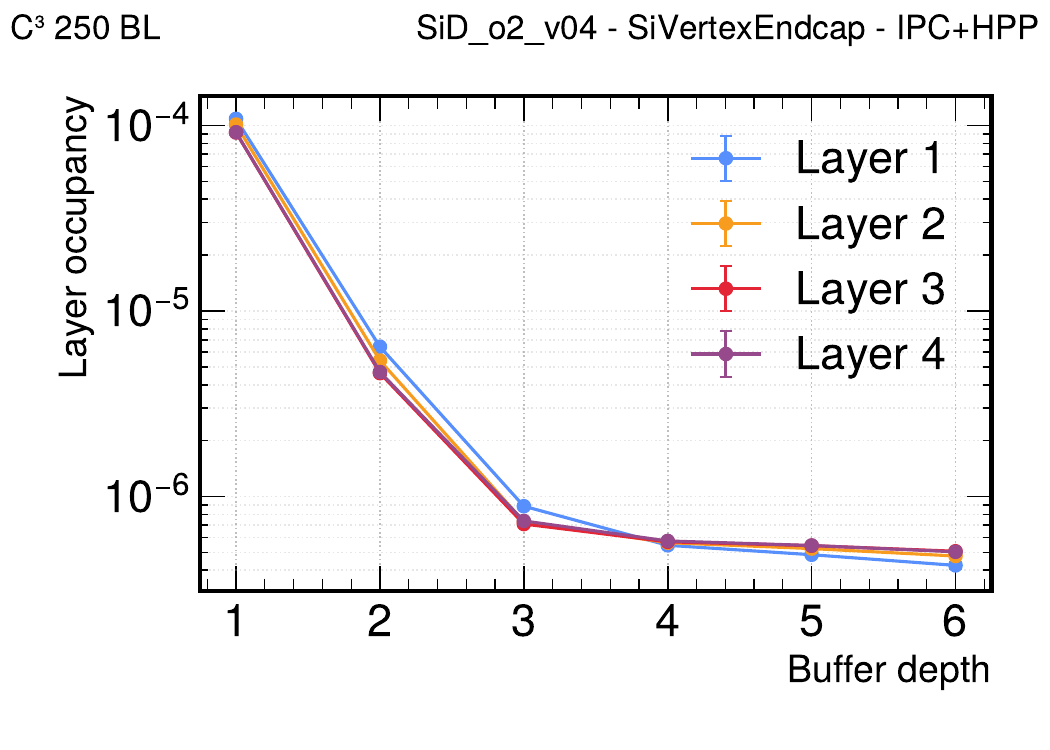}
             \caption{}
         \end{subfigure}%
     ~
     \begin{subfigure}[b]{0.33\textwidth}
         \centering

         \includegraphics[width=\textwidth]{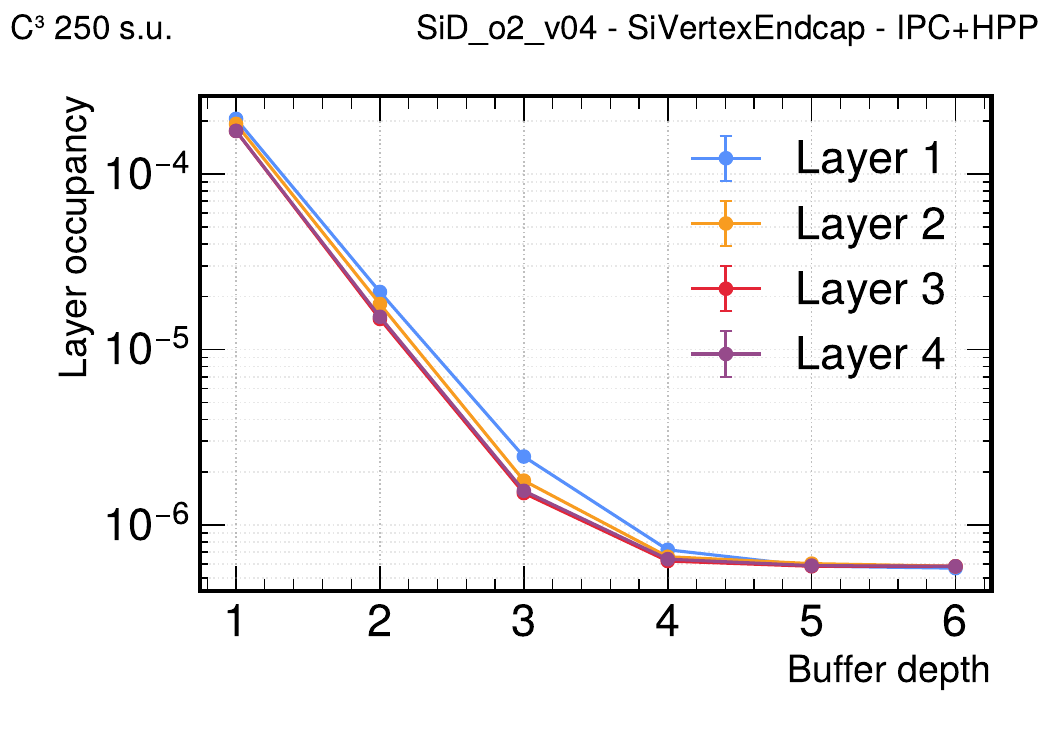}
         \caption{}
     \end{subfigure}%
     ~
    \begin{subfigure}[b]{0.33\textwidth}
         \centering
         \includegraphics[width=\textwidth]{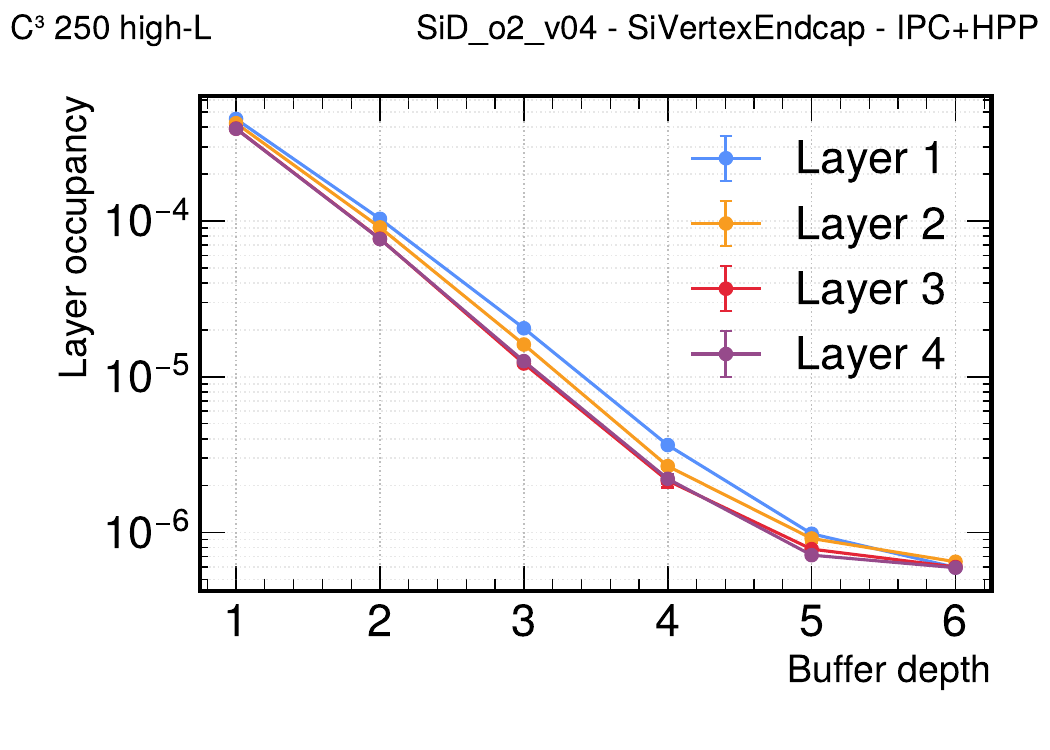}
         \caption{}
     \end{subfigure}
     \hfill
     \begin{subfigure}[b]{0.33\textwidth}
         \centering

         \includegraphics[width=\textwidth]{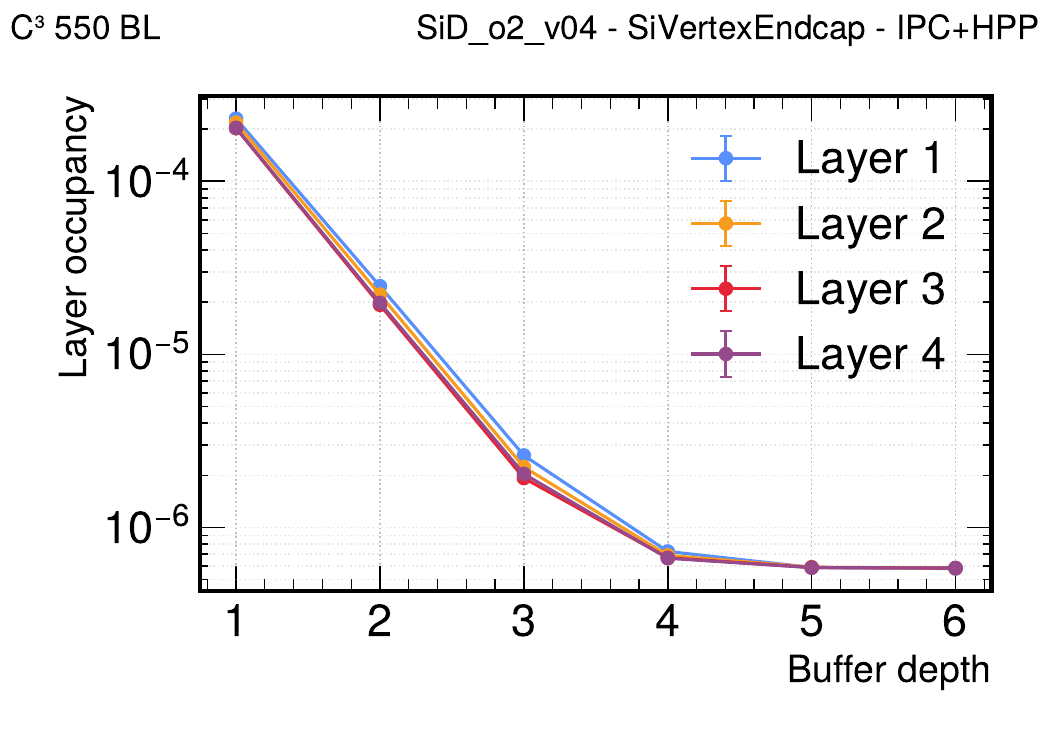}
         \caption{}
     \end{subfigure}%
     ~
    \begin{subfigure}[b]{0.33\textwidth}
         \centering
         \includegraphics[width=\textwidth]{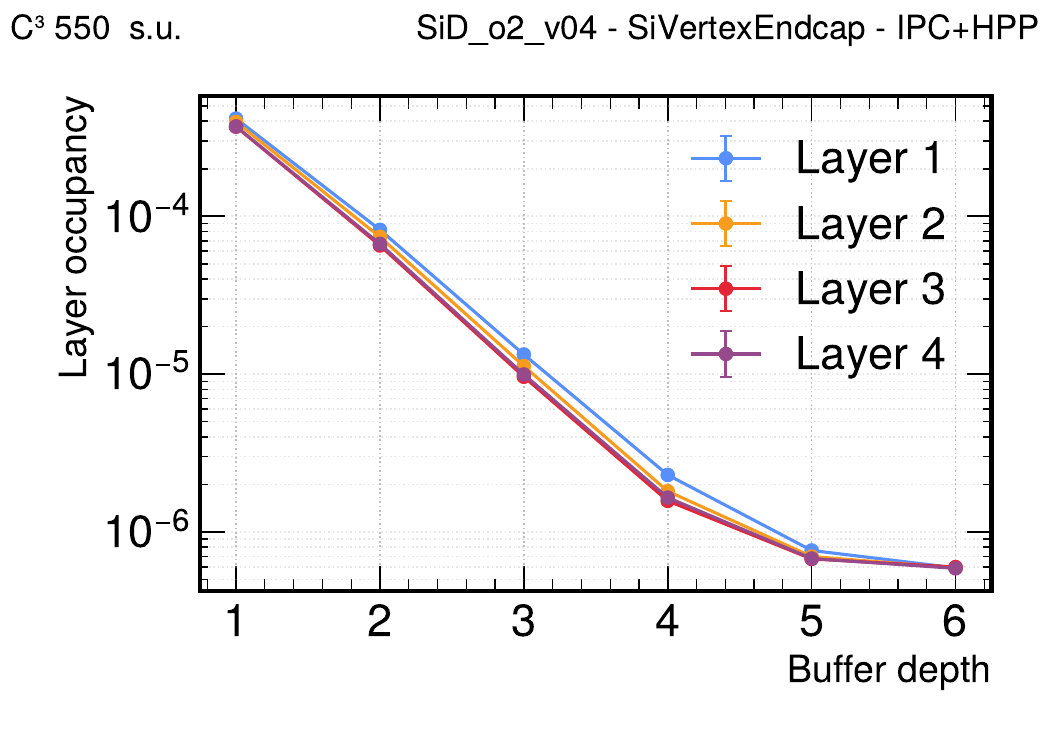}
         \caption{}
     \end{subfigure}%
     ~
     \begin{subfigure}[b]{0.33\textwidth}
         \centering

         \includegraphics[width=\textwidth]{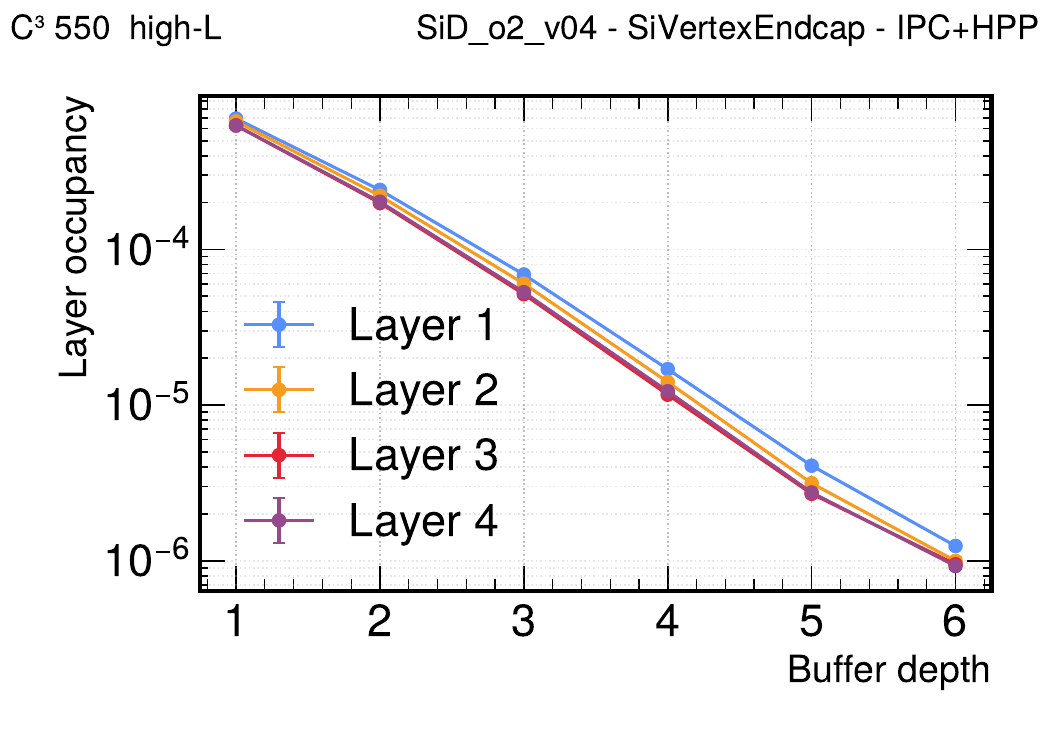}
         \caption{}
     \end{subfigure}
    \caption{Layer occupancy plots for all \CCC operating scenarios and for the vertex: (a)-(f) barrel and (g)-(l) endcap detectors. For more details see~\cref{subsec:occupancy}.}
    \label{fig:appendix_occupancy_plots}
\end{figure}

\newpage
\bibliographystyle{JHEP}
\bibliography{biblio}

\end{document}

%% file: defs.tex
\newcommand*{\GP}{{\textsc{Guinea-Pig}} }

\newcommand*{\GPnospace}{{\textsc{Guinea-Pig}}}
\newcommand*{\GPPPnospace}{{\textsc{Guinea-Pig++}}}

\newcommand*{\circetwo}{{\textsc{Circe} 2} }

\newcommand*{\circenospace}{{\textsc{Circe}}}

\newcommand*{\ee}{\ensuremath{e^{+}e^{-}}\xspace}

\newcommand*{\geant}{\ensuremath{\textsc{Geant4 }}}
\newcommand*{\ddhep}{\ensuremath{\textsc{DD4}\text{hep }}}
\newcommand*{\whizthree}{{\textsc{Whizard} 3.1.5 }}

\newcommand*{\pythia}{{\textsc{Pythia}  }}
\newcommand*{\keyhep}{\ensuremath{\text{Key4hep }}}

\newcommand*{\geantnospace}{\ensuremath{\textsc{Geant4}}}
\newcommand*{\ddhepnospace}{\ensuremath{\textsc{DD4}\text{hep}}}

\newcommand*{\edmhep}{\ensuremath{\textsc{EDM4}\text{hep }}}

\newcommand*{\whizthreenospace}{{\textsc{Whizard} 3.1.5}}
\newcommand*{\whiznospace}{{\textsc{Whizard}}}
\newcommand*{\pythianospace}{{\textsc{Pythia}}}
\newcommand*{\keyhepnospace}{\ensuremath{\text{Key4hep}}}

\newcommand*{\pt}{\ensuremath{p_{\mathrm{T}}} }
\newcommand*{\ptnospace}{\ensuremath{p_{\mathrm{T}}}}

\def\CCC{C$^{3}$~}

\def\CCCnospace{C$^{3}$}

\def\CCCnospace{C$^{3}$}

\def\CCCtwo{C$^{3}$-250 }
\def\CCCfive{C$^{3}$-550 }

%% file: biblio.bib
@article{c3_lumi,
  title = {{Luminosity and Beam-Induced Background Studies for the Cool Copper Collider}},
  author = {Ntounis, Dimitrios and Nanni, Emilio Alessandro and Vernieri, Caterina},
  journal = {Phys. Rev. Accel. Beams},
  volume = {27},
  number = {6},
  pages = {061001},
  year = {2024},
  doi = {10.1103/PhysRevAccelBeams.27.061001},
  eprint = {2403.07093},
  archivePrefix = {arXiv},
  primaryClass = {physics.acc-ph},
  url = {https://link.aps.org/doi/10.1103/PhysRevAccelBeams.27.061001}
}

@article{Sj_strand_2006,
   title={{PYTHIA} 6.4 physics and manual},
   volume={2006},
   ISSN={1029-8479},
   url={http://dx.doi.org/10.1088/1126-6708/2006/05/026},
   DOI={10.1088/1126-6708/2006/05/026},
   number={05},
   journal={Journal of High Energy Physics},
   publisher={Springer Science and Business Media LLC},
   author={Sjöstrand, Torbjörn and Mrenna, Stephen and Skands, Peter},
   year={2006},
   month=may, pages={026–026} }

@article{vernieri2023cool,
  title = {{A "Cool" Route to the Higgs Boson and Beyond: The Cool Copper Collider}},
  author = {Vernieri, Caterina and Nanni, Emilio A and Dasu, Sridhara and Peskin, Michael E and Barklow, Tim and Bartoldus, Rainer and Bhat, Pushpalatha C and Black, Kevin and Brau, James E and Breidenbach, Martin and others},
  journal = {JINST},
  volume = {18},
  number = {7},
  pages = {P07053},
  year = {2023},
  doi = {10.1088/1748-0221/18/07/P07053}
}

@article{nanni2023status,
  title = {{Status and Future Plans for C$^3$ R\&D}},
  author = {Nanni, Emilio A and Breidenbach, Martin and Li, Zenghai and Vernieri, Caterina and Wang, Faya and White, Glen and Bai, Mei and Belomestnykh, Sergey and Bhat, Pushpalatha and Barklow, Tim and others},
  journal = {JINST},
  volume = {18},
  number = {9},
  pages = {P09040},
  year = {2023},
  doi = {10.1088/1748-0221/18/09/P09040}
}

@article{dasu2022strategy,
  title = {{Strategy for Understanding the Higgs Physics: The Cool Copper Collider}},
  author = {Dasu, Sridhara and Nanni, Emilio A and Peskin, Michael E and Vernieri, Caterina and Barklow, Tim and Bartoldus, Rainer and Bhat, Pushpalatha C and Black, Kevin and Brau, James E and Breidenbach, Martin and others},
  journal = {arXiv preprint},
  eprint = {2203.07646},
  archivePrefix = {arXiv},
  primaryClass = {hep-ex},
  year = {2022}
}

@article{ILC_Snowmass,
  author = {Aryshev, Alexander and others},
  collaboration = {ILC International Development Team},
  title = {{The International Linear Collider: Report to Snowmass 2021}},
  journal = {arXiv preprint},
  eprint = {2203.07622},
  archivePrefix = {arXiv},
  primaryClass = {physics.acc-ph},
  year = {2022}
}

@article{C3_ESPPU,
  author = {Andorf, Matthew B. and others},
  title = {{ESPPU INPUT: C$^3$ within the ``Linear Collider Vision''}},
  journal = {arXiv preprint},
  eprint = {2503.20829},
  archivePrefix = {arXiv},
  primaryClass = {physics.acc-ph},
  year = {2025}
}

@phdthesis{Schulte:1996,
  author = {Daniel Schulte},
  title = {Study of Electromagnetic and Hadronic Background in the Interaction Region of the TESLA Collider},
  school = {University of Hamburg},
  year = {1996},
  note = {\href{https://cds.cern.ch/record/331845}{TESLA-97-08}}
}

@article{Yokoya:1991qz,
    author = "Yokoya, Kaoru and Chen, Pisin",
    editor = "Dienes, Margaret and Month, Melvin and Turner, Stuart",
    title = "{{Beam-beam phenomena in linear colliders}}",
    reportNumber = "KEK-PREPRINT-91-2",
    doi = "10.1007/3-540-55250-2_37",
    journal = "Lect. Notes Phys.",
    volume = "400",
    pages = "415--445",
    year = "1992"
}

@article{Rimbault:2007wfy,
    author = "Rimbault, C. and Bambade, P. and Dadoun, O. and Le Meur, G. and Touze, F. and del Alabau, M. C. and Schulte, D.",
    editor = "Petit-Jean-Genaz, C.",
    title = "{GUINEA-PIG++ : An Upgraded Version of the Linear Collider Beam Beam Interaction Simulation Code GUINEA-PIG}",
    reportNumber = "PAC07-THPMN010",
    doi = "10.1109/PAC.2007.4440556",
    journal = "Conf. Proc. C",
    volume = "070625",
    pages = "2728",
    year = "2007"
}

@article{Kilian:2007gr,
  author = {Wolfgang Kilian and Thorsten Ohl and Jurgen Reuter},
  title = {{WHIZARD}: Simulating Multi-Particle Processes at {LHC} and {ILC}},
  journal = {Eur. Phys. J. C},
  volume = {71},
  pages = {1742},
  year = {2011},
  doi = {10.1140/epjc/s10052-011-1742-y},
  eprint = {0708.4233},
  archivePrefix = {arXiv},
  journal = {arXiv preprint},
  primaryClass = {hep-ph}
}

@article{Ohl:1996fi,
    author = "Ohl, Thorsten",
    title = "{CIRCE version 1.0: Beam spectra for simulating linear collider physics}",
    eprint = "hep-ph/9607454",
    archivePrefix = "arXiv",
    reportNumber = "IKDA-96-13",
    doi = "10.1016/S0010-4655(96)00167-1",
    journal = "Comput. Phys. Commun.",
    volume = "101",
    pages = "269--288",
    year = "1997"
}

@article{Agostinelli:2002hh,
  author = {S. Agostinelli and others [GEANT4 Collaboration]},
  title = {{GEANT4}: A Simulation Toolkit},
  journal = {Nucl. Instrum. Meth. A},
  volume = {506},
  pages = {250-303},
  year = {2003},
  doi = {10.1016/S0168-9002(03)01368-8}
}

@article{Frank:2014zya,
  author = {Frank Gaede and others},
  title = {DD4hep: A Detector Description Toolkit for High Energy Physics Experiments},
  journal = {J. Phys. Conf. Ser.},
  volume = {513},
  pages = {022010},
  year = {2014},
  doi = {10.1088/1742-6596/513/2/022010}
}

@article{Sailer:2020fah,
    author = "Sailer, Andr{\'e} and Ganis, Gerardo and Mato, Pere and Petri{\v{c}}, Marko and Stewart, Graeme A.",
    editor = "Doglioni, C. and Kim, D. and Stewart, G. A. and Silvestris, L. and Jackson, P. and Kamleh, W.",
    title = "{Towards a Turnkey Software Stack for HEP Experiments}",
    doi = "10.1051/epjconf/202024510002",
    journal = "EPJ Web Conf.",
    volume = "245",
    pages = "10002",
    year = "2020"
}

@article{behnke2013internationallinearcollidertechnical,
    author = "Abramowicz, Halina and others",
    editor = "Behnke, Ties and Brau, James E. and Burrows, Philip N. and Fuster, Juan and Peskin, Michael and Stanitzki, Marcel and Sugimoto, Yasuhiro and Yamada, Sakue and Yamamoto, Hitoshi",
    title = "{The International Linear Collider Technical Design Report - Volume 4: Detectors}",
    year = "2013",
    eprint = "1306.6329",
    archivePrefix = "arXiv",
    journal = {arXiv preprint},
    primaryClass = "physics.ins-det",
    Number = {},
}

@article{adolphsen2013internationallinearcollidertechnical,
    author = "Adolphsen, Chris and others",
    title = "{The International Linear Collider Technical Design Report - Volume 3.II: Accelerator Baseline Design}",
    year = "2013",
    eprint = "1306.6328",
    journal = {arXiv preprint},
    archivePrefix = "arXiv",
    primaryClass = "physics.acc-ph",
}

@misc{k4geo,
  author       = {Shaojun Lu and
                  Frank Gaede and
                  Andre Sailer and
                  Dan Protopopescu and
                  Daniel Jeans and
                  Nikiforos Nikiforou and
                  Alvaro Tolosa Delgado and
                  TiborILD and
                  Juan Miguel Carceller and
                  Brieuc Francois and
                  Ete Remi and
                  Thomas Madlener and
                  Leonhard Reichenbach and
                  SwathiSasikumar and
                  Mahmoud Ali and
                  mmlynari and
                  Bogdan Pawlik and
                  Giovanni Marchiori and
                  Voutsi and
                  Valentin Volkl and
                  Sungwon Kim and
                  Zhibo Wu and
                  Thorben Quast and
                  Jan Strube and
                  aidanrobson and
                  Emilia Leogrande and
                  Armin Ilg and
                  MarkusFrankATcernch and
                  Rosa Simoniello and
                  Nazar Bartosik},
  title        = {key4hep/k4geo: v00-21},
  month        = oct,
  year         = 2024,
  publisher    = {Zenodo},
  version      = {v00-21},
howpublished         = {\href{https://doi.org/10.5281/zenodo.13896842}{10.5281/zenodo.13896842}}
}

@article{Breidenbach:2021sdo,
    author = "Breidenbach, M. and Brau, J. E. and Burrows, P. and Markiewicz, T. and Stanitzki, M. and Strube, J. and White, A. P.",
    title = "{Updating the SiD Detector concept}",
    eprint = "2110.09965v1",
    journal = {arXiv preprint},
    archivePrefix = "arXiv",
    primaryClass = "physics.ins-det",
    month = "10",
    year = "2021"
}

@article{Aihara:2009ad,
      author={H. Aihara and P. Burrows and M. Oreglia},
    title = "{SiD Letter of Intent}",
    eprint = "0911.0006",
    archivePrefix = "arXiv",
    journal = {arXiv preprint},
    primaryClass = "physics.ins-det",
    reportNumber = "SLAC-R-989, FERMILAB-LOI-2009-01, FERMILAB-PUB-09-681-E",
    month = "11",
    year = "2009"
}

@article{warpx,
    author = "Vay, J. L. and others",
    editor = "Cianchi, Alessandro and Assmann, Ralph and Ferrario, Massimo and Holzer, Bernhard and Nghiem, Phi and Delerue, Nicolas and Schaper, Lucas and Walczak, Roman and Gschwendtner, Edda and Walker, Paul Andreas",
    title = "{Toward the modeling of chains of plasma accelerator stages with WarpX}",
    doi = "10.1088/1742-6596/1596/1/012059",
    journal = "J. Phys. Conf. Ser.",
    volume = "1596",
    number = "1",
    pages = "012059",
    year = "2020"
}

@article{CAIN,
    author = "Chen, P. and Horton-Smith, G. and Ohgaki, T. and Weidemann, A. W. and Yokoya, K.",
    title = "{CAIN: Conglomerat d'ABEL et d'interactions nonlineaires}",
    Number = "SLAC-PUB-6583",
    doi = "10.1016/0168-9002(94)01186-9",
    journal = "Nucl. Instrum. Meth. A",
    volume = "355",
    pages = "107--110",
    year = "1995"
}

@article{Rimbault:2006,
  author    = {C. Rimbault and P. Bambade and K. M{\"o}nig and D. Schulte},
  title     = {Incoherent pair generation in a beam-beam interaction simulation},
  journal   = {Phys. Rev. ST Accel. Beams},
  volume    = {9},
  pages     = {034402},
  year      = {2006},
  doi       = {10.1103/PhysRevSTAB.9.034402},
  url       = {https://journals.aps.org/prab/abstract/10.1103/PhysRevSTAB.9.034402}
}

@article{Esberg:2014,
  author    = {J. Esberg and U. I. Uggerh{\o}j and A. Nystr{\"o}m and P. S{\o}rensen},
  title     = {Strong field processes in beam-beam interactions at the interaction point of future linear colliders},
  journal   = {Phys. Rev. ST Accel. Beams},
  volume    = {17},
  pages     = {051003},
  year      = {2014},
  doi       = {10.1103/PhysRevSTAB.17.051003},
  url       = {https://journals.aps.org/prab/abstract/10.1103/PhysRevSTAB.17.051003}
}

@techreport{Schulte:1999,
  author      = {Daniel Schulte},
  title       = {Beam–Beam Simulations with {GUINEA-PIG}},
  institution = {CERN},
  number      = {CERN-PS-99-014-LP},
  year        = {1999},
  url         = {https://cds.cern.ch/record/382453/files/ps-99-014.pdf}
}

@article{Stienemeier:2021,
      title={WHIZARD 3.0: Status and News}, 
       author    = {Pascal Stienemeier and Simon Bra{\ss} and Pia Bredt and Wolfgang Kilian and Nils Kreher and Thorsten Ohl and J{\"u}rgen Reuter and Vincent Rothe and Tobias Striegl},
      year={2021},
      journal={arXiv preprint},
      eprint={2104.11141},
      archivePrefix={arXiv},
      primaryClass={hep-ph},
      url={https://arxiv.org/abs/2104.11141}, 
}

@article{Bierlich:2022,
  author  = {Christian Bierlich and Smita Chakraborty and Nishita Desai and Ilkka Helenius and Philip Ilten and Leif L{\"o}nnblad and Stephen Mrenna and Stefan Prestel and Peter Skands and Torbj{\"o}rn Sj{\"o}strand and others},
  title   = {A comprehensive guide to the physics and usage of {PYTHIA}~8.3},
  journal = {SciPost Phys. Codebases},
  number  = {8},
  year    = {2022},
  eprint  = {2203.11601},
  archivePrefix = {arXiv},
  url     = {https://pythia.org/pdfdoc/pythia8300.pdf},
  doi     = {10.21468/SciPostPhysCodeb.8}
}

@article{Chen:1993,
  author  = {Pisin Chen and Timothy L. Barklow and Michael E. Peskin},
  title   = {Hadron production in $\gamma\gamma$ collisions as a background for $e^+e^-$ linear colliders},
  journal = {Phys. Rev. D},
  volume  = {49},
  pages   = {3209--3227},
  year    = {1994},
  eprint  = {hep-ph/9305247},
  url     = {https://arxiv.org/abs/hep-ph/9305247},
  doi     = {10.1103/PhysRevD.49.3209}
}

@article{Brau:2022sxr,
    author = "Brau, James E. and Breidenbach, Martin and Habib, Alexandre and Rota, Lorenzo and Vernieri, Caterina",
    title = "{The SiD Digital ECal Based on Monolithic Active Pixel Sensors}",
    doi = "10.3390/instruments6040051",
    journal = "Instruments",
    volume = "6",
    number = "4",
    pages = "51",
    year = "2022"
}

@article{Andronic:2025hcm,
    author = "Andronic, Anton and others",
    title = "{Detection efficiency and spatial resolution of Monolithic Active Pixel Sensors bent to different radii}",
    eprint = "2502.04941",
    archivePrefix = "arXiv",
    journal = {arXiv preprint},
    primaryClass = "physics.ins-det",
    month = "2",
    year = "2025"
}

@article{Rinella:2022htp,
    author = "Rinella, Gianluca Aglieri and others",
    title = "{Digital pixel test structures implemented in a 65 nm CMOS process}",
    eprint = "2212.08621",
    archivePrefix = "arXiv",
    primaryClass = "physics.ins-det",
    doi = "10.1016/j.nima.2023.168589",
    journal = "Nucl. Instrum. Meth. A",
    volume = "1056",
    pages = "168589",
    year = "2023"
}

@article{Tan:2025rss,
    author = "Tan, Wei-Hou and White, Glen and Ntounis, Dimitrios and Li, Zenghai and Kim, Dongsung and Xu, Haoran and Simakov, Evgenya and Nanni, Emilio A.",
    title = "{{Emittance preservation in the C$^3$ main linear accelerator}}",
    doi = "10.1016/j.nima.2025.170660",
    journal = "Nucl. Instrum. Meth. A",
    volume = "1080",
    pages = "170660",
    year = "2025"
}

@article{Ganis:2021vgv,
    author = {Ganis, Gerardo and Helsens, Cl{\'e}ment and V{\"o}lkl, Valentin},
    title = "{Key4hep, a framework for future HEP experiments and its use in FCC}",
    eprint = "2111.09874",
    archivePrefix = "arXiv",
    primaryClass = "hep-ex",
    doi = "10.1140/epjp/s13360-021-02213-1",
    journal = "Eur. Phys. J. Plus",
    volume = "137",
    number = "1",
    pages = "149",
    year = "2022"
}

@article{Gaede:2021izq,
    author = "Gaede, Frank and Ganis, Gerardo and Hegner, Benedikt and Helsens, Clement and Madlener, Thomas and Sailer, Andre and Stewart, Graeme A. and Volkl, Valentin and Wang, Joseph",
    title = "{EDM4hep and podio - The event data model of the Key4hep project and its implementation}",
    doi = "10.1051/epjconf/202125103026",
    journal = "EPJ Web Conf.",
    volume = "251",
    pages = "03026",
    year = "2021"
}

@article{Petric:2017psf,
    author = "Petri{\v{c}}, M. and Frank, M. and Gaede, F. and Lu, S. and Nikiforou, N. and Sailer, A.",
    editor = "Mount, Richard and Tull, Craig",
    title = "{Detector simulations with DD4hep}",
    reportNumber = "CLICdp-Conf-2017-001",
    doi = "10.1088/1742-6596/898/4/042015",
    journal = "J. Phys. Conf. Ser.",
    volume = "898",
    number = "4",
    pages = "042015",
    year = "2017"
}

@misc{k4geo_sid_o2_v04,
  author = "Dan Protopopescu and others",
  title = {{k4geo: SiD\_o2\_v04 detector geometry}},
  howpublished = {\url{https://github.com/key4hep/k4geo/tree/main/SiD/compact/SiD_o2_v04}},
}

@article{Groettvik:2024onw,
    author = "Groettvik, Ola",
    collaboration = "ALICE",
    title = "{ALICE ITS3: a bent stitched MAPS-based vertex detector}",
    eprint = "2401.04629",
    archivePrefix = "arXiv",
    primaryClass = "physics.ins-det",
    doi = "10.1088/1748-0221/19/02/C02048",
    journal = "JINST",
    volume = "19",
    number = "02",
    pages = "C02048",
    year = "2024"
}

@article{Bane:2018fzj,
    author = "Bane, Karl L. and others",
    title = "{An Advanced NCRF Linac Concept for a High Energy e$^+$e$^-$ Linear Collider}",
    eprint = "1807.10195",
    archivePrefix = "arXiv",
    primaryClass = "physics.acc-ph",
    journal = {arXiv preprint},
    month = "7",
    year = "2018"
}

@article{Grudiev2010DesignOT,
      author        = "Grudiev, A and Wuensch, W",
      title={{Design of the CLIC Main Linac Accelerating Structure for CLIC Conceptual Design Report}},
      reportNumber  = "EuCARD-CON-2010-073, CERN-ATS-2010-212",
      year          = "2010",
      journal="\href{https://cds.cern.ch/record/1346987}{CERN-ATS-2010-212}",
      note = {\url{https://cds.cern.ch/record/1346987}}

}

@article{ALICE:2023udb,
    author = "Acharya, Shreyasi and others",
    collaboration = "ALICE",
    title = "{ALICE upgrades during the LHC Long Shutdown 2}",
    eprint = "2302.01238",
    archivePrefix = "arXiv",
    primaryClass = "physics.ins-det",
    reportNumber = "CERN-EP-2023-009",
    doi = "10.1088/1748-0221/19/05/P05062",
    journal = "JINST",
    volume = "19",
    number = "05",
    pages = "P05062",
    year = "2024"
}

@article{Arominski:2704642,
      author        = "Arominski, Dominik and Sailer, Andre and Latina, Andrea",
      title         = "{Beam-induced backgrounds in CLICdet}",
      year          = "2019",
      journal="\href{https://cds.cern.ch/record/2704642}{CLICdp-Note-2019-007}"
}

@article{Boscolo:2023grv,
    author = "Boscolo, Manuela and Ciarma, Andrea",
    title = "{Characterization of the beamstrahlung radiation at the future high-energy circular collider}",
    eprint = "2307.15597",
    archivePrefix = "arXiv",
    primaryClass = "hep-ex",
    doi = "10.1103/PhysRevAccelBeams.26.111002",
    journal = "Phys. Rev. Accel. Beams",
    volume = "26",
    number = "11",
    pages = "111002",
    year = "2023"
}

@phdthesis{Schutz:2018ynd,
    author       = {Schütz, Anne},
    title        = {{O}ptimizing the design of the {F}inal-{F}ocus region for
                      the {I}nternational {L}inear {C}ollider},
    school       = {Karlsruhe Institute of Technology (KIT)},
    year         = {2018},
    note= {\href{https://doi.org/10.5445/IR/1000083323}{10.5445/IR/1000083323}},

}

@article{LinearColliderVision:2025hlt,
    author = "Abramowicz, H. and others",
    collaboration = "Linear Collider Vision",
    title = "{A Linear Collider Vision for the Future of Particle Physics}",
    eprint = "2503.19983",
    archivePrefix = "arXiv",
    journal = {arXiv preprint},
    primaryClass = "hep-ex",
    reportNumber = "FERMILAB-PUB-25-0216-CSAID-TD",
    month = "3",
    year = "2025"
}

@article{LinearCollider:2025lya,
    author = "Subba, A. and others",
    collaboration = "Linear Collider",
    title = "{The Linear Collider Facility (LCF) at CERN}",
    eprint = "2503.24049",
    archivePrefix = "arXiv",
    journal = {arXiv preprint},
    primaryClass = "hep-ex",
    reportNumber = "DESY-25-054, FERMILAB-PUB-25-0239-CSAID",
    month = "3",
    year = "2025"
}
